\documentclass[onecolumn,showpacs,11pt,nofootinbib]{revtex4}
\pdfoutput=1
\usepackage{axodraw}
\usepackage{graphicx}
\usepackage{dcolumn}
\usepackage{bm}
\usepackage{color}
\usepackage{eurosym}
\usepackage{multirow}
\usepackage{amsfonts}
\usepackage{amsmath}
\usepackage{hyperref}
\usepackage{amssymb}
\usepackage[english]{babel}
\usepackage{graphicx}
\usepackage{epsfig}
\usepackage{subfigure}
\usepackage{unicode}
\begin{document}
\newcommand{\hs}{\hspace*{0.2cm}}
\newcommand{\hsp}{\hspace*{0.5cm}}
\newcommand{\vs}{\vspace*{0.5cm}}
\newcommand{\be}{\begin{equation}}
\newcommand{\ee}{\end{equation}}
\newcommand{\bea}{\begin{eqnarray}}
\newcommand{\eea}{\end{eqnarray}}
\newcommand{\ben}{\begin{enumerate}}
\newcommand{\een}{\end{enumerate}}
\newcommand{\bde}{\begin{widetext}}
\newcommand{\ede}{\end{widetext}}
\newcommand{\nn}{\nonumber}
\newcommand{\crn}{\nonumber \\}
\newcommand{\Tr}{\mathrm{Tr}}
\newcommand{\non}{\nonumber}
\newcommand{\noi}{\noindent}
\newcommand{\al}{\alpha}
\newcommand{\la}{\lambda}
\newcommand{\bet}{\beta}
\newcommand{\ga}{\gamma}
\newcommand{\va}{\varphi}
\newcommand{\om}{\omega}
\newcommand{\pa}{\partial}
\newcommand{\+}{\dagger}
\newcommand{\fr}{\frac}
\newcommand{\sq}{\sqrt}
\newcommand{\bc}{\begin{center}}
\newcommand{\ec}{\end{center}}
\newcommand{\Ga}{\Gamma}
\newcommand{\de}{\delta}
\newcommand{\De}{\Delta}
\newcommand{\ep}{\epsilon}
\newcommand{\varep}{\varepsilon}
\newcommand{\ka}{\kappa}
\newcommand{\La}{\Lambda}
\newcommand{\si}{\sigma}
\newcommand{\Si}{\Sigma}
\newcommand{\ta}{\tau}
\newcommand{\up}{\upsilon}
\newcommand{\Up}{\Upsilon}
\newcommand{\ze}{\zeta}
\newcommand{\ps}{\psi}
\newcommand{\Ps}{\Psi}
\newcommand{\ph}{\phi}
\newcommand{\vph}{\varphi}
\newcommand{\Ph}{\Phi}
\newcommand{\Om}{\Omega}
\newcommand{\Vien}[1]{{#1}}
\newcommand{\Long}[1]{{#1}}
\newcommand{\Antonio}[1]{{#1}}
\newcommand{\Revised}[1]{{#1}}
\title{Fermion masses and mixings in a $U(1)_X$ model \\
based on the $\Sigma(18)$ discrete symmetry}
\author{V. V. Vien$^{a,b}$}
\email{vovanvien@tdtu.edu.vn}
\author{A. E. C\'arcamo Hern\'andez$^{c,d,e}$}
\email{antonio.carcamo@usm.cl}
\author{H. N. Long$^{f}$}
\email{hnlong@iop.vast.ac.vn}
\affiliation{$^a$Theoretical Particle Physics and Cosmology Research Group, Advanced Institute of Materials Science,
Ton Duc Thang University, Ho Chi Minh City, Vietnam\\
$^b$Faculty of Applied Sciences, Ton Duc Thang University, Ho Chi Minh City, Vietnam\\
$^c$Departamento de F\'{\i}sica, Universidad T\'{e}cnica Federico Santa Mar\'{\i}a,\\
 Casilla 110-V, Valpara\'{\i}so, Chile\\
$^d$Centro Cient\'{\i}fico-Tecnol\'ogico de Valpara\'{\i}so, Casilla 110-V,
Valpara\'{\i}so, Chile\\
$^e$Millennium Institute for Subatomic Physics at High-Energy Frontier
(SAPHIR), Fern\'andez Concha 700, Santiago, Chile\\
$^f$Institute of Physics, \Revised{Vietnam Academy of Science and Technology,} 10 Dao Tan, Ba Dinh, Hanoi, Vietnam.}
\begin{abstract}
We have built a renormalizable $U(1)_X$ model with a $\Sigma (18)\times Z_4$ symmetry, whose spontaneous breaking yields the observed SM fermion masses and fermionic mixing parameters. The tiny masses of the light active neutrinos are produced by the
 type I seesaw mechanism mediated by very heavy right handed Majorana neutrinos. To the best of our knowledge, this
model is the first implementation of the $\Sigma (18)$ flavor symmetry in a renormalizable $U(1)_X$ model.
Our model allows a successful fit for the SM fermion masses,
fermionic mixing angles and CP phases for both quark and lepton sectors. The obtained values for the physical observables of both quark and lepton sectors are in accordance with the experimental data. We obtain an effective neutrino mass parameter of $\langle m_{ee}\rangle=1.51\times 10^{-3}\, \mathrm{eV}$ for normal ordering and $\langle m_{ee}\rangle =4.88\times 10^{-2} \, \mathrm{eV}$ for inverted ordering which are well consistent with the recent experimental limits on neutrinoless double beta decay.
\end{abstract}
\date{\today}
\maketitle
\section{\label{intro} Introduction}
In recent years neutrino oscillation experiments have confirmed that the leptonic mixing angles and neutrino mass squared differences are measured with high precision which require us to extend the standard model (SM) to successfully explain the current pattern of lepton masses and mixing angles. Among the possible extensions of the SM, the versions with an extra $U(1)_{X}$ gauge symmetry \cite{U1x1, U1x2, U1x3, U1X0, U1X1, U1X2, U1X3, U1X4, U1X5, U1X6, U1X7,U1X8,U1X9,U1X10,U1X11,U1X12,U1X13,U1X14,U1X15,U1X16,U1X17, U1X18, U1X19,U1X20} are promising scenarios since the simplest possibility is to introduce three right-handed neutrinos that we need to incorporate the neutrino masses in the SM. In this type of model, many phenomena including neutrino masses \cite{U1X9, U1X6, U1X7,U1X8}, dark matter \cite{U1X14, U1X11, U1X10, U1X12, U1X13, U1X9, U1X15}, the muon anomalous magnetic moment \cite{U1X16}, inflation \cite{U1X17}, leptogenesis \cite{U1X18, U1X19}, gravitational wave radiation \cite{U1X20} are explained, however, the most minimal versions of $U(1)_{X}$ models do not include a description of SM fermion masses and mixings.

In order to explain the pattern on fermion masses and mixings, many extensions of the SM have been proposed with the inclusion of non-Abelian discrete
groups, which have brought many outstanding advantages, see for instance, $S_3$ \cite{S31,S32,S33,S34,S35,S36,S37, S3DLNV,S3VL,Hernandez:2013hea1,Hernandez:2013hea2,Hernandez:2014vta,Hernandez:2015zeh,
Hernandez:2015hrt,CarcamoHernandez:2016pdu,Hernandez:2015dga,Arbelaez:2016mhg,
CarcamoHernandez:2018vdj,Garces:2018nar,Pramanick:2019oxb,CarcamoHernandez:2020pxw,Garcia-Aguilar:2020vsy}, $T^{'}$
\cite{Tp2,Tp3,Tp4,Tp5,CarcamoHernandez:2019vih, Tpvla2019}, $D_4$ \cite{D41, D42, D43,  D4VL, D4V, D4VLq,CarcamoHernandez:2020ney},
$Q_6$ \cite{Q61, Q62, Q63, Q64, Q65, Q66, Q67, Q68, Q69}, $A_4$ \cite{A401,A41, A42, A43, A44, A45,A46, A47, A48, A49, A410,
A411, A412, A413, A414,
 A415, Ishimori12b, A4Dong2010, A4VL2015, A4Hue}, $Q_8$ \cite{Q8Dev2010}, etc. However, there are substantial differences between our present work and others since in most of previous works the lepton and/or quark masses and mixings are generated (1)  by the texture zero mass matrices \cite{S31, Q8Dev2010},
 (2) via non-renormalizable terms \cite{S32, S33, S34,S35,S36, Hernandez:2013hea1, Hernandez:2013hea2,Hernandez:2014vta,Hernandez:2015zeh,Hernandez:2015hrt,CarcamoHernandez:2016pdu,Hernandez:2015dga,Arbelaez:2016mhg,
 CarcamoHernandez:2018vdj,Garcia-Aguilar:2020vsy,Tp2,Tp3,CarcamoHernandez:2019vih, D42,D43, CarcamoHernandez:2020ney,
 Q62,Q69,A401,A41,A42,A43,A44,A46,A47,A48,A49,
 A411,A412,A413,A414, A415, Ishimori12b, A4Hue}; (3) at loop levels \cite{Pramanick:2019oxb, D41, A43,A410,A413}, and (4) by combining with other gauge
 symmetries and/or supplementing other discrete symmetries \cite{S35, S37,S3DLNV,S3VL,Hernandez:2013hea1,Hernandez:2013hea2,Hernandez:2014vta, Hernandez:2015zeh, Garces:2018nar,
CarcamoHernandez:2020pxw,Tp2,
Tp4,Tp5,Tpvla2019,D4VL, D4V, D4VLq, Q61,A45,A48, A4Dong2010,A4VL2015}.
 The $U(1)_{B-L}$ extension of the SM based on $S_3, D_4$ and $Q_6$ has been studied in Refs. \cite{S3BLGM2019, D4BL2020, Q6BL2020} in
 which the fermion masses and mixings are obtained at the first order of perturbation theory.

In this work, we propose a $\mathrm{U}(1)_{X}$ renormalizable theory based on the $\Sigma (18)$ flavor symmetry, supplemented by the $Z_4$ discrete group capable of reproducing the SM fermion masses and mixings \emph{at tree-level}. We use the $\Sigma(18)$ discrete group, since it is the simplest non-trivial group of the type $\Sigma(2N^2)$ with $N=3$ which is isomorphic to $(Z_3 \times Z^\prime_3)\rtimes Z_2$. The $\Sigma(18)$ discrete group has 18 elements which are divided into nine conjugacy classes and has nine irreducible representations: the six singlets $1_{+0}$,
$1_{+1}$, $1_{+2}$, $1_{-0}$, $1_{-1}$, $1_{-2}$ and the three doublets $2_{10}$, $2_{20}$ and $2_{21}$. Mathematical properties of the $\Sigma (18)$ discrete group are discussed in detail in Ref. \cite{Ishi}. However, for convention, we present briefly the tensor products of $\Si(18)$ in Appendix \ref{Si18CG}. The reason for adding the auxiliary
symmetry $\mathrm{U}(1)_X$ was introduced in Ref. \cite{HeU1X} in another different multiHiggs model based on the $A_4$ discrete symmetry where the global $U(1)_X$ symmetry is softly broken in the scalar potential in order to prevent the appearance of a Goldstone boson; thus, we do not further discuss on this issue here. Let us note that the $\Si(18)$ symmetry has not been considered before in this type of models and to the best of our knowledge the model proposed in this work is the first implementation of the $\Sigma (18)$ flavor symmetry in a renormalizable $U(1)_X$ model\footnote{In this model, fermion masses and mixing angles are generated from renormalizable Yukawa interactions. Non-Abelian discrete groups $S_3, T^{'}, Q_4, D_4, Q_6$ contain one-and two-dimensional representations, however, their singlet/doublet components are combined in different ways. Furthermore, $\Si(18)$ contains three two dimensional representations $2_{10}, 2_{20}, 2_{21}$ where $2^*_{10}=2_{20}$ and $2^*_{20}=2_{10}$ while $2_{21}$ is a real representation together with its tensor products presented in Appendix \ref{Si18CG} make $\Si(18)$ group has some advantages compared to the other discrete groups and our proposed model is completely different from previous works.}.

The layout of the remainder of the paper is as follows. In Section \ref{model} we describe our proposed SM extension by adding the $U(1)_{X}$, $\Si(18)$ and $Z_4$ symmetries and considering an extended scalar sector and right handed Majorana neutrinos. In Section \ref{lepton} we describe the implications of our model in lepton masses and mixings. Section \ref{quark} deals with quark masses and mixings. The implications of our model in $K-\bar{K}$ and $B-\bar{B}$ mixings are discussed in Section \ref{secKKbar}. \Antonio{The consequences of our model in charged lepton flavor violation are analyzed in section \ref{secclfv}.} We conclude in Section \ref{conclusion}. A brief description of the Clebsch Gordan coefficients for the $\Sigma(18)$ group is presented in Appendix \ref{Si18CG}.

\section{The model \label{model}}
The electroweak gauge group of the SM is supplemented by a $\Sigma (18)\times Z_4$ discrete symmetry and a global symmetry $\mathrm{U}(1)_{X}$ where $\psi_{i L}, l_{i R} \, (i=1,2,3)$ and $\varphi, \, \varphi^{'}$ carry $X = 1$ while all other fields have $X = 0$. In addition to the SM model particle content, three right-handed neutrinos ($\nu_{1 R}, \nu_{\al R}$), one $SU(2)_L$ doublet $\phi$ with $X=0$ are assigned as $2_{10}$, two $SU(2)_L$ doublets $\varphi, \varphi^{'}$ with $X=1$, respectively, put in $1_{-0}$ and $2_{20}$ under $\Sigma(18)$ and two $SU(2)_L$ singlets $\chi, \rho$ with $X=0$ respectively put in $2_{10}$ and $1_{+1}$ under $\Sigma(18)$ are introduced. The particle content of the model are summarized in Tables \ref{partcont} and \ref{scalarcont}.

\begin{table}[h]
\caption{\label{partcont}Fermion assignments under the symmetry $SU(2)_L\times U(1)_Y\times U(1)_{X}\times \Si(18)\times Z_4 \equiv \mathrm{G}$. Here $\al=2,3$ and $\beta=1,2$.}
\begin{center}
\begin{tabular}{|c||c|c|c|c|c|c|c|c|c|c|c|c|c|c|c|c|c|c|c|}
\hline
   Fields & $\psi_{1L}$& $\psi_{\alpha L}$  &$l_{1R}$&$l_{\alpha R}$&\,\,$\nu_{1R}$&\,\,$\nu_{\alpha R}$\,\, &$Q_{\beta L}$&$Q_{3 L}$&$u_{\beta R}$&$u_{3 R}$&$d_{\beta R}$&$d_{3 R}$\\ \hline\hline
$\mathrm{SU}(2)_L$ & $2$ & $2$ &$1$&$1$&$1$&$1$ &$2$&$2$&$1$&$1$&$1$&$1$  \\
$\mathrm{U}(1)_Y$  & $-1$& $-1$ &$-2$&$-2$&$0$&$0$ &$\frac{1}{3}$&$\frac{1}{3}$&$\frac{4}{3}$&$\frac{4}{3}$&$-\frac{2}{3}$&$-\frac{2}{3}$\\
$\mathrm{U}(1)_X$ &$1$ & $1$ &$1$&$1$&   $0$&   $0$  &$0$&$0$&$0$&$0$&$0$&$0$ \\
$\Sigma(18)$&$1_{+0}$&  $2_{10}$  &$1_{+1}$&$2_{21}$&$1_{+1}$&$2_{10}$ &$2_{10}$&$1_{+2}$&$2_{20}$&$1_{+1}$&$2_{21}$&$1_{+0}$ \\
$Z_4$&$i$&  $i$  &$i$&$i$&$i$&$i$ &$-i$&$i$&$-i$&$i$&$-i$&$i$ \\\hline
\end{tabular}
\end{center}
\vspace*{-0.3cm}
\end{table}
\begin{table}[h]
\caption{\label{scalarcont}Scalar assignments under $\mathrm{G}$ symmetry.}
\begin{center}
\begin{tabular}{|c||c|c|c|c|c|c|c|c|c|c|c|c|c|c|c|c|c|c|c|}
\hline
   Fields & \,\,$H$\,\,&\,\,$\phi$&\,\,$\phi^{'}$\,\,&\,\,$\varphi$&\,\,$\varphi^{'}$\,\,&\,\,$\chi$ &\,\,$\rho$\\ \hline\hline
$\mathrm{SU}(2)_L$ &$2$  &  $2$ &  $2$  &    $2$     &$2$ & $1$  & $1$  \\
$\mathrm{U}(1)_Y$  &$1$  &$1$  & $1$   & $-1$&$-1$&$0$&$0$   \\
$\mathrm{U}(1)_X$ &$0$  & $0$&$0$ &$1$ & $1$ & $0$& $0$ \\
$\Sigma(18)$&$1_{+2}$  &$2_{10}$&$2_{10}$  & $1_{-0}$ & $2_{20}$&$2_{10}$ &$1_{+1}$ \\
$Z_4$&$1$  &$1$&$-1$  & $1$ & $1$&$-1$ &$-1$ \\\hline
\end{tabular}
\end{center}
\vspace*{-0.3cm}
\end{table}
The charged lepton masses can arise from the couplings of
$\bar{\psi}_{(1,\al) L} l_{(1,\al)R}$ to scalars and the neutrino masses are generated by the couplings of $\bar{\psi}_{(1,\al) L} \nu_{(1,\al) R}$ and $\bar{\nu}^c_{(1,\al) R}\nu_{(1,\al) R}$ to scalars while quarks masses can arise from couplings of $\bar{Q}_{(\beta, 3) L} u_{(\beta,3) R}$ and $\bar{Q}_{(\beta, 3) L} d_{(\beta,3) R}$ to scalars. Under $\mathrm{G}$ symmetry these couplings are summarized in Table \ref{fermioncoups}.
\begin{table}[h]
\vspace*{-0.5cm}
\bc
\caption{\label{fermioncoups} List of couplings which can give masses to the fermions}
\begin{tabular}{|c|c|c|}
\hline Couplings& $\left[SU(2)_L, U(1)_Y, U(1)_{X}, \Sigma(18), Z_4\right]$\\ \hline\hline
$\overline{\psi}_{1L} l_{1R}$&$\left(2, -1, 0, \underline{1}_{+1}, 1\right)$\\
$\overline{\psi}_{1L} l_{\al R}$&$\left(2, -1, 0, \underline{2}_{21}, 1\right)$\\
$\overline{\psi}_{\al L} l_{1 R}$&$\left(2, -1, 0, \underline{2}_{10}, 1\right)$\\
$\overline{\psi}_{\al L} l_{\al R}$&$\left(2, -1, 0, \underline{1}_{+1}\oplus \underline{1}_{-1}\oplus \underline{2}_{20}, 1\right)$\\\hline
\hline
$\overline{\psi}_{1 L} \nu_{1 R}$&$\left(2, 1, -1, \underline{1}_{+1}, 1\right)$\\
$\overline{\psi}_{1 L} \nu_{\al R}$&$\left(2, 1, -1, \underline{2}_{10}, 1\right)$\\
$\overline{\psi}_{\al L} \nu_{1 R}$&$\left(2, 1, -1, \underline{2}_{10}, 1\right)$\\
$\overline{\psi}_{\al L} \nu_{\al R}$&$\left(2, 1, -1, \underline{1}_{+0}\oplus\underline{1}_{-0}\oplus \underline{2}_{21}, 1\right)$\\
\hline\hline
$\overline{\nu}^c_{1 R} \nu_{1 R}$&$\left(1, 0, 0, \underline{1}_{+2}, -1\right)$\\
$\overline{\nu}^c_{1 R} \nu_{\al R}$&$\left(1, 0, 0, \underline{2}_{21}, -1\right)$\\
$\overline{\nu}^c_{\al R} \nu_{1 R}$&$\left(1, 0, 0, \underline{2}_{21}, -1\right)$\\
$\overline{\nu}^c_{\al R} \nu_{\al R}$&$\left(1, 0, 0, \underline{1}_{+1}\oplus \underline{1}_{-1}\oplus \underline{2}_{20}, -1\right)$\\ \hline
 \hline
$\overline{Q}_{\beta L} u_{\beta R}$&$\left(2, 1, 0, \underline{1}_{+2} + \underline{1}_{-2}+\underline{2}_{10}, 1\right)$\\
$\overline{Q}_{\beta L} u_{3 R}$&$\left(2, 1, 0, \underline{2}_{10}, -1\right)$\\
$\overline{Q}_{3 L} u_{\beta R}$&$\left(2, 1, 0, \underline{2}_{10}, -1\right)$\\
$\overline{Q}_{3 L} u_{3 R}$&$\left(2, 1, 0, \underline{1}_{+2}, 1\right)$\\\hline
\hline
$\overline{Q}_{\beta L} d_{\beta R}$&$\left(2, -1, 0, \underline{1}_{+1} + \underline{1}_{-1}+\underline{2}_{20}, 1\right)$\\
$\overline{Q}_{\beta L} d_{3 R}$&$\left(2, -1, 0, \underline{2}_{20},-1\right)$\\
$\overline{Q}_{3 L} d_{\beta R}$&$\left(2, -1, 0, \underline{2}_{20},-1\right)$\\
$\overline{Q}_{3 L} d_{3 R}$&$\left(2, -1, 0, \underline{1}_{+1},1\right)$ \\ \hline
 \end{tabular}
\ec
\vspace*{-0.5cm}
\end{table}
In order to generate all SM fermion masses, we introduce seven 
scalars as shown in Table \ref{scalarcont}
where $H$, $\phi$ and $\phi^{'}$ give the charged-lepton and quarks masses, whereas $\varphi, \, \varphi^{'}$ are responsible for generating the Dirac mass terms and $\chi,\, \rho$ yield the Majorana mass terms. The Yukawa interactions for leptons and quarks invariant under all the symmetries of the model are\footnote{Here, $\widetilde{\phi}, \widetilde{\phi^{'}}$ and $\widetilde{H}$ are respectively the complex conjugate fields of $\phi, \phi^{'}$ and $H$, i.e., $\widetilde{\phi} = i\sigma_2 \phi^*=(\phi^0_2 \hs -\phi^-_1)^T \sim [2,-1,0, \underline{2}_{20}, 1]$, $\widetilde{\phi^{'}}\sim  [2,-1,0, \underline{2}_{20}, -1]$,\, $\widetilde{H} \sim [2,-1,0, \underline{1}_{+1}, 1]$.}:
\bea -\mathcal{L}^{l}_{Y}&=& h_1\bar{\psi}_{1L} H l_{1R} + h_2 \left(\bar{\psi}_{\al L} l_{\al R}\right)_{1_{+1}} H + h_3 \left(\bar{\psi}_{\al L} l_{\al R}\right)_{2_{20}} \phi\crn
 &+&\frac{x_1}{2} \left(\bar{\psi}_{\al L}\nu_{\al  R} \right)_{\underline{1}_{-0}} \varphi+ \frac{x_2}{2}\bar{\psi}_{1L}\left(\varphi^{'} \nu_{\al R}\right)_{\underline{1}_{+0}}
+ \frac{x_3}{2}\left(\bar{\psi}_{\al L}\varphi^{'} \right)_{\underline{1}_{+2}}\nu_{1 R}
\crn
&+&\frac{y_1}{2} (\bar{\nu}^c_{1 R} \nu_{1 R})_{\underline{1}_{+2}}\rho  +\frac{y_2}{2} (\bar{\nu}^c_{\al R}\nu_{\al R})_{\underline{1}_{+1}}\rho^*
+\frac{y_3}{2} (\bar{\nu}^c_{\al R}\nu_{\al R})_{\underline{2}_{20}}\chi +\mathrm{H.c},\label{Lylep0}\\
-\mathcal{L}_q &=&  h_{1u} (\bar{Q}_{\beta L} u_{\beta R})_{\underline{1}_{+2}} \widetilde{H} + h_{2u} (\bar{Q}_{3 L}\widetilde{H})_{\underline{1}_{+0}} u_{3 R} +h_{3u} (\bar{Q}_{\beta L} u_{\beta R})_{\underline{2}_{10}} \widetilde{\phi}\crn
&+& h_{4u} (\bar{Q}_{\beta L} u_{3 R})_{\underline{2}_{10}} \widetilde{\phi^{'}}+h_{5u} (\bar{Q}_{3 L} u_{\beta R})_{\underline{2}_{10}} \widetilde{\phi^{'}}\crn
&+& h_{1d} (\bar{Q}_{\beta L} d_{\beta R})_{\underline{1}_{+1}} H  + h_{2d} (\bar{Q}_{3 L} H)_{\underline{1}_{+0}} d_{3 R}+ h_{3d} (\bar{Q}_{\beta L} d_{\beta R})_{\underline{2}_{20}} \phi \crn
&+& h_{4d} (\bar{Q}_{\beta L} d_{3 R})_{\underline{2}_{20}} \phi^{'}+h_{5d} (\bar{Q}_{3 L} d_{\beta R})_{\underline{2}_{20}} \phi^{'} +\mathrm{H.c.} \label{Lquark0}\eea

It is important to note that the $U(1)_X$ and $\Si(18)$ symmetries forbid some Yukawa interactions thus giving rise to the desired textures for the lepton and quark sectors as shown in Eqs. (\ref{Mclep}), (\ref{MDR}) and (\ref{Mud}) and this is an interesting feature of these symmetries. For instance, for the known scalars in Table \ref{scalarcont}, in the charged lepton sector, the following interactions $(\overline{\psi}_{1L} l_{1R}) \phi $, $(\overline{\psi}_{1L} l_{\al R}) \phi $, $(\overline{\psi}_{\al L} l_{1R}) \phi$, $(\overline{\psi}_{1L} l_{\al R}) H $, $(\overline{\psi}_{\al L} l_{1R}) H $ are forbidden by the $\Si(18)$ symmetry; in the neutrino sector, the following interactions $(\overline{\psi}_{1L} \nu_{1R}) \varphi$, $(\overline{\psi}_{1L} \nu_{1R}) \varphi^{'}$, $(\overline{\psi}_{1L} \nu_{\al R}) \varphi$, $(\overline{\psi}_{\al L} \nu_{1 R}) \varphi$  and $(\overline{\psi}_{\al L} \nu_{\al R}) \varphi^{'}$, $(\nu^c_{1 R} \nu_{1 R})\chi,\, (\nu^c_{1 R} \nu_{\al R})\chi,\, (\nu^c_{\al R} \nu_{1 R})\chi,\, (\nu^c_{1 R} \nu_{\al R})\rho$ and $(\nu^c_{\al R} \nu_{1 R})\rho$ are forbidden by the $\Si(18)$ symmetry; in quark sector,  $(\bar{Q}_{\beta L} u_{3 R})\widetilde{H}$, $(\bar{Q}_{3 L} u_{\beta R})\widetilde{H}$, $(\bar{Q}_{\beta L} u_{3 R})\widetilde{\phi}$, $(\bar{Q}_{3 L} u_{\beta R})\widetilde{\phi}$, $(\bar{Q}_{\beta L} d_{3 R})H$, $(\bar{Q}_{3 L} d_{\beta R})H$, $(\bar{Q}_{\beta L} d_{3 R})\phi$, $(\bar{Q}_{3 L} d_{\beta R})\phi$ are prevented by the $\Si(18)$ symmetry, whereas the following interactions $(\overline{\psi}_{\al L} l_{\al R}) \widetilde{\varphi}^{'} $,  $(\overline{\psi}_{\al L} \nu_{\al R}) \widetilde{\phi}$, $(\overline{\psi}_{1 L} \nu_{\al R}) \widetilde{\phi}$, $(\overline{\psi}_{\al L} \nu_{1 R}) \widetilde{\phi}$, $(\bar{Q}_{\beta L} u_{3 R})\varphi^{'}$ and $(\bar{Q}_{\beta L} d_{3 R})\widetilde{\varphi}^{'}$ are prevented by the $U(1)_X$ symmetry.

In order to generate the observed pattern of SM fermion masses and mixing angles, from the potential minimization condition, we consider the following VEV configuration for the scalar fields:
\bea
\langle H \rangle &=& \left(%
\begin{array}{c}
 0  \\
 v_H  \\
\end{array}%
\right),\hs \langle \phi \rangle = (\langle \phi_1 \rangle \hs\hs \langle \phi_2 \rangle), \hs \langle \phi_i \rangle = \left(%
\begin{array}{c}
 0  \\
 v_{i}  \\
\end{array}%
\right)\hs (i=1,2), \crn
 \langle \phi^{'} \rangle &=& (\langle \phi^{'}_1 \rangle \hs\hs \langle \phi^{'}_1 \rangle), \hs \langle \phi^{'}_1 \rangle = \left(%
\begin{array}{c}
 0  \\
 v^{'}  \\
\end{array}%
\right), \hs \langle \varphi \rangle = \left(%
\begin{array}{c}
 v_\varphi  \\
 0  \\
\end{array}%
\right), \crn
\langle \varphi^{'} \rangle &=& (\langle \varphi^{'}_1 \rangle \hs\hs \langle \varphi^{'}_1 \rangle),\hs \langle  \varphi^{'}_1 \rangle = \left(%
\begin{array}{c}
 v_{\varphi^{'}}  \\
 0  \\
\end{array}%
\right), \,\,
 \langle \chi \rangle = (0 \hs\hs \langle \chi_2 \rangle), \hs \langle \chi_2 \rangle= v_\chi,\hs \langle \rho \rangle =v_\rho. \label{scalarvev}
\eea
In order to proof that the scalar fields with the VEV alignments as chosen in Eq. (\ref{scalarvev}) is obtained from the
minimization condition of $\mathcal{V}_{\mathrm{total}}$ in Appendix \ref{potential}, let us put \bea
&&v_{\phi^'_2}=v_{\phi^'_1}=v^',\, v_{\varphi^'_2}=v_{\varphi^'_1}=v_{\varphi^'}, \, v_{\chi_1}=0, \, v_{\chi_2}=v_{\chi},\\
&& v^*_{H}=v_{H}, \,v^*_1=v_1, \, v^*_2=v_2,\, v^{'*}=v^{'},\, v^*_{\varphi}=v_{\varphi},\, v^*_{\varphi^{'}}=v_{\varphi^'},\, v^*_{\chi}=v_{\chi},\, v^*_{\rho}=v_{\rho}, \eea
 which leads to
  \bea
 &&\frac{\partial \mathcal{V}_{\mathrm{total}}}{\partial v^*_j} =\frac{\partial \mathcal{V}_{\mathrm{total}}}{\partial v_j}, \hspace{0.1 cm}
\frac{\partial^2 \mathcal{V}_{\mathrm{total}}}{\partial v^{*2}_j}=\frac{\partial^2 \mathcal{V}_{\mathrm{total}}}{\partial v^2_j} \,\, (v_j=v_H, v_1, v_2, v^', v_\varphi, v_{\varphi^'}, v_\chi, v_\rho), \label{eqv1}
 \eea
and  the minimization condition of $\mathcal{V}_{\mathrm{total}}$ become
\bea
\frac{\partial \mathcal{V}_{\mathrm{total}}}{\partial v_j} &=&0,\hs
\frac{\partial^2 \mathcal{V}_{\mathrm{total}}}{\partial v^2_j} >0 \,\, (v_j=v_H, v_1, v_2, v^', v_\varphi, v_{\varphi^'}, v_\chi, v_\rho). \label{conditionv}
\eea
Furthermore, for simplicity and without loss of generality, we consider the \Vien{following} benchmark point of the Yukawa coupling\Vien{s:}
\Vien{\bea
&&\lambda^{\phi}_1 =\lambda^{\phi}_2 =\lambda^{\phi}_3=\lambda^{\phi}, \hspace{0.1 cm} \lambda^{\phi^'}_1 =\lambda^{\phi^'}_2 =\lambda^{\phi^'}_3=\lambda^{\phi^'}, \hspace{0.1 cm} \lambda_1^{\varphi^'} =\lambda_2^{\varphi^'} =\lambda_3^{\varphi^'} = \lambda^{\varphi^'}, \\
&&
\lambda_1^{\chi} = \lambda_2^{\chi} =\lambda_3^{\chi} = \lambda^{\chi}, \hspace{0.05 cm}
\lambda_1^{H\phi} = \lambda_2^{H\phi} = \lambda_3^{H\phi} = \lambda_4^{H\phi} =\lambda^{H\phi},  \\
&&\lambda_1^{H\phi^'} = \lambda_2^{H\phi^'} = \lambda_3^{H\phi^'} = \lambda_4^{H\phi^'} =\lambda^{H\phi^'}, \hspace{0.05 cm}
\lambda_1^{H\varphi} =\lambda_2^{H\varphi} =\lambda^{H\varphi},  \\
&&
\lambda_1^{H\varphi^'} =\lambda_2^{H\varphi^'} = \lambda^{H\varphi^'},\hspace{0.1 cm}
\lambda_1^{H\chi} = \lambda_2^{H\chi} = \lambda^{H\chi}, \hspace{0.1 cm}\lambda_1^{H\rho} =\lambda_2^{H\rho} =\lambda^{H\rho}, \\
&&
\lambda_1^{\phi\phi^'} =
\lambda_2^{\phi\phi^'} =\lambda_3^{\phi\phi^'} =\lambda_4^{\phi\phi^'}= \lambda_5^{\phi\phi^'} =\lambda_6^{\phi\phi^'} =\lambda^{\phi\phi^'}, \hspace{0.1 cm}
\lambda_1^{\phi\varphi} =\lambda_2^{\phi\varphi} = \lambda^{\phi\varphi},  \\
 &&\lambda_1^{\phi\varphi^'}=\lambda_2^{\phi\varphi^'}=\lambda_3^{\phi\varphi^'} =\lambda_4^{\phi\varphi^'}
 =\lambda_5^{\phi\varphi^'} =\lambda_6^{\phi\varphi^'}=\lambda^{\phi\varphi^'},\\
&&
\lambda_1^{\phi\chi}=\lambda_2^{\phi\chi}= \lambda_3^{\phi\chi} = \lambda_4^{\phi\chi} =\lambda_5^{\phi\chi}=\lambda_6^{\phi\chi}= \lambda^{\phi\chi}, \hspace{0.1 cm}\lambda_1^{\phi\rho} =\lambda_2^{\phi\rho} =\lambda^{\phi\rho},  \\
&&
\lambda_1^{\phi^'\varphi} =\lambda_2^{\phi^'\varphi} =\lambda^{\phi^'\varphi}, \hspace{0.1 cm}
\lambda_1^{\phi^'\varphi^'} =\lambda_2^{\phi^'\varphi^'} =\lambda_3^{\phi^'\varphi^'}=\lambda_4^{\phi^'\varphi^'}=\lambda_5^{\phi^'\varphi^'}=\lambda_6^{\phi^'\varphi^'}=\lambda^{\phi^'\varphi^'},  \\
&&
\lambda_1^{\phi^'\chi}=\lambda_2^{\phi^'\chi}=\lambda_3^{\phi^'\chi}=\lambda_4^{\phi^'\chi}
=\lambda_5^{\phi^'\chi}=\lambda_6^{\phi^'\chi}=\lambda^{\phi^'\chi}, \hspace{0.1 cm}
\lambda_1^{\phi^'\rho}=\lambda_2^{\phi^'\rho} = \lambda^{\phi^'\rho}, \\
&&
\lambda_1^{\varphi\varphi^'}=\lambda_2^{\varphi\varphi^'}= \lambda^{\varphi\varphi^'}, \hspace{0.1 cm}
 \lambda_1^{\varphi\chi}=\lambda_2^{\varphi\chi}=\lambda^{\varphi\chi}, \hspace{0.1 cm}
\lambda_1^{\varphi\rho}=\lambda_2^{\varphi\rho}=\lambda^{\varphi\rho},  \\
&& \lambda_1^{\varphi^'\chi}=\lambda_2^{\varphi^'\chi}=\lambda_3^{\varphi^'\chi}
=\lambda_4^{\varphi^'\chi}=\lambda_5^{\varphi^'\chi}=\lambda_6^{\varphi^'\chi}=\lambda^{\varphi^'\chi}, \\
&&
\lambda_1^{\varphi^'\rho}\lambda_2^{\varphi^'\rho}=\lambda^{\varphi^'\rho}, \hspace{0.1 cm}
\lambda_1^{\chi\rho}=\lambda_2^{\chi\rho}=\lambda^{\chi\rho}, \hspace{0.1 cm}
\lambda_1^{H\phi\phi^'}=\lambda_2^{H\phi\phi^'}=\lambda^{H\phi\phi^'}, \\
&&
\lambda_1^{H\phi\varphi^'}=\lambda_2^{H\phi\varphi^'}=\lambda^{H\phi\varphi^'}, \hspace{0.1 cm}
\lambda_1^{H\phi\chi}=\lambda_2^{H\phi\chi}=\lambda^{H\phi\chi}, \hspace{0.1 cm}
\lambda_1^{\varphi\varphi^'\chi\rho} = \lambda_2^{\varphi\varphi^'\chi\rho}=\lambda^{\varphi\varphi^'\chi\rho}.
\eea}The expressions of the scalar potential minimum equations in Eq. (\ref{conditionv}) thus reduce to the expressions in Appendix \ref{minimumeq} \Vien{in which t}he system of Eqs. (\ref{eq1})--(\ref{eq8}) always have the solution
\bea
\lambda^{H}&=&\frac{\beta_H}{2 (v_1 - v_2) (2 v_1^2 + v_1 v_2 + 2 v_2^2) v_H^4}, \hspace{0.05 cm}
\lambda^{\phi}=\frac{\beta_{\phi}}{2 (v_1 - v_2) (2 v_1^2 + v_1 v_2 + 2 v_2^2)}, \label{sollaHandphi}\\
 \lambda^{\phi^'}&=&-\frac{\beta_{\phi^'}}{12v^{'2}},\hs
 \lambda^{\varphi}=-\frac{\beta_{\varphi}}{2v^{3}_\varphi},\hs
  \lambda^{\varphi^'} = -\frac{\beta_{\varphi^'}}{24v^{3}_{\varphi^'}},\hs
   \lambda^{\chi} = -\frac{\beta_{\chi}}{4v^{3}_{\chi}},\hs  \lambda^{\rho} = -\frac{\beta_{\rho}}{v^{3}_{\rho}},\\
   \lambda^{H\phi}&=&\frac{\beta_{H\phi}}{2 (2 v_1^4 + v_1^3 v_2 - v_1 v_2^3 - 2 v_2^4) v_H^2}, \label{sollaHphi}
 \eea
 where \Vien{$\beta_\Phi \, (\Phi=H, \phi, \phi^', \varphi, \varphi^', \chi, \rho)$ }
 and $\beta_{H\phi}$ are defined in Appendix \ref{betaexpression}.

\begin{figure}[ht]
\begin{center}
\vspace*{-1.0cm}
\hspace*{-0.25cm}\includegraphics[width=0.825\textwidth]{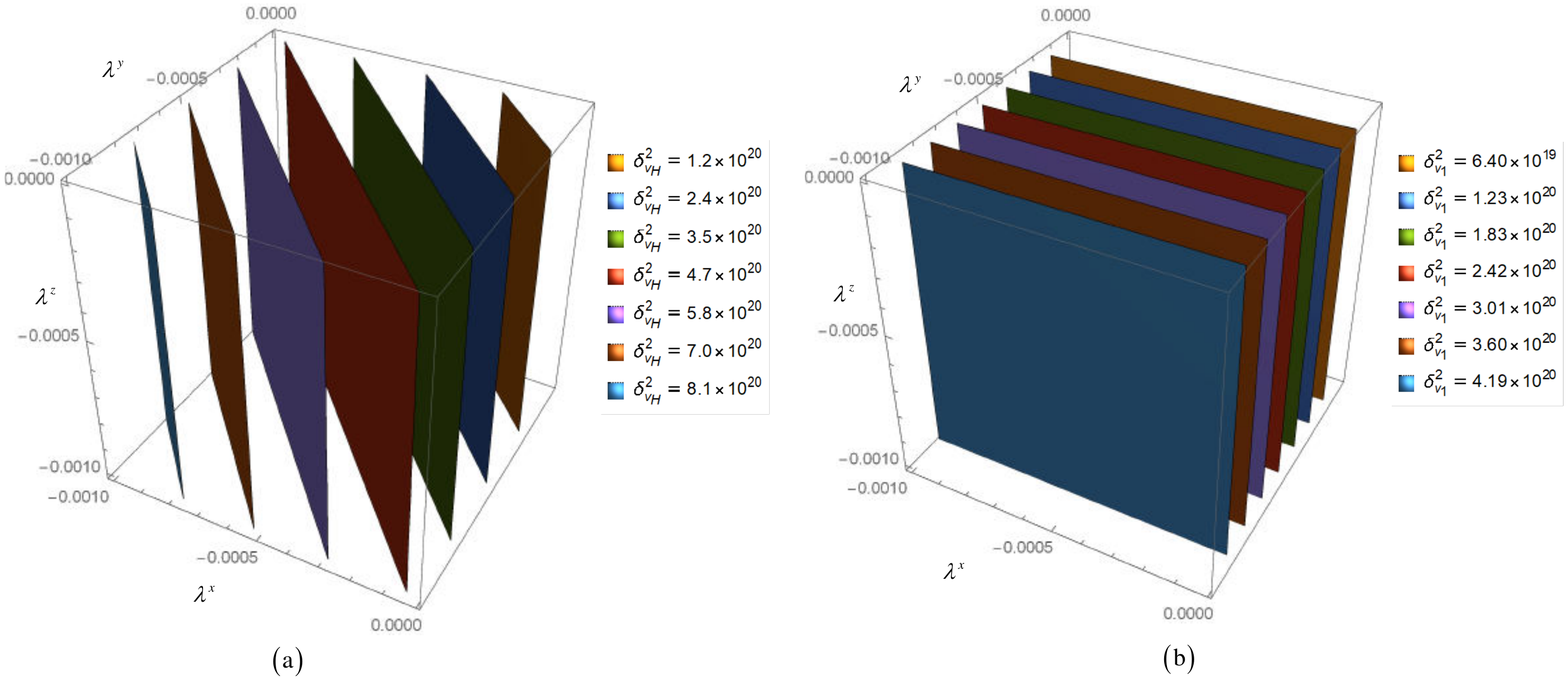}\hspace*{-2.0cm}\\
\vspace*{-4.5cm}
\hspace*{-0.5cm}\includegraphics[width=0.825\textwidth]{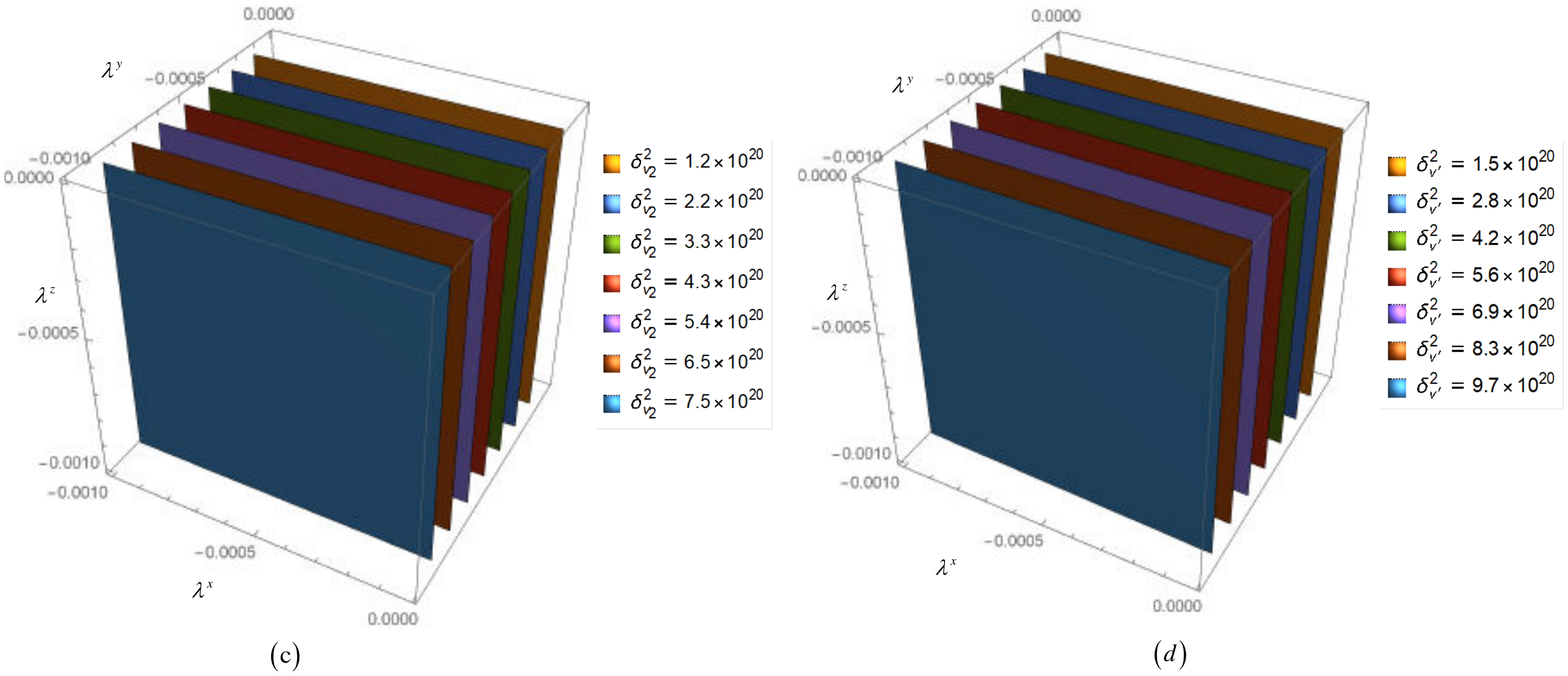}\hspace*{-2.0cm}\\
\vspace*{-4.5cm}
\hspace*{-0.25cm}\includegraphics[width=0.825\textwidth]{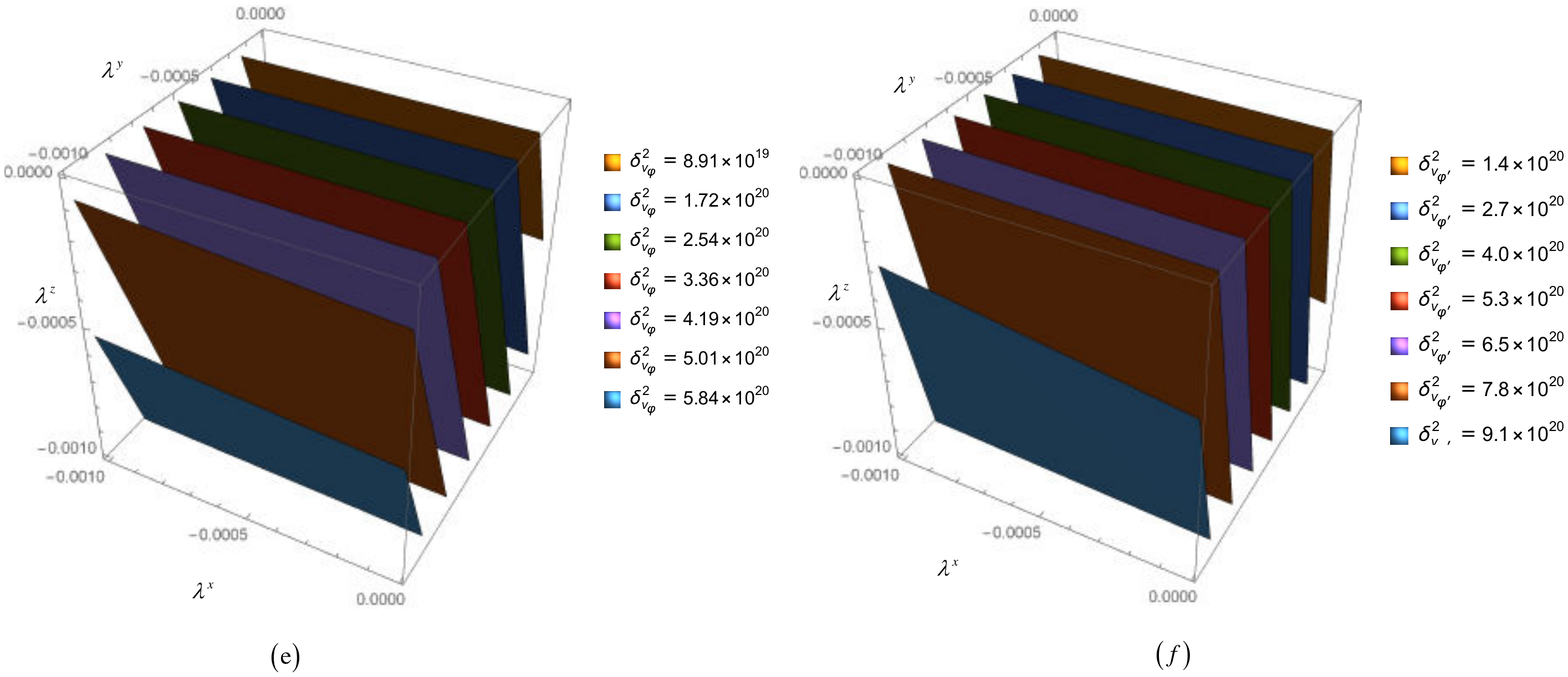}\hspace*{-2.0cm}\\
\vspace*{-4.5cm}
\hspace*{-0.25cm}\includegraphics[width=0.825\textwidth]{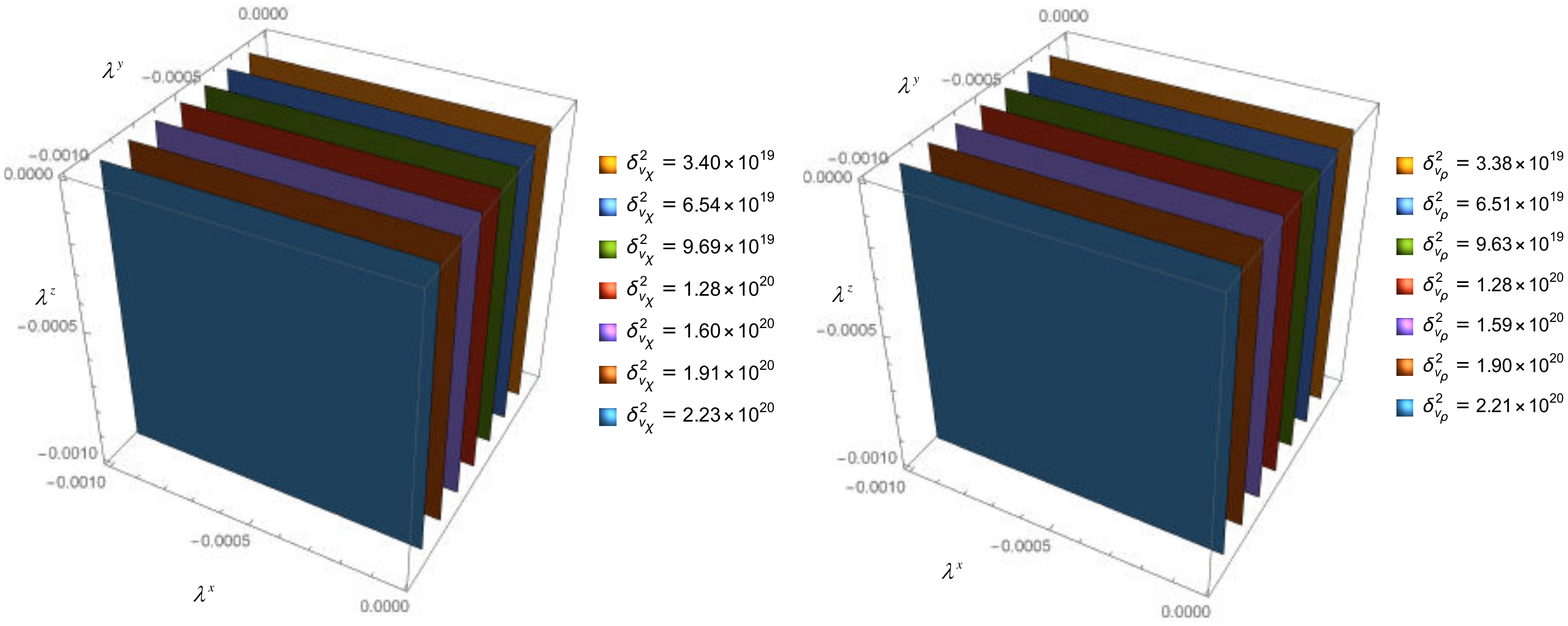}\hspace*{-2.0cm}\\
\vspace*{-3.5cm}
\caption[$\delta^2_{v_h}, \delta^2_{v_1}, \delta^2_{v_2}, \delta^2_{v^'}, \delta^2_{v_\varphi}, \delta^2_{v_{\varphi^'}}, \delta^2_{v_\chi}$ and $\delta^2_{v_\rho}$ verses $\lambda^x, \lambda^y$ and $\lambda^z$ with $\lambda^x\in (-10^{-3}, -10^{-5}),\, \lambda^y\in (-10^{-3}, -10^{-5})$ and $\lambda^z\in (-10^{-3}, -10^{-5})$.]{$\delta^2_{v_h}, \delta^2_{v_1}, \delta^2_{v_2}, \delta^2_{v^'}, \delta^2_{v_\varphi}, \delta^2_{v_{\varphi^'}}, \delta^2_{v_\chi}$ and $\delta^2_{v_\rho}$ verses $\lambda^x, \lambda^y$ and $\lambda^z$ with $\lambda^x\in (-10^{-3}, -10^{-5}),\, \lambda^y\in (-10^{-3}, -10^{-5})$ and $\lambda^z\in (-10^{-3}, -10^{-5})$.}\label{Figur1}
\end{center}
\end{figure}

We will show that, with $\lambda_{\Vien{\Phi}}$ 
\Vien{a}nd $\lambda_{H\phi}$ in Eqs. (\ref{sollaHandphi})-- (\ref{sollaHphi}), there exist Yukawa couplings such that expressions in (\ref{conditionv}) are always give the solution as chosen in Eq. (\ref{scalarvev}). For instance, for the following benchmark
\bea
&&v_1= 10^{10}\, \mathrm{eV}, \hs v_\chi = v_\rho =1.725\times 10^{11}\, \mathrm{eV},  \\
&&v_H = v^' = v_\varphi = v_{\varphi^'} =v_2=1.5\times 10^{10}\, \mathrm{eV}, \\
&&\mu_H = \mu_\phi = \mu_{\phi^'} = \mu_\varphi =\mu_{\varphi^'}=\mu_\chi = \mu_\rho =10^8 \, \mathrm{eV}, \\
&&\lambda^{H\phi^'} = \lambda^{H\varphi} =
 \lambda^{H\varphi^'} = \lambda^{\phi\phi^'} =
\lambda^{\phi\varphi} = \lambda^{\phi\varphi^'} = \lambda^{\varphi\varphi^'} \crn
&&\hspace{0.9 cm}= \lambda^{\phi^'\varphi^'} = \lambda^{\phi^'\varphi}=\lambda^{H\chi} = \lambda^{H\rho} = \lambda^{x}\Vien{,\hs } \lambda^{\varphi\varphi^'\chi\rho} = \lambda^{H\phi\varphi^'} = \lambda^{z}, \\
&&\lambda^{\phi\chi} =\lambda^{\phi\rho} = \lambda^{\varphi\chi} =
\lambda^{\varphi^'\chi} = \lambda^{\varphi^'\rho} = \lambda^{\varphi\rho} = \lambda^{\phi^'\chi} =\lambda^{\phi^'\rho} = \lambda^{\chi\rho} =\lambda^{y},
\eea
the expressions in (\ref{conditionv}) are always satisfied in the case of $\lambda^{x}<0, \lambda^{y} <0$ and $\lambda^{z}<0$, for instance, $\lambda^{x,y,z} \in (-10^{-3}, -10^{-5})$ which are shown in Fig. \ref{Figur1}\footnote{Here, we have used the notations: $\delta^2_{v_H}\equiv \frac{\partial^2 \mathcal{V}_{\mathrm{total}}}{\partial v^2_H},\, \delta^2_{v_1}\equiv\frac{\partial^2 \mathcal{V}_{\mathrm{total}}}{\partial v^2_1},\, \delta^2_{v_2}\equiv\frac{\partial^2 \mathcal{V}_{\mathrm{total}}}{\partial v^2_2},\, \delta^2_{v^'}\equiv\frac{\partial^2 \mathcal{V}_{\mathrm{total}}}{\partial v^{'2}},\, \delta^2_{v_\varphi}\equiv \frac{\partial^2 \mathcal{V}_{\mathrm{total}}}{\partial v^2_{\varphi}},\, \delta^2_{v_{\varphi^'}}\equiv \frac{\partial^2 \mathcal{V}_{\mathrm{total}}}{\partial v^2_{\varphi^'}},\, \delta^2_{v_\chi}\equiv \frac{\partial^2 \mathcal{V}_{\mathrm{total}}}{\partial v^2_{\chi}},\, \delta^2_{v_\rho}\equiv \frac{\partial^2 \mathcal{V}_{\mathrm{total}}}{\partial v^2_{\rho}}.$}. Therefore, the VEV alignments in Eq. (\ref{scalarvev}) is the natural solution of the potential minimum condition.

In models with more than one $SU(2)_L$ Higgs doublet, as in the present model, the Flavor Changing Neutral Current (FCNC) processes exist however they can be suppressed by adding discrete symmetries  which have been presented in Refs. \cite{AME07, PJ08, Kubo13, PPich17, Bento18, S3BLGM2019, Diaz19, WU19}. In addition, the large amount of parametric freedom allows to find a suitable region of parameter space where these FCNC can be suppressed. A numerical analysis of the FCNC, along with other phenomenological aspects in a multiHiggs doublet model with the $D_4$ discrete symmetry is presented in \cite{CarcamoHernandez:2020ney}. The implications
of our model in the FCNC interactions are discussed in section \ref{KKbar}.
\section{Lepton masses and mixings \label{lepton}}
From the lepton Yukawa terms given by Eq. (\ref{Lylep0}) and the tensor product of $\Si (18)$ in Appendix \ref{Si18CG}, we can rewrite the Yukawa interactions in the lepton sector:
\bea -\mathcal{L}^{l}_{Y}&=& h_1\bar{\psi}_{1L} H l_{1R} + h_2 \left(\bar{\psi}_{2 L} H l_{2 R}+\bar{\psi}_{3 L} H l_{3 R}\right) + h_3 \left(\bar{\psi}_{2 L} \phi_2 l_{3 R} + \bar{\psi}_{3 L} \phi_1 l_{2 R}\right)\crn
&+&\frac{x_1}{2} \left(\bar{\psi}_{2 L}\nu_{2  R} -\bar{\psi}_{3 L}\nu_{3 R}\right) \varphi + \frac{x_2}{2}\left(\bar{\psi}_{1L} \varphi^{'}_1 \nu_{2 R} +\bar{\psi}_{1L} \varphi^{'}_2 \nu_{3 R}\right)
+ \frac{x_3}{2}\left(\bar{\psi}_{2 L}\varphi^{'}_2 \nu_{1 R}+\bar{\psi}_{3 L}\varphi^{'}_1 \nu_{1 R}\right)
\crn
&+&\frac{y_1}{2} (\bar{\nu}^c_{1 R} \nu_{1 R}) \rho  +\frac{y_2}{2} (\bar{\nu}^c_{2 R}\nu_{3 R}+\bar{\nu}^c_{3 R}\nu_{2 R})\rho^*
+\frac{y_3}{2} (\bar{\nu}^c_{2 R}\chi_1\nu_{2 R}+\bar{\nu}^c_{3 R}\chi_2\nu_{3 R})+ H.c. \label{Lclep}\eea
With the help of Eq.(\ref{scalarvev}), we get the mass terms for leptons as follows:
\bea
-\mathcal{L}^{mass}_{lep}&=&h_1 v_H \bar{l}_{1L} l_{1R} + h_2 v_H(\bar{l}_{2 L} l_{2 R}+\bar{l}_{3 L} l_{3 R}) + h_3 (v_2 \bar{l}_{2 L} l_{3 R} + v_1 \bar{}_{3 L}l_{2 R})\crn
&+&\frac{x_1 v_{\varphi}}{2}\bar{\nu}_{2 L} \nu_{2 R}-\frac{x_1 v_{\varphi}}{2}\bar{\nu}_{3 L} \nu_{3 R}+\frac{x_2 v_{\varphi^{'}}}{2}\bar{\nu}_{1 L} \nu_{2  R} +\frac{x_2 v_{\varphi^{'}}}{2} \bar{\nu}_{1 L} \nu_{3 R}\crn
&+&\frac{x_3 v_{\varphi^{'}}}{2} \bar{\nu}_{2L} \nu_{1 R}+\frac{x_3 v_{\varphi^{'}}}{2} \bar{\nu}_{3L}\nu_{1 R}
+\frac{y_1 v_\rho}{2} \bar{\nu}^c_{1 R}\nu_{1 R} \crn
 &+&\frac{y_2 v^*_\rho}{2}\left( \bar{\nu}^c_{2 R} \nu_{3 R}+ \bar{\nu}^c_{3 R} \nu_{2 R}\right)+\frac{y_3 v_{\chi}}{2} \bar{\nu}^c_{3 R} \nu_{3 R} + H.c,
\eea\
which can be written in the matrix form:
\bea
-\mathcal{L}^{\mathrm{mass}}_{lep}=\bar{l}_L\mathcal{M}_{cl} l_{R} +\fr 1 2
\bar{n}^c_L \mathcal{M}_\nu n_L+ H. c, \eea
where
\bea
l_l&=& (l_{1L},l_{2L},l_{3L})^T, \hs l_{R}= (l_{1R},l_{2R},l_{3R})^T, \hs \mathcal{M}_{cl}=\left(%
\begin{array}{ccc}
  a_1 & 0 & 0 \\
   0 &  a_2& b_2\\
  0 & b_1  & a_2\\
\end{array}%
\right), \label{Mclep}\\
n_L&=& \left(\nu^c_L ,\,  \nu_R \right)^T,\hs \mathcal{M}_\nu =\left(%
\begin{array}{cc}
  0 & \mathcal{M}_D \\
  \mathcal{M}^T_D & \mathcal{M}_R \\
\end{array}%
\right), \eea
with $\nu^c_L =(\nu^c_{1L}, \nu^c_{2L}, \nu^c_{3L})^T,\hs
\nu_R=(\nu_{1R}, \nu_{2R}, \nu_{3R})^T$ and $\mathcal{M}_D, \mathcal{M}_R$ are respectively Dirac and Majorana neutrino mass matrices,
 \bea \mathcal{M}_D &=&\left(%
\begin{array}{ccc}
 0      &\,\, b_{D} & b_{D} \\
 c_{D}  &\,\, a_{D} & 0 \\
 c_{D}  &\,\, 0     & -a_{D} \\
\end{array}%
\right),\hs \mathcal{M}_R=
\left(%
\begin{array}{ccc}
a_R &0       & 0 \\
0     & 0      & b_R \\
0     & b_R  & c_{R} \\
\end{array}%
\right),  \label{MDR}\eea
and
\bea
a_i&=& h_i v_H,\hs b_i= h_3 v_{i} \hs (i=1,2), \\
a_D&=&x_1 v_{\varphi},\hs b_D = x_2 v_{\varphi^{'}}, \hs c_D = x_3 v_{\varphi^{'}}, \crn
a_R&=& y_1 v_\rho, \hs b_R =  y_2 v^*_\rho,\hs c_{R}=y_3 v_{\chi}.\label{abc}
\eea
Let us define a Hermitian matrix $\mathrm{m}_l$ as follows
\bea
\mathcal{M}_l&=& \mathcal{M}_{cl} \mathcal{M}^+_{cl} = \left(%
\begin{array}{ccc}
  |a_1|^2 & 0 & 0 \\
   0 &   |a_2|^2+ |b_2|^2 & a_2 b^*_1 + a^*_2 b_2\\
  0 & (a_2 b^*_1 + a^*_2 b_2)^*  & |a_2|^2+ |b_1|^2\\
\end{array}%
\right), \label{ml}
\eea
which 
can be diagonalised by $\mathcal{U}_{L, R}$ satisfying $\mathcal{U}^+_{L} \mathcal{M}_l \mathcal{U}_R=\mathrm{diag} (m^2_e, m^2_\mu, m^2_\tau)$, where
\bea
\mathcal{U}_L&=&\mathcal{U}_R=\left(%
\begin{array}{ccc}
  1 & 0 & 0 \\
   0 & \mathrm{cos} \theta_l &\,\,\,\,\,\, -\mathrm{sin} \theta_l . e^{-i \al}\\
  0 & \hs \mathrm{sin} \theta_l . e^{i \al} & \mathrm{cos} \theta_l  \\
\end{array}%
\right), \label{Uclepal}\\
m^2_e &=&|a_1|^2, \,\, m^2_{\mu, \tau}= \frac{1}{2}\left(\ga_1 \mp \ga_2 \right), \label{memt}
\eea
with
\bea
\ga_2&=&
\sqrt{(|b_1|^2-|b_2|^2)^2 + 4 |a_2|^2 \left(|b_1|^2 +|b_2|^2\right) +8 |a_2|^2 |b_1||b_2| \cos (2\al_2 -\beta_1-\beta_2)},\crn
\ga_1&=& 2 |a_2|^2 + |b_1|^2 + |b_2|^2, \hs \al_2 = \arg(a_2),\hs \beta_i = \arg(b_i) \hs (i=1,2), \label{g1g2}\\
\al &=&\frac{i}{2}\log \left[\frac{a_2 b^*_1 + a^*_2 b_2}{(a_2 b^*_1 + a^*_2 b_2)^*}\right], \hs \theta_l=\arctan\left[\frac{(a^*_2 b_1 + a_2 b^*_2) e^{-i \al}}{|a_2|^2 + |b_2|^2 - m^2_\tau}\right]. \label{altheta}
\eea
Comparing the result in Eq. (\ref{memt}) with the experimental values of the charged lepton masses given in Ref. \cite{PDG2020}, $m_e \simeq 0.51999 \,\mathrm{MeV},  m_\mu \simeq  105.65837\,\mathrm{MeV}, m_\tau = 1776.86 \,\mathrm{MeV}$, we obtain:
\bea
|a_1| = 0.510999 \times 10^{6}\, \mathrm{eV}, \ga_1=3.1684 \times 10^{18}\, \mathrm{eV}^2,\, \ga_2=3.14607 \times 10^{18} \, \mathrm{eV}^2.
\eea
In the case $\al_{2} =\beta_1=\beta_2$ and $|v_1|\sim |v_2|$, we get:
\bea
|h_1|&\sim &\frac{5 \times 10^{5}}{|v_H|}, \hs |h_2|\sim \frac{8 \times 10^{8}}{|v_H|},\hs |h_3|\sim \frac{9 \times 10^{8}}{|v_2|}. \label{h1h2h3}
\eea
As we will see below, since the charged lepton mixing matrix $\mathcal{U}_L$ is non trivial in our model, it can contribute to the final leptonic mixing matrix, defined by $U=\mathcal{U}^+_{L} \mathcal{U}_{\nu}$ where $\mathcal{U}_{L}$ refers to the left-handed charged-lepton mixing matrix and $\mathcal{U}_{\nu}$ is the neutrino mixing matrix.

Regarding the neutrino sector, from Eq. (\ref{MDR}), the light active neutrino mass matrix arises from the type-I seesaw
mechanism as follows:
 \bea
\mathcal{M}_\nu &=&- \mathcal{M}_D \mathcal{M}_R^{-1} \mathcal{M}^T_D=\left(%
\begin{array}{ccc}
\frac{b_D^2 (c_R-2 b_R)}{b_R^2}  & \frac{a_D b_D (c_R-b_R)}{b_R^2}& \frac{a_D b_D}{b_R}\\
\frac{a_D b_D (c_R-b_R)}{b_R^2} & \frac{a_D^2 c_R}{b_R^2}-\frac{c_D^2}{a_R}      & \frac{a_D^2}{b_R}-\frac{c_D^2}{aR} \\
\frac{a_D b_D}{b_R}                         & \frac{a_D^2}{b_R}-\frac{c_D^2}{aR}             & -\frac{c_D^2}{a_R} \\
\end{array}%
\right), \label{Meff}\eea
which has three exact eigenvalues
\bea m_1&=&0,\hs m_{2,3}=\kappa_1 \mp \kappa_2,\label{m123}\eea
where
\bea
\kappa_1&=&\frac{(a_D^2 + b_D^2) c_R}{2 b_R^2}-\frac{b_D^2}{a_R}-\frac{c_D^2}{a_R}, \hs \kappa_2=\frac{\sqrt{\Bbbk}}{2 a_R b_R^4}, \label{kapal2k}\\
\Bbbk &=& b_R^4 \left\{4 b_R^2 \left[a_R (a_D^2 + b_D^2)-b_R c_D^2\right]^2 -
   4 a_R b_D^2 b_R  \left[a_R (a_D^2 + b_D^2)- b_Rc_D^2\right] c_R \right.\crn
   &+&\left. a_R^2 (a_D^2 + b_D^2)^2 cR^2\right\}, \nn
\eea
and the corresponding mixing matrix is:
\bea
\mathbb{R}=\left( \begin{array}{ccc}
\frac{\mathbb{K}}{\sqrt{\mathbb{K}^2+2}}&\frac{\mathbb{K}_{-}}{\sqrt{\mathbb{K}^2_{-}+\mathbb{N}^2_{-}+1}}&\frac{\mathbb{K}_{+}}{\sqrt{\mathbb{K}^2_{+}+\mathbb{N}^2_{+}+1}}\\
-\frac{1}{\sqrt{\mathbb{K}^2+2}} &\frac{\mathbb{N}_{-}}{\sqrt{\mathbb{K}^2_{-}+\mathbb{N}^2_{-}+1}}&\frac{\mathbb{N}_{+}}{\sqrt{\mathbb{K}^2_{+}+\mathbb{N}^2_{+}+1}}\\
\frac{1}{\sqrt{\mathbb{K}^2+2}} &\frac{1}{\sqrt{\mathbb{K}^2_{-}+\mathbb{N}^2_{-}+1}}      &\frac{1}{\sqrt{\mathbb{K}^2_{+}+\mathbb{N}^2_{+}+1}}
\end{array}\right). \mathbb{P},\label{neumix}
\eea
where $\mathbb{P}=\mathrm{diag}(1,\,1,\,i)$ and $\mathbb{K}, \mathbb{K}_{\mp}, \mathbb{N}_{\mp}$ are defined
\bea
\mathbb{K} &=& \frac{a_D}{b_D}, \hs \mathbb{K}_{\mp} =\kappa_{11} \mp \kappa_{12},  \hs \mathbb{N}_{\mp} =\epsilon_{11} \mp \epsilon_{12},\label{KKiNi} \eea
where
\bea
\kappa_{11}&=& \frac{b_D}{2a_D} \left\{\frac{(a_D^2 + b_D^2) a_R c_R}{b_R \left[(a_D^2 + b_D^2) a_R- b_R c_D^2 \right]}-2\right\}, \,
\kappa_{12}= \frac{b_D \sqrt{\Bbbk}}{2 a_D b_R^3 \left[(a_D^2 + b_D^2)a_R-c_D^2 b_R \right]},  \crn
 \epsilon_{11}&=&\frac{(a_D^2 + b_D^2) a_R c_R}{2 b_R \left[a_R (a_D^2 + b_D^2)-c_D^2 b_R\right]},  \hs\hs\hs\hs\hs\hs\hs\hs\hs\hs\,
 \epsilon_{12}= \frac{\sqrt{\Bbbk}}{2 b_R^3 \left[(a_D^2 + b_D^2)a_R -c_D^2 b_R\right]}, \label{k11kl2ep11ep12}
\eea
and $a_{D}, b_{D}, c_D, a_R, b_R, c_{ R}$ are given in Eq. (\ref{abc}).

 From the explicit expressions of $m_{2,3}$, $\mathbb{K}, \mathbb{K}_{\mp}$ and $\mathbb{N}_{\mp}$ in Eqs. (\ref{m123}), (\ref{kapal2k}), (\ref{KKiNi}) and (\ref{k11kl2ep11ep12}), the following relations hold:
 \bea
 && 1+\mathbb{K} \mathbb{K}_{-} - \mathbb{N}_{-}=0,\,\, 1+\mathbb{K} \mathbb{K}_{+} - \mathbb{N}_{+}=0,\,\,  1+\mathbb{K}_{-} \mathbb{K}_{+}+\mathbb{N}_{-} \mathbb{N}_{+}=0,  \label{relation}\\
 && a_D=\frac{(\mathbb{N}_{-} + \mathbb{N}_{+}-2) b_D}{\mathbb{K}_{-} + \mathbb{K}_{+}}, \crn
 && a_R= \frac{ (\La_\mathrm{N}+2)\La_\mathrm{K}^2 b_R c_D^2}{\left[\La_\mathrm{K}^2+\La_\mathrm{N}(\La_\mathrm{N}-2)\right] (\La_\mathrm{N}-2)b_D^2- (m_2 + m_3)\La_\mathrm{K}^2b_R},\crn
 && c_R=  \frac{\left\{2  [2 \La_\mathrm{K}^2 + (\La_\mathrm{N}-2)^2] b_D^2+ \La_\mathrm{K}^2 (m_2 + m_3) b_R \right\} \La_\mathrm{N} b_R }{\left[\La_\mathrm{K}^2 + (\La_\mathrm{N}-2)^2\right] (\La_\mathrm{N}+ 2) b_D^2}. \label{relation1}
 \eea
 with
 \bea
 \La_\mathrm{K} &=&\mathbb{K}_{-} + \mathbb{K}_{+},\hs \La_\mathrm{N} =\mathbb{N}_{-} + \mathbb{N}_{+}.\label{LaKN}
 \eea
 The effective neutrino mass matrix  $\mathcal{M}_{\nu}$ in Eq. (\ref{Meff}) is diagonalized as
 \begin{equation}
\mathcal{U}_{\nu }^\mathrm{\+} \mathcal{M}_{\nu} \mathcal{U}_{\nu }=\left\{
\begin{array}{l}
\left(
\begin{array}{ccc}
0 & 0 & 0 \\
0 & m_{2} & 0 \\
0 & 0 & m_{3}%
\end{array}%
\right) ,\hspace{0.1cm} \mathcal{U}_{\nu }=\left( \begin{array}{ccc}
\frac{\mathbb{K}}{\sqrt{\mathbb{K}^2+2}}&\frac{\mathbb{K}_{-}}{\sqrt{\mathbb{K}^2_{-}+\mathbb{N}^2_{-}+1}}&\frac{i\mathbb{K}_{+}}{\sqrt{\mathbb{K}^2_{+}+\mathbb{N}^2_{+}+1}}\\
-\frac{1}{\sqrt{\mathbb{K}^2+2}} &\frac{\mathbb{N}_{-}}{\sqrt{\mathbb{K}^2_{-}+\mathbb{N}^2_{-}+1}}&\frac{i\mathbb{N}_{+}}{\sqrt{\mathbb{K}^2_{+}+\mathbb{N}^2_{+}+1}}\\
\frac{1}{\sqrt{\mathbb{K}^2+2}} &\frac{1}{\sqrt{\mathbb{K}^2_{-}+\mathbb{N}^2_{-}+1}}      &\frac{i}{\sqrt{\mathbb{K}^2_{+}+\mathbb{N}^2_{+}+1}}
\end{array}\right) \hspace{0.1cm}\mbox{for NO,}\ \  \\
\left(
\begin{array}{ccc}
m_{2} & 0 & 0 \\
0 & m_{3} & 0 \\
0 & 0 & 0
\end{array}%
\right) ,\hspace{0.1cm} \mathcal{U}_{\nu }=\left( \begin{array}{ccc}
\frac{\mathbb{K}_{-}}{\sqrt{\mathbb{K}^2_{-}+\mathbb{N}^2_{-}+1}}&\frac{\mathbb{K}_{+}}{\sqrt{\mathbb{K}^2_{+}+\mathbb{N}^2_{+}+1}}&\frac{i \mathbb{K}}{\sqrt{\mathbb{K}^2+2}}\\
\frac{\mathbb{N}_{-}}{\sqrt{\mathbb{K}^2_{-}+\mathbb{N}^2_{-}+1}}&\frac{\mathbb{N}_{+}}{\sqrt{\mathbb{K}^2_{+}+\mathbb{N}^2_{+}+1}}& -\frac{i}{\sqrt{\mathbb{K}^2+2}}\\
\frac{1}{\sqrt{\mathbb{K}^2_{-}+\mathbb{N}^2_{-}+1}}      &\frac{1}{\sqrt{\mathbb{K}^2_{+}+\mathbb{N}^2_{+}+1}}&\frac{i}{\sqrt{\mathbb{K}^2+2}}
\end{array}\right) \hspace{0.1cm}\mbox{for IO,}%
\end{array}%
\right.  \label{Unu}
\end{equation}
where $m_{2,3}$ and $\mathbb{K}, \mathbb{K}_{\mp}, \mathbb{N}_{\mp}$ are respectively given in Eqs. (\ref{m123}) and (\ref{KKiNi}).

The final leptonic mixing matrix then reads:
\begin{equation}
U=\mathcal{U}^+_{L} \mathcal{U}_{\nu}=\left\{
\begin{array}{l}
\left( \begin{array}{ccc}
\frac{\mathbb{K}}{\sqrt{\mathbb{K}^2+2}} &\frac{\mathbb{K}_{-}}{\sqrt{\mathbb{K}^2_{-}+\mathbb{N}^2_{-}+1}}&\frac{i\mathbb{K}_{+}}{\sqrt{\mathbb{K}^2_{+}+\mathbb{N}^2_{+}+1}}\\
\frac{e^{-i\al}\sin\theta_l-\cos\theta_l}{\sqrt{\mathbb{K}^2+2}} &\frac{\cos\theta_l \mathbb{N}_{-}  +e^{-i \al}\sin\theta_l }{\sqrt{\mathbb{K}^2_{-}+\mathbb{N}^2_{-}+1}}&\frac{i(\cos\theta_l \mathbb{N}_{+}  +  e^{-i \al}\sin\theta)}{\sqrt{\mathbb{K}^2_{+}+\mathbb{N}^2_{+}+1}}\\
\frac{e^{i\al}\sin\theta_l+\cos\theta_l}{\sqrt{\mathbb{K}^2+2}} &\frac{ \cos\theta_l -  e^{i \al}\sin\theta_l \mathbb{N}_{-}}{\sqrt{\mathbb{K}^2_{-}+\mathbb{N}^2_{-}+1}} &\frac{i(\cos\theta_l - e^{i \al}\sin\theta_l \mathbb{N}_{+})}{\sqrt{\mathbb{K}^2_{+}+\mathbb{N}^2_{+}+1}}\\
\end{array}\right) \,\,\,\,\mbox{for \, NO},    \\
\left(%
\begin{array}{ccc}
\frac{\mathbb{K}_{-}}{\sqrt{\mathbb{K}_{-}^2+\mathbb{K}_{-}^2+1}}&\frac{\mathbb{K}_{+}}{\sqrt{\mathbb{K}_{+}^2+\mathbb{K}_{+}^2+1}}& \frac{i \mathbb{K}}{\sqrt{\mathbb{K}^2+2}}\\
\frac{\cos\theta_l \mathbb{N}_{-}  +e^{-i \al} \sin\theta_l}{\sqrt{\mathbb{K}_{-}^2+\mathbb{K}_{-}^2+1}}&\frac{\cos\theta_l \mathbb{N}_{+}  +e^{-i \al} \sin\theta_l}{\sqrt{\mathbb{K}_{+}^2+\mathbb{K}_{+}^2+1}}& \frac{i(e^{-i\al} \sin\theta_l-\cos\theta_l)}{\sqrt{\mathbb{K}^2+2}}\\
\frac{\cos\theta_l - e^{i \al} \sin\theta_l \mathbb{N}_{-}}{\sqrt{\mathbb{K}_{-}^2+\mathbb{K}_{-}^2+1}} &\frac{\cos\theta_l - e^{i \al} \sin\theta_l \mathbb{N}_{+}}{\sqrt{\mathbb{K}_{+}^2+\mathbb{K}_{+}^2+1}} &\frac{i(e^{i\al} \sin\theta_l+\cos\theta_l)}{\sqrt{\mathbb{K}^2+2}}  \\
\end{array}%
\right) \hspace{0.15cm}\mbox{for \ IO}.
\end{array}%
\right.  \label{ULep}
\end{equation}
In the three neutrino oscillation picture, the lepton mixing  matrix can be parametrized as\footnote{Here, $\delta$ is the Dirac CP violating phase and
$c_{ij}=\cos \theta_{ij}$, $s_{ij}=\sin \theta_{ij}$ with
$\theta_{12}$, $\theta_{23}$ and $\theta_{13}$ being the solar, atmospheric and reactor angle, respectively. $P$ contains two Majorana phases ($\alpha_{21}, \alpha_{31}$) which play no role in neutrino oscillations, $P=\mathrm{diag}\left(1, e^{i \frac{\alpha_{21}}{2}}, e^{i \frac{\alpha_{31}}{2}}\right)$, and thus will be ignored.} \cite{PDG2020}
\bea
       U_{\mathrm{PMN}S} = \begin{pmatrix}
    c_{12} c_{13}                                                     & s_{12} c_{13}                    & s_{13} e^{-i \delta}\\
-s_{12} c_{23}-c_{12} s_{23} s_{13}e^{i \delta} & c_{12} c_{23}-s_{12} s_{23} s_{13} e^{i \delta} & s_{23} c_{13}\\
 s_{12} s_{23}-c_{12} c_{23} s_{13}e^{i \delta} &- c_{12} s_{23}-s_{12} c_{23} s_{13} e^{i \delta} & c_{23} c_{13} \\
     \end{pmatrix}.P, \label{Ulepg}
\eea
whereby, $\theta_{12}, \theta_{23}, \theta_{13}$ can be defined via the elements of the leptonic mixing
matrix:
\bea  s^2_{13}&=& |U_{13}|^2, \hs
 t^2_{12} =\left|\frac{U_{12}}{U_{11}}\right|^2, \hs
  t^2_{23}=\left|\frac{U_{23}}{U_{33}}\right|^2.\label{thetaijsq}
\eea
As it is well known, the neutrino mass spectrum is currently unknown and it can be NO or IO depending on the sign of $\De m^2_{32}$ \cite{PDG2020} which will be presented in the next section.

\subsection{\label{NHsector} Normal spectrum}
In NO, the Jarlskog invariant $J_{CP}$ which determines the magnitude of CP violation in
neutrino oscillations \cite{PDG2020}, determined from Eqs. (\ref{relation}) and (\ref{ULep}), takes the form
\bea
J^N_{CP}&=&\mathrm{Im} (U_{23} U^*_{13} U_{12} U^*_{22})
=\frac{\mathbb{K}^2_{+} (1-\mathbb{N}^2_{+}) \cos \theta_l\sin \theta_l \sin\alpha}{\left[2\mathbb{K}_{+}^2 + (\mathbb{N}_{+}-1)^2\right] (1 + \mathbb{K}_{+}^2 + \mathbb{N}_{+}^2)}. \label{JN}
\eea
Comparing Eq. (\ref{JN}) with its corresponding expression in the standard parametrization of the neutrino
mixing matrix given in Ref. \cite{PDG2020}, $J_{CP}=s_{13} c_{13}^2 s_{12} c_{12} s_{23} c_{23} \sin\delta$, we get:
\bea
\sin\delta^N&=&\frac{\mathbb{K}^2_{+} (1-\mathbb{N}^2_{+})}{\left[2\mathbb{K}_{+}^2 + (\mathbb{N}_{+}-1)^2\right] (1 + \mathbb{K}_{+}^2 + \mathbb{N}_{+}^2)}\frac{\cos \theta_l\sin \theta_l \sin\alpha}{s_{13} c_{13}^2 s_{12} c_{12} s_{23} c_{23}}.\label{sdN} \eea
Furthermore, from Eqs. (\ref{ULep}) and (\ref{thetaijsq}), for NO, we get:
 \bea  s^2_{13}&=&\frac{\mathbb{K}^2_{+}}{1 + \mathbb{K}^2_{+} + \mathbb{N}^2_{+}},\hs
 t^2_{12}=\frac{\mathbb{K}_{+}^2 (\mathbb{N}_{+}+1)^2}{(\mathbb{N}_{+}-1)^2 (1 + \mathbb{K}_{+}^2 + \mathbb{N}_{+}^2)}, \crn
  t^2_{23}&=&\frac{\sin^2\theta_l + \cos^2\theta_l\mathbb{N}_{+}^2 +\sin (2\theta_l) \cos\al.\mathbb{N}_{+}}{\cos^2\theta_l  +  \sin^2\theta_l \mathbb{N}_{+}^2 - \sin(2\theta_l) \cos\al.\mathbb{N}_{+}}.\label{thetaijsqN}
\eea
Combining  Eqs. (\ref{relation}) and (\ref{thetaijsqN}) yields:
\bea
\mathbb{K}_{+}&=&\frac{\sqrt{2} s_{13} \sqrt{s_{13}^2 + t_{12}^2}}{c_{13}( t_{12}-s_{13})}, \,\,\, \mathbb{N}_{+}=1+\frac{2 s_{13}}{t_{12}-s_{13}}, \label{K2N2N}\\
\cos\al&=&\frac{(1-t^2_{23})(s^2_{13} + t_{12}^2) + 2(1 - 2 \sin^2\theta_l)(1 + t^2_{23})s_{13} t_{12}}{(s_{13}^2 - t_{12}^2) (1 + t^2_{23})\sin(2\theta_l)}. \label{al}
\eea
Next, substituting Eq. (\ref{K2N2N}) into Eq. (\ref{sdN}) yields:
\bea
\sin\delta&=&-\frac{\sin (2\theta_l)  \sin\alpha}{\sin (2\theta_{23})}. \label{sd}
\eea
 We note that the elements $U_{1i} \, (i=1,2,3)$ depend only on $\theta_{12}$ and $\theta_{13}$ while $U_{2i}$ and $U_{3i} \, (i=1,2,3)$ depend on all lepton mixing angles $\theta_{ij}\, (ij=12,23,13)$ and $\theta_l$.\\
For NO, by taking the best-fit values of leptonic mixing angles $\theta_{ij}\,\, (i,j=1,2,3)$ \cite{PDG2020}, $\sin^2 \theta_{12}=0.307,\,\, \sin^2 \theta_{13}=2.18 \times 10^{-2},\,\,
\sin^2 \theta_{23}= 0.545$, we obtain $U^N_{11} =0.823$, $U^N_{12}=-0.548$, $U^N_{13}=0.148 i$ and $U^N_{ij} \, (i=2,3; j=1,2,3)$ depend only on $\sin \theta_l$ that is plotted in Fig. \ref{UijN} with $\sin\theta_l\in (0.8, 0.9)$. In the case $\delta =1.36 \pi$ \cite{PDG2020}, from Eq. (\ref{sd}) we obtain the model parameters as shown in Tab. \ref{paraN}.
\begin{center}
\begin{figure}[h]
\begin{center}
\vspace{-1.0cm}
\includegraphics[width=0.85\textwidth]{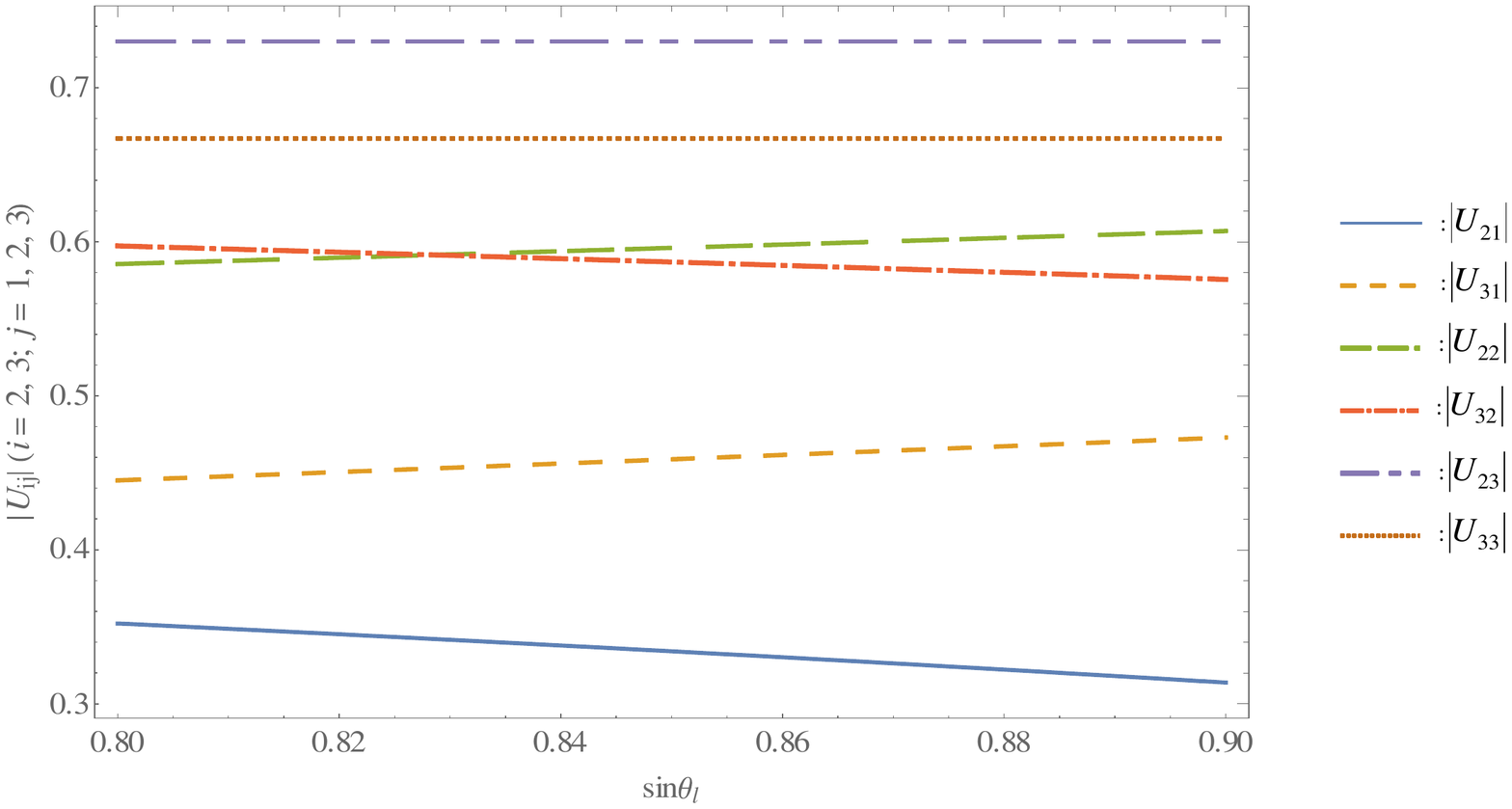}
\vspace*{-3.85cm}
\caption[$|U^N_{ij}|  \, (i=2,3; j=1,2,3)$ as functions of
 $\sin\theta_l$ with $\sin\theta_l\in (0.8, 0.9)$  for NO]{$|U^N_{ij}|  \, (i=2,3; j=1,2,3)$ as functions of
 $\sin\theta_l$ with $\sin\theta_l\in (0.8, 0.9)$  for NO.}\label{UijN}
\end{center}
\end{figure}
\vspace{-0.35cm}
\end{center}
\begin{table}[!h]
\caption{\label{paraN} The model parameters in the case\cite{PDG2020} $\delta = 1.36 \pi$ in the NO}
\centering
\begin{tabular}{|c||c|}
\hline
 Parameters &\hs The derived values \\ \hline
\,\,$\mathbb{K}$ &\, $2.05 $\\
\hline
 \,\,$\mathbb{K}_{-}$ &\, $-0.734$\\ \hline
   \,\,$\mathbb{K}_{+}$ &\, $0.278 $\\ \hline
   \,\,$\mathbb{N}_{-}$ &\, $-0.507$\\ \hline
  \,\, $\mathbb{N}_{+}$ &\, $1.57$\\ \hline
  \,\, $\theta_l$ &\, $55.1^\circ$\\ \hline
    \,\, $\alpha$ &\, $73.9^\circ$\\ \hline
\end{tabular}
\end{table}
The lepton mixing matrix in Eq.(\ref{ULep}) then takes the form:
\bea
U^N=\left(
\begin{array}{ccc}
0.823  & -0.548 & 0.148i  \\
-0.138 - 0.316i & -0.046 - 0.588i & 0.419+ 0.598i \\
0.321 + 0.316 i & 0.513 + 0.298 i&0.657 + 0.113 i
\end{array}\right), \label{ULN}
\eea
which is unitary and consistent with the constraint on the absolute values of the
entries of the lepton mixing matrix given in Ref. \cite{Ivan2019}.

Now, by using the recent best-fit values for the squared-neutrino mass differences \cite{PDG2020}$, \Delta m^2_{21}=7.53\times 10^{-5}\, \mathrm{eV}^2, \,\, \Delta m^2_{32}=2.453\times 10^{-3}\, \mathrm{eV}^2$ for the NO, we get a solution
\bea
\kappa_1 &=&2.95\times 10^{-2}, \hs \kappa_2 = 2.08\times 10^{-2}, \crn
m_1&=&0\,\mathrm{eV},\hs m_2=8.68\times 10^{-3}\, \mathrm{eV},\hs m_3=5.03\times 10^{-2}\, \mathrm{eV}. \label{m1m2m3exp}\eea
The absolute neutrino mass, defined as the
sum of the mass of the three neutrino mass eigenstates, is found to be $\sum_{i=1}^3 m^N_{\nu_i}=5.90\times10^{-2} \,\mathrm{eV}$. At present, the sum of the three neutrino
masses has not been precisely determined, however, the result obtained from our model is well consistent
with the strongest bound from cosmology, $\sum m_\nu < 0.078 \, \mathrm{eV}$ \cite{nubound}.

 Next, substituting explicit expressions of $b_{D}, b_{D}, c_D, a_R, b_R, c_{ R}$ from Eq. (\ref{abc}) and the obtained values of $\mathbb{K}_{\mp}, \mathbb{N}_{\mp}$ in Tab. \ref{paraN} and $m_{2, 3}$ in Eq. (\ref{m1m2m3exp}) into Eq. (\ref{relation1}), we get the following relations:
 \bea
 x_1&=& 2.05 \left(\frac{  v_{\varphi^{'}}}{v_{\varphi}}\right)x_2,\hs
 y_1 = 
 \frac{1}{1.16 \left(\frac{v_{\rho}}{ v^*_{\rho}}\right)\frac{x^2_2}{x^2_3 y_2}-0.0192 \left(\frac{v_{\rho}}{ v^2_{\varphi^{'}}}\right)\frac{1}{x^2_3}},\crn
 y_3&=& \frac{0.827 v^*_\rho y_2}{v_\chi }+\frac{0.00393 (v^*_\rho)^2 y^2_2}{v_\chi v^2_{\varphi^{'}} x^2_2}.
 \eea
\subsection{\label{IHsector} Inverted spectrum}
Similar to the NO, from Eqs. (\ref{ULep}) and (\ref{thetaijsq}) for IO, we get a solution:
\bea
\mathbb{K}_{+}&=&-\frac{\sqrt{2} c_{13} t_{12}}{1+s_{13} t_{12}}, \,\,\, \mathbb{N}_{+}=\frac{2}{1+ s_{13} t_{12}}-1,\hs
\cos\al=\frac{1}{\sin(2\theta)}\frac{1-t^2_{23}}{1 + t^2_{23}}, \label{aK2N2I}
\eea
and the Jarlskog invariant $J_{CP}$ determined from (\ref{ULep}) and $\sin\delta$ take the form:
\bea
J^I_{CP}&=&-\frac{\mathbb{K} \mathbb{K}_{+} (1+\mathbb{N}_{+}) \cos \theta_l\sin \theta_l \sin\alpha}{(\mathbb{K}^2+ 2)(\mathbb{K}^2_{+}+\mathbb{N}^2_{+}+1)}, \label{Ji}\\
\sin\delta^I&=&\frac{\mathbb{K} \mathbb{K}_{+} (1+\mathbb{N}_{+})}{(\mathbb{K}_{+}^2 + 2) (1 + \mathbb{K}_{+}^2 + \mathbb{N}_{+}^2)}
\frac{\cos \theta_l\sin \theta_l \sin\alpha}{s_{13} c_{13}^2 s_{12} c_{12} s_{23} c_{23}}.\label{sdI}
\eea
With the help of Eq. (\ref{relation}), it is easy to show that $J^N_{CP}=J^I_{CP}$ and $\sin\delta^N=\sin\delta^I$ thus the relations in Eqs. (\ref{JN}) and (\ref{sdN}) are satisfied for both normal and inverted orderings and the differences start from Eqs. (\ref{K2N2N}) and (\ref{aK2N2I}).

Next, by taking the best-fit values of leptonic mixing angles $\theta_{ij}\,\, (i,j=1,2,3)$ for IO \cite{PDG2020}, $s^2_{12} = 0.307, \, s^2_{23} = 0.547$ and $s^2_{13} = 2.18\times 10^{-2}$, we get $U^I_{11} =0.823$, $U^I_{12}=-0.548$, $U^I_{13}=0.148i$ and the other elements of $U$ depend only on $\sin \theta_l$ that is plotted in Fig. \ref{UijI}.
\begin{center}
\begin{figure}[h]
\begin{center}
\vspace{-1.0cm}
\hspace*{0.01cm}
\includegraphics[width=0.81\textwidth]{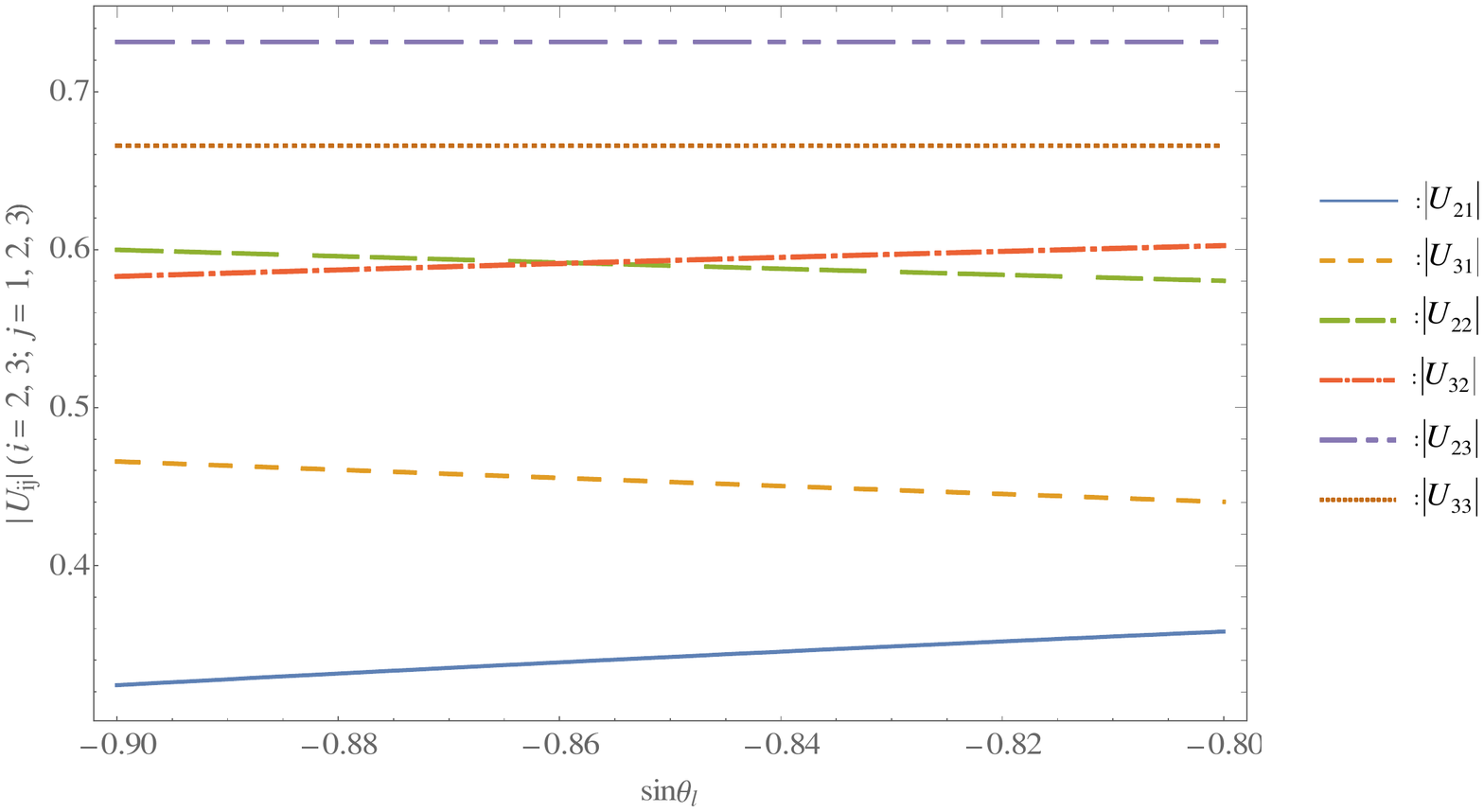}
\vspace*{-3.5cm}
\caption[$|U^I_{ij}|  \, (i=2,3; j=1,2,3)$ as functions of
 $\sin\theta_l$ with $\sin\theta_l\in (-0.9, -0.8)$  for IO]{$|U^I_{ij}|  \, (i=2,3; j=1,2,3)$ as functions of
 $\sin\theta_l$ with $\sin\theta_l\in (-0.9, -0.8)$  for IO.}\label{UijI}
\end{center}
\end{figure}
\vspace*{-0.35cm}
\end{center}
In the case the CP violating phase takes the best-fit values \cite{PDG2020}, $\delta =1.36 \pi$ we find $\sin\theta_l =-0.537\, (\theta_l=327.5^\circ)$ and the other model parameters are explicitly given in Tab. \ref{paraI}.
\begin{table}[!h]
\caption{\label{paraI} The model parameters in the case\cite{PDG2020} $\delta = 1.36 \pi$ in IO}
\centering
\begin{tabular}{|c||c|}
\hline
 Parameters &\hs The derived values \\ \hline
 \,\,$\mathbb{K}$ &\, $0.211$\\ \hline
  \,\,$\mathbb{K}_{-}$ &\, $2.7$\\ \hline
   \,\,$\mathbb{K}_{+}$ &\, $-0.848$\\ \hline
   \,\,$\mathbb{N}_{-}$ &\, $1.57$\\ \hline
  \,\, $\mathbb{N}_{+}$ &\, $0.821$\\ \hline
  \,\, $\alpha$ &\, $84.0^\circ$\\ \hline
\end{tabular}
\end{table}
\newline
The PMNS leptonic mixing matrix of Eq.(\ref{ULep}) takes the form:
\bea
U^I=\left(
\begin{array}{ccc}
0.823               &-0.548               & 0.148 i \\
0.387 + 0.163 i & 0.412+0.345 i   & -0.373 - 0.629i \\
0.284+0.256 i   & 0.575 + 0.283 i & 0.373 + 0.551 i
\end{array}\right), \label{ULI}
\eea
which is unitary and consistent with the constraint on the absolute values of the
entries of the lepton mixing matrix given in Ref. \cite{Ivan2019}.

Now, by using the recent best-fit values for the squared-neutrino mass differences \cite{PDG2020}$, \Delta m^2_{21}=7.53\times 10^{-5}\, \mathrm{eV}^2, \,\, \Delta m^2_{32}=-2.546\times 10^{-3}\, \mathrm{eV}^2$ for IO, we get a solution
\bea
\kappa_1 &=&5.01\times 10^{-2}, \hs \kappa_2 = 3.76\times 10^{-4}, \crn
m_1&=&4.97\times 10^{-2}\,\mathrm{eV},\hs m_2=5.05\times 10^{-2}\, \mathrm{eV},\hs m_3=0\, \mathrm{eV}. \label{m1m2m3expI}\eea
Using our best fit results given above, we find that the sum of three light neutrino masses is given by $\sum_{i=1}^3 m^I_{\nu_i}=0.1 \,\mathrm{eV}$. Currently, the cosmological data set limits on the sum of light active neutrino masses \cite{nubound}, $\sum m_{\nu} < 0.152$ in the minimal $\Lambda \mathrm{CDM } + \sum m_{\nu}$ model, $\sum m_{\nu} < 0.118$ eV in in the minimal $\Lambda \mathrm{CDM} + \sum m_{\nu}$ model with the high-$l$ polarization data, $\sum m_{\nu} < 0.305$ eV in the DDE model with TT + BAO + PAN + $\tau 0p055$, $\sum m_{\nu} < 0.305$ eV in the DDE model with TT + BAO + PAN + $\tau 0p055$, $\sum m_{\nu} < 0.247$ eV in the DDE model with TTTEEE + BAO + PAN + R16 + $\tau 0p055$ and $\sum m_{\nu} < 0.101$ eV in NPDDE model with TTTEEE + BAO + PAN + $\tau 0p055$. Thus, our obtained value for the sum of the light active neutrino masses is well consistent with the aforementioned bounds arising from cosmology.

 By substituting explicit expressions of $b_{D}, b_{D}, c_D, a_R, b_R, c_{ R}$ from Eq. (\ref{abc}) and the obtained values of $\mathbb{K}_{\mp}, \mathbb{N}_{\mp}$ in Tab. \ref{paraI} and $m_{1, 2}$ in Eq. (\ref{m1m2m3expI}) into Eq. (\ref{relation1}), we get the relations:
 \bea
 x_1&=& \frac{0.211 v_{\varphi^{'}} x_2}{v_{\varphi}},\hs
 y_1 = \frac{1}{\frac{0.113 v_\rho x^2_2}{v^*_\rho x^2_3 y_2}-\frac{0.0228 v_\rho}{v^2_{\varphi^{'}} x^2_3}},\crn
 y_3&=& \frac{2.13 v^*_\rho y_2}{v_\chi}+\frac{0.052 (v^*_\rho)^2 y^2_2}{v_\chi v^2_{\varphi^{'}} x^2_2}. \label{xyzI}
 \eea
\subsection{Effective neutrino mass parameters}
The effective neutrino mass parameters governing the beta decay and neutrinoless double beta decay are defined as
$m_{\beta } =\sqrt{\sum^3_{k=1} \left|U_{ek}\right|^2 m_k^2},\hs \langle m_{ee}\rangle = \left| \sum^3_{k=1} U_{ek}^2 m_k \right|$,
   where $U_{ek}\, (k=1,2,3)$ are the leptonic mixing matrix elements
and $m_{k}$ correspond to the masses of three light neutrinos.
Using the model parameters obtained in subsections \ref{NHsector} and \ref{IHsector}, we find the following numerical values for the above mentioned mass parameters:
\bea
\langle m_{ee}\rangle=\left\{
\begin{array}{l}
1.51\times 10^{-3} \, \mbox{eV}\ \ \ \ \ \  \mbox{for \ \ \ \ NO,} \\
4.88\times 10^{-2} \, \mbox{eV}\ \ \ \ \ \  \mbox{for \ \ \ \ IO,}%
\end{array}%
\right.   \label{meetvalues}
\eea
and \bea
m_{\beta}=\left\{
\begin{array}{l}
8.82\times 10^{-3} \, \mbox{eV}\ \ \ \ \ \  \mbox{for \ \ \ \ NO,} \\
4.94\times 10^{-2} \, \mbox{eV}\ \ \ \ \ \  \mbox{for \ \ \ \ IO.}%
\end{array}%
\right.   \label{mbetvalues}
\eea
The resulting effective neutrino mass parameters in Eqs. (\ref{meetvalues}) and (\ref{mbetvalues}), for both  normal and inverted orderings, are below all the upper bounds arising from present $0\nu \beta \beta $ decay experiments, such as, KamLAND-Zen\cite{KamLAND} $\langle m_{ee} \rangle <0.05\div0.16 \,\mathrm{eV}$, GERDA \cite{Agostini18}$\langle m_{ee} \rangle < 0.12 \div 0.26 \,\mathrm{eV}$, MAJORANA\cite{MAJO} $\langle m_{ee} \rangle < 0.24 \div 0.53 \,\mathrm{eV}$, EXO\cite{EXO1, EXO2, Albert18} $\langle m_{ee} \rangle < 0.17 \div 0.49 \,\mathrm{eV}$, CUORE\cite{CUORE1, CUORE2} $\langle m_{ee} \rangle < 0.11 \div 0.5 \,\mathrm{eV}$. Hence, our obtained effective neutrino mass parameter is beyond the reach of the present and forthcoming $0\nu \beta \beta $-decay experiments.
\section{Quark masses and mixings \label{quark}}
In this section, we show that our model is able to successfully reproduce the observed pattern of SM quark masses and mixing parameters. Indeed, from the quark Yukawa terms given by Eq. (\ref{Lquark0}) and the tensor product of $\Si (18)$ in Appendix \ref{Si18CG}, we can rewrite the Yukawa interactions in the quark sector in the form:
\bea -\mathcal{L}_q &=&  h_{1u} (\bar{Q}_{1 L} \widetilde{H} u_{2 R}+ \bar{Q}_{2 L} \widetilde{H}u_{1 R})+ h_{2u} \bar{Q}_{3 L}\widetilde{H} u_{3 R}+h_{3u} (\bar{Q}_{1 L} \widetilde{\phi}_1 u_{1 R} + \bar{Q}_{2 L} \widetilde{\phi}_2 u_{2 R})\crn
&+& h_{4u} (\bar{Q}_{1 L} \widetilde{\phi}^{'}_2 u_{3 R} + \bar{Q}_{2 L} \widetilde{\phi}^{'}_1 u_{3 R}) + h_{5u} (\bar{Q}_{3 L} \widetilde{\phi}^{'}_2 u_{1 R} + \bar{Q}_{3 L} \widetilde{\phi}^{'}_1 u_{2 R}) \crn
&+& h_{1d} (\bar{Q}_{1 L}H d_{1 R} + \bar{Q}_{2 L} H d_{2 R}) + h_{2d} \bar{Q}_{3 L} H d_{3 R} + h_{3d} (\bar{Q}_{1 L} \phi_2 d_{2 R} +
\bar{Q}_{2 L} \phi_1 d_{1 R})\crn
&+& h_{4d} (\bar{Q}_{1 L} \phi^{'}_1 d_{3 R} + \bar{Q}_{2 L} \phi^{'}_2 d_{3 R}) + h_{5d} (\bar{Q}_{3 L} \phi^{'}_2 d_{1 R} + \bar{Q}_{3 L} \phi^{'}_1 d_{2 R}) +\mathrm{H.c.}\eea
With the VEV alignments of $H$ and $\phi$ as chosen in Eq. (\ref{scalarvev}), the mass Lagrangian of quarks reads
\bea -\mathcal{L}^{\mathrm{mass}}_q &=& h_{1u} v_H (\bar{u}_{1 L} u_{2 R}+ \bar{u}_{2 L} u_{1 R}) + h_{2u} v_H \bar{u}_{3 L} u_{3 R}+h_{3u} (v^*_1\bar{u}_{1 L} u_{1 R} + v^*_2 \bar{u}_{2 L} u_{2 R})\crn
&+& h_{4u}  (v^{'*}_2\bar{u}_{1 L} u_{3 R} + v^{'*}_1\bar{u}_{2 L} u_{3 R}) + h_{5u}  (v^{'*}_2\bar{u}_{3 L}  u_{1 R} + v^{'*}_1\bar{u}_{3 L} u_{2 R})\crn
&+& h_{1d} v_H (\bar{d}_{1 L} d_{1 R} + \bar{d}_{2 L} d_{2 R})  + h_{2d} v_H \bar{d}_{3 L} d_{3 R} + h_{3d} (v_2 \bar{d}_{1 L} d_{2 R} +
v_1 \bar{d}_{2 L} d_{1 R})\crn
&+& h_{4d} (v^{'}_1\bar{d}_{1 L} d_{3 R} +v^{'}_2\bar{d}_{2 L} d_{3 R})+h_{5d} (v^{'}_2\bar{d}_{3 L} d_{1 R} +v^{'}_1\bar{d}_{3 L} d_{2 R})+\mathrm{H.c.}\crn
&\equiv &  (\bar{u}_{1L},\bar{u}_{2L},\bar{u}_{3L}) M_u (u_{1R}, u_{2R},
u_{3R})^T+(\bar{d}_{1L},\bar{d}_{2L},\bar{d}_{3L}) M_d (d_{1R},
d_{2R}, d_{3R})^T \label{Lqmass}
+ \mathrm{H.c.,} \nn\eea
where the up-and down-quark mass matrices are
\bea M_{u} &=&
\left(%
\begin{array}{ccc}
  b_{1u} & a_{1u} & c_{2u} \\
  a_{1u} & b_{2u} & c_{1u}  \\
  g_{2u} & g_{1u} & a_{2u}\\
\end{array}%
\right),  \hs
M_{d} =
\left(%
\begin{array}{ccc}
  a_{1d} & b_{1d} & c_{1d} \\
  b_{2d} & a_{1d} & c_{2d}  \\
  g_{2d} & g_{1d} & a_{2d}\\
\end{array}%
\right),  \label{Mud}
\eea
with
\bea
a_{(1,2)u}&=& h_{(1,2)u} v_H, \hs b_{(1,2)u} =  h_{3u} v^*_{(1,2)}, \hs c_{(1,2)u}=h_{4u} v^{'*}_{(1,2)}, \hs g_{(1,2)u}=h_{5u} v^{'*}_{(1,2)},\crn
a_{(1,2)d}&=& h_{(1,2)d} v_H, \hs b_{(1,2)d} =  h_{3d} v_{(2,1)}, \hs\, c_{(1,2)d}=h_{4d} v^{'}_{(1,2)}, \hs g_{(1,2)d}=h_{5d} v^{'}_{(1,2)}.  \label{abcgud}\eea

Now we turn our attention to the experimental values of the SM quark masses and CKM parameters \cite{Xing:2019vks,Tanabashi:2018oca}:
\begin{eqnarray}
&& m_{u}(MeV)=1.24\pm 0.22, \hspace{3mm}
m_{d}(MeV)=2.69\pm 0.19, \hspace{3mm} m_{s}(MeV)=53.5\pm 4.6,
\label{eq:Qsector-observables}\notag \\
&&m_{c}(GeV)=0.63\pm 0.02,\hspace{3mm} m_{t}(GeV)=172.9\pm 0.4,\hspace{%
3mm} m_{b}(GeV)=2.86\pm 0.03,\hspace{3mm}  \notag \\
&&\sin\theta_{12}=0.2245\pm 0.00044,\hspace{3mm} \sin \theta_{23}=0.0421\pm 0.00076,\hspace{3mm}
\sin \theta_{13}=0.00365\pm 0.00012,\notag\\
&&J=\left(3.18\pm 0.15\right)\times 10^{-5}\,.
\label{eq:Qsector-observables}
\end{eqnarray}%
We look for the eigenvalue problem solutions reproducing the experimental values of the quark masses and the CKM
parameters given by Eq.~(\ref{eq:Qsector-observables}), finding the following solution:
\begin{eqnarray}
&&M_u=\left(
\begin{array}{ccc}
 -27.6375-62.7392 i & -31.3158-66.0758 i & 24.5526\, -34.1468 i \\
 -31.3158-66.0758 i & -36.0052-69.6769 i & 25.5349\, -37.6027 i \\
 24.5526\, -34.1468 i & 25.5349\, -37.6027 i & 25.9351\, +2.79953 i \\
\end{array}
\right) \mathrm{GeV}, \notag\\
&&M_d=\left(
\begin{array}{ccc}
 1.24842 & 1.22041 -0.00504449 i & 0.37787 -0.563654 i \\
 1.22041 +0.00504449 i & 1.24842 & 0.368746 -0.599054 i \\
 0.37787 +0.563654 i & 0.368746+0.599054 i & 0.419347 \\
\end{array}
\right) \mathrm{GeV}.\notag\\
\end{eqnarray}
This show that our model is consistent with and successfully accommodate the experimental values of the physical
observables of the quark sector: the six quark masses, the quark mixing angles and the CP violating phase in the quark sector.

\section{$K-\bar{K}$ and $B-\bar{B}$ mixings.}
\label{secKKbar}
In this section we discuss the implications of our model in the 
FCNC interactions in the down type quark sector. The FCNC Yukawa interactions in the down type quark sector give rise to meson oscillations. Here we focus on the $K-\bar{K}$ mixing, whose corresponding $\Delta M_K$ parameter arises from the following effective Hamiltonian:
\begin{equation}
\mathcal{H}_{eff}^{\left( \Delta S=2\right) }\mathcal{=}\frac{%
G_{F}^{2}m_{W}^{2}}{16\pi ^{2}}\sum_{i}C_{i}\left( \mu \right) O_{i}\left(
\mu \right).
\end{equation}
In our analysis of the $K-\bar{K}$ mixing we follow the approach of \cite{Dedes:2002er,Aranda:2012bv}. As in Ref. \cite{Dedes:2002er,Aranda:2012bv}, the $K-\bar{K}$ mixing in our model mainly arise from the tree level exchange of neutral CP even and CP odd scalars, thus giving rise to the following operators:
\bigskip
\begin{equation}
O_{1}^{LL}=\left( \overline{s}P_{L}d\right) \left( \overline{s}P_{L}d\right)
,\hspace{0.5cm}O_{1}^{RR}=\left( \overline{s}P_{R}d\right) \left( \overline{s%
}P_{R}d\right), \hspace{0.5cm}O_{1}^{LR}=\left( \overline{s}P_{L}d\right)
\left( \overline{s}P_{R}d\right),  \label{op3f}
\end{equation}
where the corresponding Wilson coefficient are given by:
\begin{eqnarray}
C_{1}^{LL} &=&\frac{16\pi ^{2}}{G_{F}^{2}m_{W}^{2}}\left( \sum_{i=1}^{N}%
\frac{y_{H_{i}^{0}\overline{s}_{R}d_{L}}^{2}}{m_{H_{i}^{0}}^{2}}%
-\sum_{i=1}^{N-4}\frac{y_{A_{i}^{0}\overline{s}_{R}d_{L}}^{2}}{%
m_{H_{i}^{0}}^{2}}\right) =\frac{16\pi ^{2}}{G_{F}^{2}m_{W}^{2}}\widetilde{C}%
_{1}^{LL}, \\
C_{1}^{RR} &=&\frac{16\pi ^{2}}{G_{F}^{2}m_{W}^{2}}\left( \sum_{i=1}^{N}%
\frac{y_{H_{i}^{0}\overline{s}_{L}d_{R}}^{2}}{m_{H_{i}^{0}}^{2}}%
-\sum_{i=1}^{N-4}\frac{y_{A_{i}^{0}\overline{s}_{L}d_{R}}^{2}}{%
m_{H_{i}^{0}}^{2}}\right) =\frac{16\pi ^{2}}{G_{F}^{2}m_{W}^{2}}\widetilde{C}%
_{1}^{RR}, \\
C_{2}^{LR} &=&\frac{16\pi ^{2}}{G_{F}^{2}m_{W}^{2}}\left( \sum_{i=1}^{N}%
\frac{y_{H_{i}^{0}\overline{s}_{R}d_{L}}y_{H_{i}^{0}\overline{s}_{L}d_{R}}}{%
m_{H_{i}^{0}}^{2}}-\sum_{i=1}^{N-4}\frac{y_{A_{i}^{0}\overline{s}%
_{R}d_{L}}y_{A_{i}^{0}\overline{s}_{L}d_{R}}}{m_{H_{i}^{0}}^{2}}\right) =%
\frac{16\pi ^{2}}{G_{F}^{2}m_{W}^{2}}\widetilde{C}_{2}^{LR},
\end{eqnarray}
with
\begin{eqnarray}
\widetilde{C}_{1}^{LL} &=&\sum_{i=1}^{N}\frac{y_{H_{i}^{0}\overline{s}%
_{R}d_{L}}^{2}}{m_{H_{i}^{0}}^{2}}-\sum_{i=1}^{N-4}\frac{y_{A_{i}^{0}%
\overline{s}_{R}d_{L}}^{2}}{m_{A_{i}^{0}}^{2}}, \\
\widetilde{C}_{1}^{RR} &=&\sum_{i=1}^{N}\frac{y_{H_{i}^{0}\overline{s}%
_{L}d_{R}}^{2}}{m_{H_{i}^{0}}^{2}}-\sum_{i=1}^{N-4}\frac{y_{A_{i}^{0}%
\overline{s}_{L}d_{R}}^{2}}{m_{A_{i}^{0}}^{2}}, \\
\widetilde{C}_{2}^{LR} &=&\sum_{i=1}^{N}\frac{y_{H_{i}^{0}\overline{s}%
_{R}d_{L}}y_{H_{i}^{0}\overline{s}_{L}d_{R}}}{m_{H_{i}^{0}}^{2}}%
-\sum_{i=1}^{N-4}\frac{y_{A_{i}^{0}\overline{s}_{R}d_{L}}y_{A_{i}^{0}%
\overline{s}_{L}d_{R}}}{m_{A_{i}^{0}}^{2}}.
\end{eqnarray}%
Here $N=11$ is the number of CP even scalars of our model, whereas $N-4=7$ is the number of CP odd scalars. Let us note that our model is an extended 8HDM where the scalar sector is enlarged by the inclusion of 3 real gauge singlet scalars.

\begin{figure}[h]
\vspace{-0.75cm}
\hspace{-0.25cm}
\includegraphics[width=0.6\textwidth]{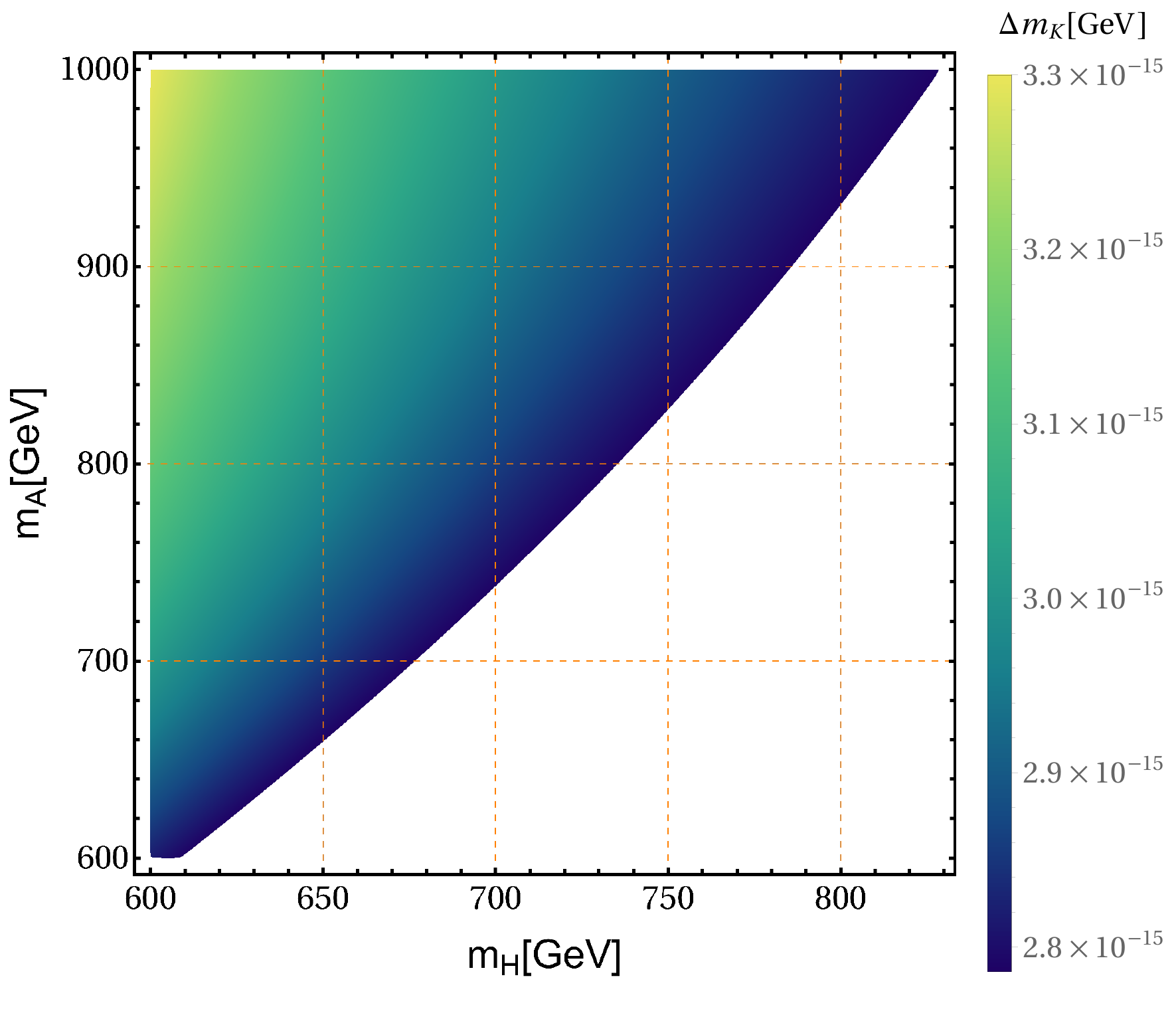}
\vspace{-0.75cm}
\caption{Allowed region in the $m_H-m_A$ plane consistent with the constraint arising from $K-\bar{K}$ mixing.}
\label{KKbar}
\end{figure}

\begin{figure}[h]
\includegraphics[width=0.52\textwidth]{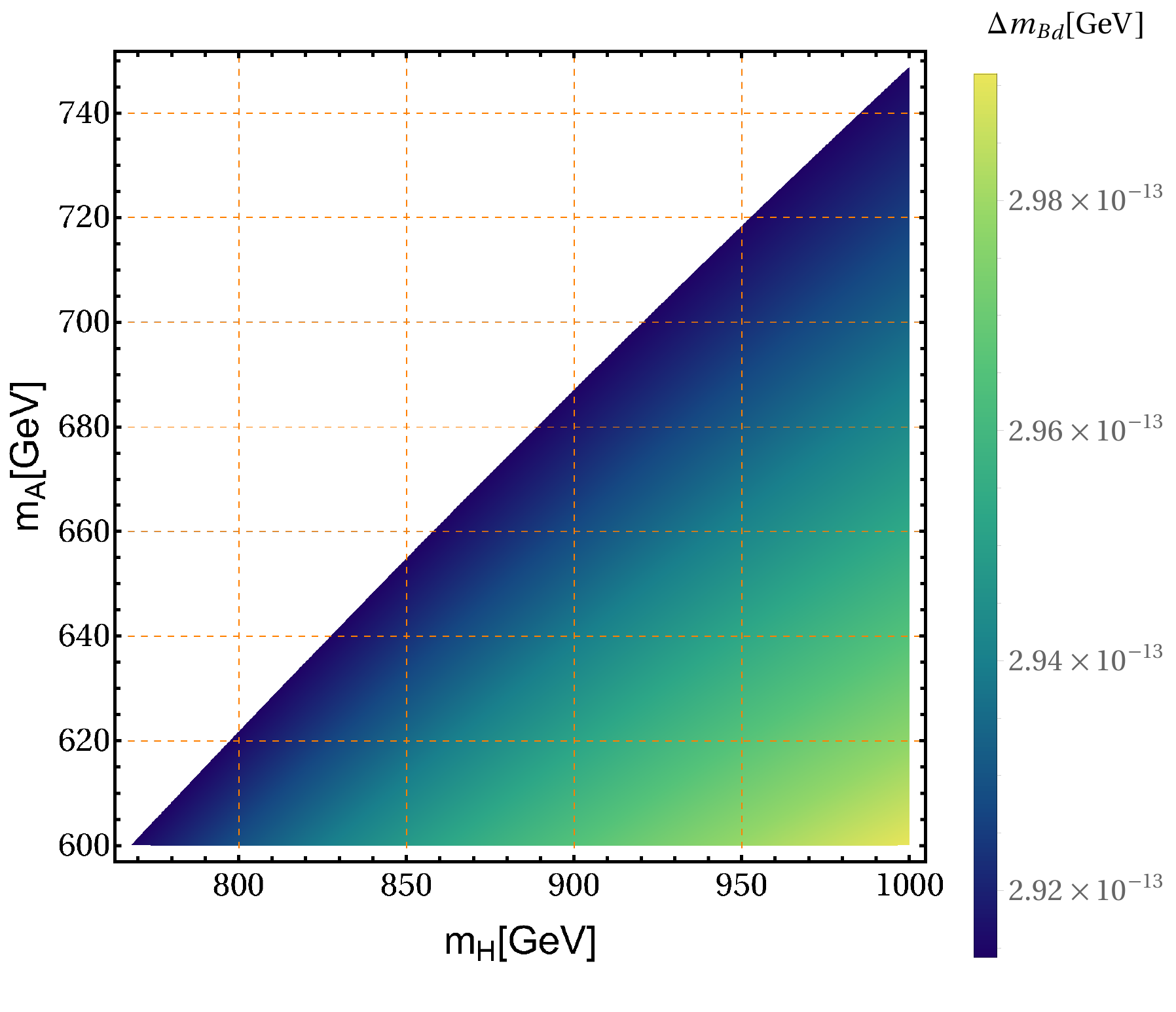}\includegraphics[width=0.52\textwidth]{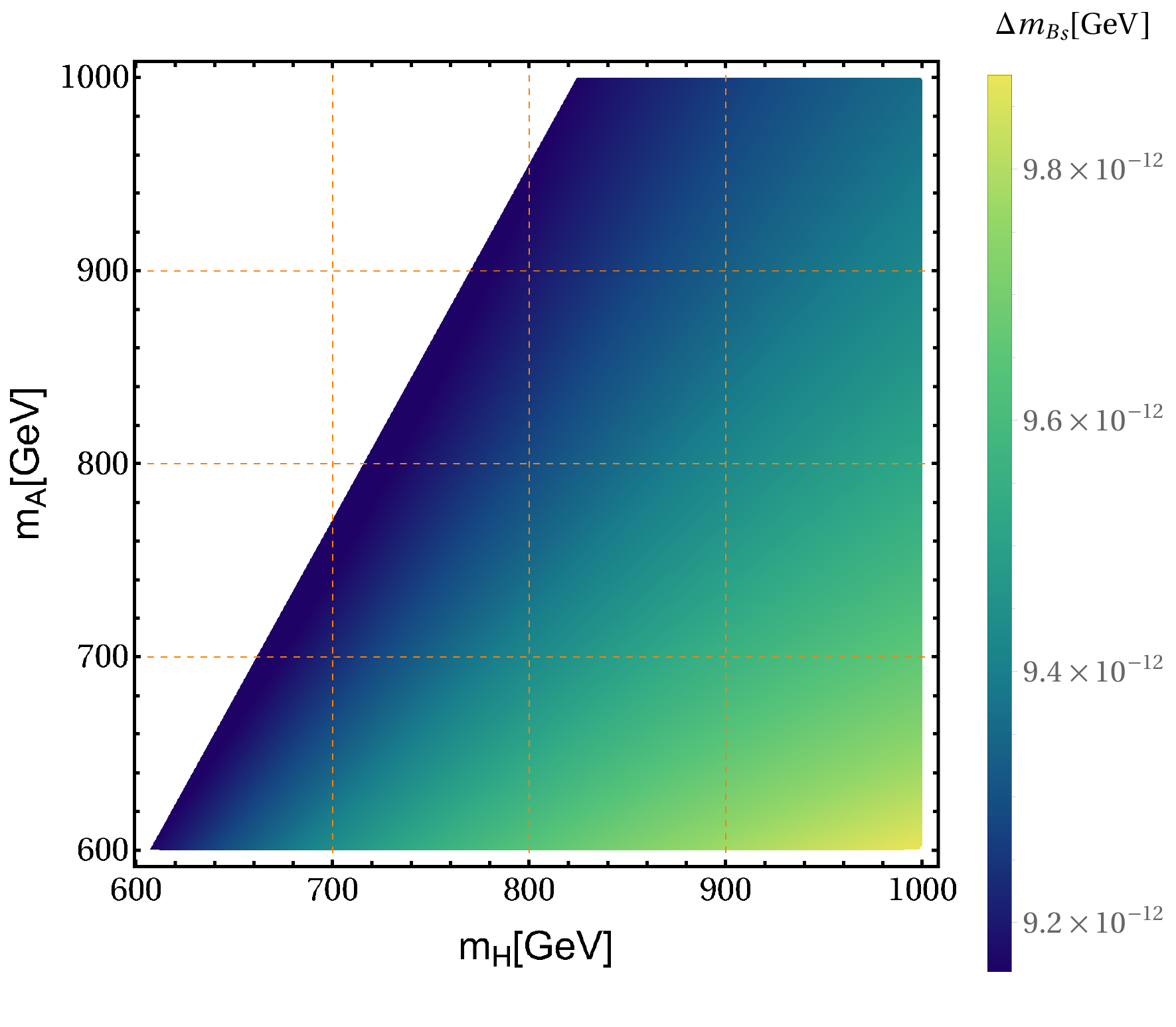}
\vspace{-0.75cm}
\caption{Allowed region in the $m_H-m_A$ plane consistent with the constraint arising from $B_d-\bar{B}_d$ (left-plot) and $B_s-\bar{B}_s$ (right-plot) mixings. The couplings of the flavor changing neutral Yukawa interactions have been set to be equal to $2\times 10^{-4}$ and $10^{-3}$ for the left and right plots, respectively.}
\label{BBbar}
\end{figure}

On the other hand, the $K-\bar{K}$ mass splitting has the form:
\begin{eqnarray}
\Delta m_{K} &=&2Re\left\langle \overline{K^{0}}\left\vert \mathcal{H}%
_{eff}^{\left( \Delta S=2\right) }\right\vert K^{0}\right\rangle \nonumber \\
&=&\frac{G_{F}^{2}m_{W}^{2}}{12\pi ^{2}}m_{K}f_{K}^{2}\eta _{K}B_{K}\left[
P_{2}^{LR}C_{2}^{LR}+P_{1}^{LL}\left( C_{1}^{LL}+C_{1}^{RR}\right) \right] \nonumber\\
&=&\frac{4}{3}m_{K}f_{K}^{2}\eta _{K}B_{K}\left[ P_{2}^{LR}\widetilde{C}%
_{2}^{LR}+P_{1}^{LL}\left( \widetilde{C}_{1}^{LL}+\widetilde{C}%
_{1}^{RR}\right) \right].
\end{eqnarray}
Then, it follows that:
\begin{equation}
M_{12}=\frac{\Delta m_{K}}{m_{K}}=\frac{4}{3}f_{K}^{2}\eta _{K}B_{K}\left[
P_{2}^{LR}\widetilde{C}_{2}^{LR}+P_{1}^{LL}\left( \widetilde{C}_{1}^{LL}+%
\widetilde{C}_{1}^{RR}\right) \right].
\end{equation}
Using the following parameters \cite{Dedes:2002er,Aranda:2012bv}: 
\begin{eqnarray}
&&\Delta m_{K} =3.483\times 10^{-12}MeV,\hspace{0.25cm} m_{K}=497.614MeV, \hspace{0.25cm} f_{K} =160MeV, \crn
&&B_{K}=0.85\pm 0.15, \hspace{0.25 cm}\sqrt{B_{K}}f_{K} =135MeV,\hspace{0.25cm}\eta _{K}=0.57, P_{1}^{LL} =-9.3,\crn
&&P_{2}^{LR}=30.6,\hspace{0.25cm} M_{12} =\frac{\Delta m_{K}}{m_{K}}=7.2948\times 10^{-15}.
\end{eqnarray}
We get the following constraint arising from $K-\bar{K}$ mixing:
\begin{equation}
P_{2}^{LR}\widetilde{C}_{2}^{LR}+P_{1}^{LL}\left( \widetilde{C}_{1}^{LL}+%
\widetilde{C}_{1}^{RR}\right) \leq 4.41\times 10^{-19}MeV^{-2}.
\end{equation}
Given the large amount parametric freedom in both fermion and scalar sectors of our model, such constraint can be fulfilled. To show explicitly that the above given constraint resulting from $K-\bar{K}$ mixing is successfully fullfilled and given the large amount of parameters in our model, we consider a simplified benchmark scenario where:
\begin{eqnarray}
y_{H_{1}^{0}\overline{s}_{R}d_{L}} &=&y_{H_{1}^{0}\overline{s}%
_{L}d_{R}}=y_{H_{j}^{0}\overline{s}_{R}d_{L}}=y_{H_{j}^{0}\overline{s}%
_{L}d_{R}}=y_{A_{i}^{0}\overline{s}_{R}d_{L}}=y_{A_{i}^{0}\overline{s}%
_{L}d_{R}}=y, \\
m_{H_{1}^{0}} &=&m_{h}=126GeV,\hspace{1.5cm}m_{H_{j}^{0}}=m_{H}, \\
m_{A_{j}^{0}} &=&m_{A},\hspace{1.5cm}j=2,3,\cdots ,11,\hspace{1.5cm}%
i=1,2,\cdots ,7.
\end{eqnarray}
Here we identified $H_{1}^{0}$ with $126$ GeV SM like Higgs boson. We plot
in Figure \ref{KKbar} the allowed parameter space in the $m_{H}-m_{A}$ plane
consistent with the constraint arising from $K-\bar{K}$ mixing in the
aforementioned simplified benchmark scenario of our model. Here, for the
sake of simplicity we have set $y=2\times 10^{-5}$. Figure shows that our
model can successfully accommodate the constraint arising from $K-\bar{K}$
mixing, in a large region of parameter space.
\Antonio{It is worth mentioning that we are considering a scenario where the down type quark Yukawa couplings have been taken to be real, which implies that the  CP violation in the quark sector only arises from the up type quark sector. Consequently, in that scenario the stringent constraints that are usually imposed on any possible new contribution to the $K-\bar{K}$ mixing from CP violating processes, are not relevant in our case. Furthermore, in what regards the $B_d-\bar{B}_d$ and $B_s-\bar{B}_s$ mixings, we have numerically checked that in the aforementioned simplified benchmark scenario and above described region of parameter space with flavour violating Yukawa coupling of the order of $10^{-5}$, the obtained values for the $\Delta m_{B_d}$ and $\Delta m_{B_s}$ parameters are about two and four orders of magnitude, respectively, below their corresponding experimental values $\Delta m_{B_d} =3.337\times 10^{-10}MeV$ and $\Delta m_{B_s} =1.042\times 10^{-8}MeV$. On the other hand, in the simplified benchmark scenario, when the couplings of the flavor changing neutral Yukawa interactions responsible for the $B_d-\bar{B}_d$ and $B_s-\bar{B}_s$ mixings take values of about $2\times 10^{-4}$ and $10^{-3}$, respectively, the $B_d-\bar{B}_d$ and $B_s-\bar{B}_s$ mixings arising from these interactions reach values close to their experimental upper bounds, thus giving rise to the allowed regions in the $m_H-m_A$ plane consistent with these constraints and displayed in Figure \ref{BBbar}.}

\Antonio{\section{Charged lepton flavor violation}
\label{secclfv}
In this section we analyze the implications of our model in charged lepton flavor violation. From the charged lepton Yukawa interactions it follows that the $\mu \rightarrow e\gamma $ and $\tau \rightarrow e\gamma $ decays are absent in our model since there are no flavor changing neutral (FCN) scalar interactions involving the first family of SM charged leptons with the remaining ones. However, the are flavor changing neutral scalar interactions involving the tau and the muon that give rise to the $\tau \rightarrow \mu \gamma $ decay. The $\tau\rightarrow \mu \gamma $ decay appears at one loop level and involve the exchange of electrically neutral  CP even and CP odd scalars and the tau and muon leptons running in the internal lines of the loop. Its branching ratio is given by \cite{Lindner:2016bgg}:
\begin{eqnarray}
Br\left( \tau \rightarrow \mu \gamma \right)  &\simeq &\frac{3\left( 4\pi
\right) ^{3}\alpha _{EM}}{4G_{F}^{2}}\left( \frac{1}{16\pi ^{2}}\right)
^{2}\left\vert \sum_{l=\mu ,\tau }\sum_{i=1}^{N}\frac{y_{H_{i}^{0}\tau 
\overline{l}}y_{H_{i}^{0}\overline{l}\mu }}{m_{H_{i}^{0}}^{2}}\left\{ \frac{1%
}{6}-\frac{m_{l}}{m_{\mu }}\left[ \frac{3}{2}+\ln \left( \frac{m_{l}^{2}}{%
m_{H_{i}^{0}}^{2}}\right) \right] \right\} \right.   \notag \\
&&+\left. \sum_{l=e,\mu }\sum_{i=1}^{N-4}\frac{y_{A_{i}^{0}\tau \overline{l}%
}y_{A_{i}^{0}\overline{l}\mu }}{m_{A_{i}^{0}}^{2}}\left\{ \frac{1}{6}+\frac{%
m_{l}}{m_{\mu }}\left[ \frac{3}{2}+\ln \left( \frac{m_{l}^{2}}{%
m_{A_{i}^{0}}^{2}}\right) \right] \right\} \right\vert ^{2}.
\end{eqnarray}
To simplify our analysis we choose the benchmark scenario described in section \ref{KKbar}. We display in Figure the allowed parameter space in the $m_H-m_A$ plane consistent with the existing $\tau\to\mu\gamma$ experimental constraints. Here we set the Yukawa couplings of the FCN leptonic Yukawa interactions equal to $10^{-2}$ and $5\times 10^{-3}$ for the left and right plots, respectively. Consequently, our model is highly consistent with the constraints arising from lepton flavor violating decays for a large region of parameter space. Furthermore, it allows for charged lepton flavor violating (CLFV) processes within the reach of future experimental sensitivity.
\begin{figure}[h]
\vspace{-0.75cm}
\hspace{-0.25cm}
\includegraphics[width=0.52\textwidth]{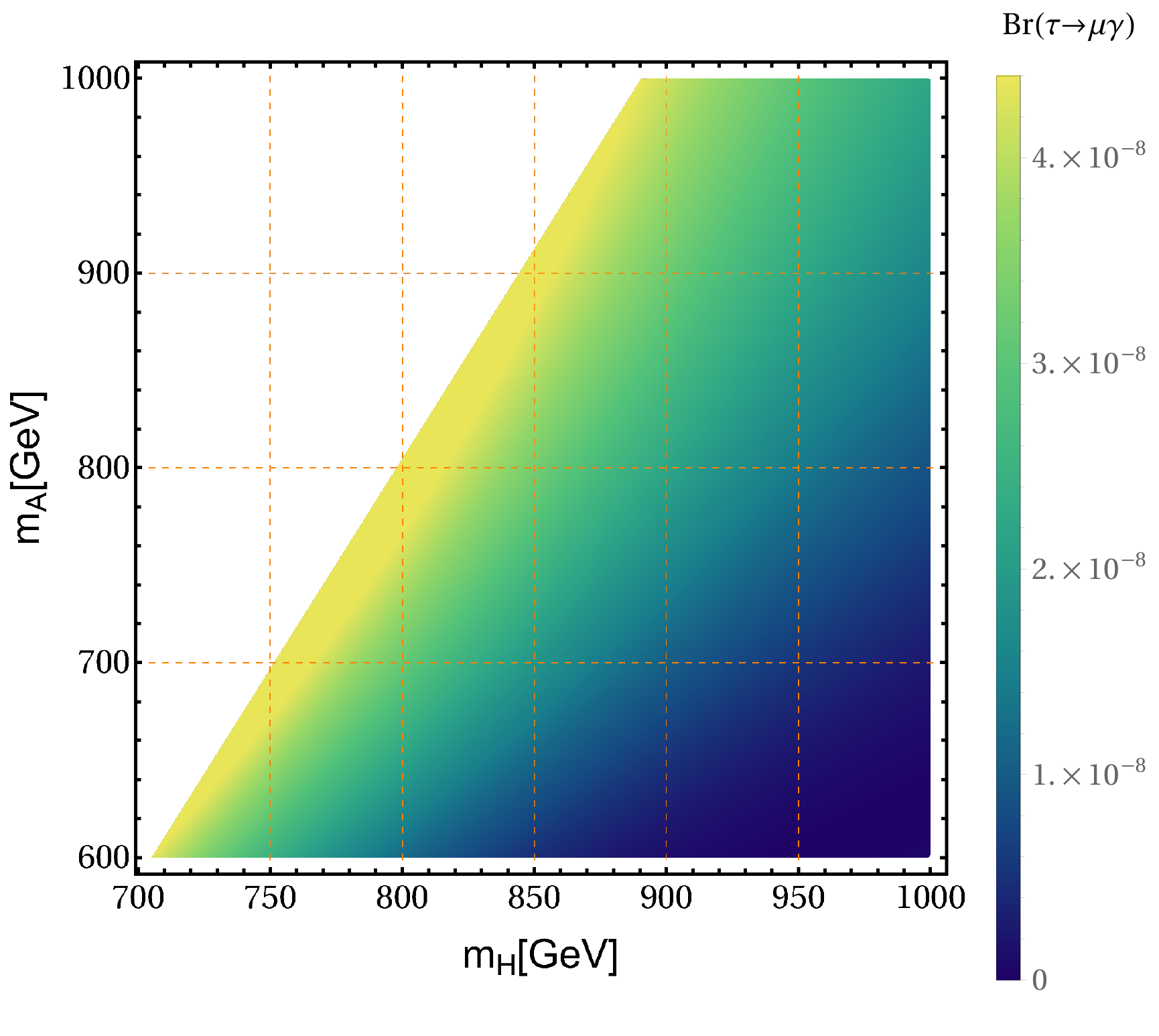}\includegraphics[width=0.52\textwidth]{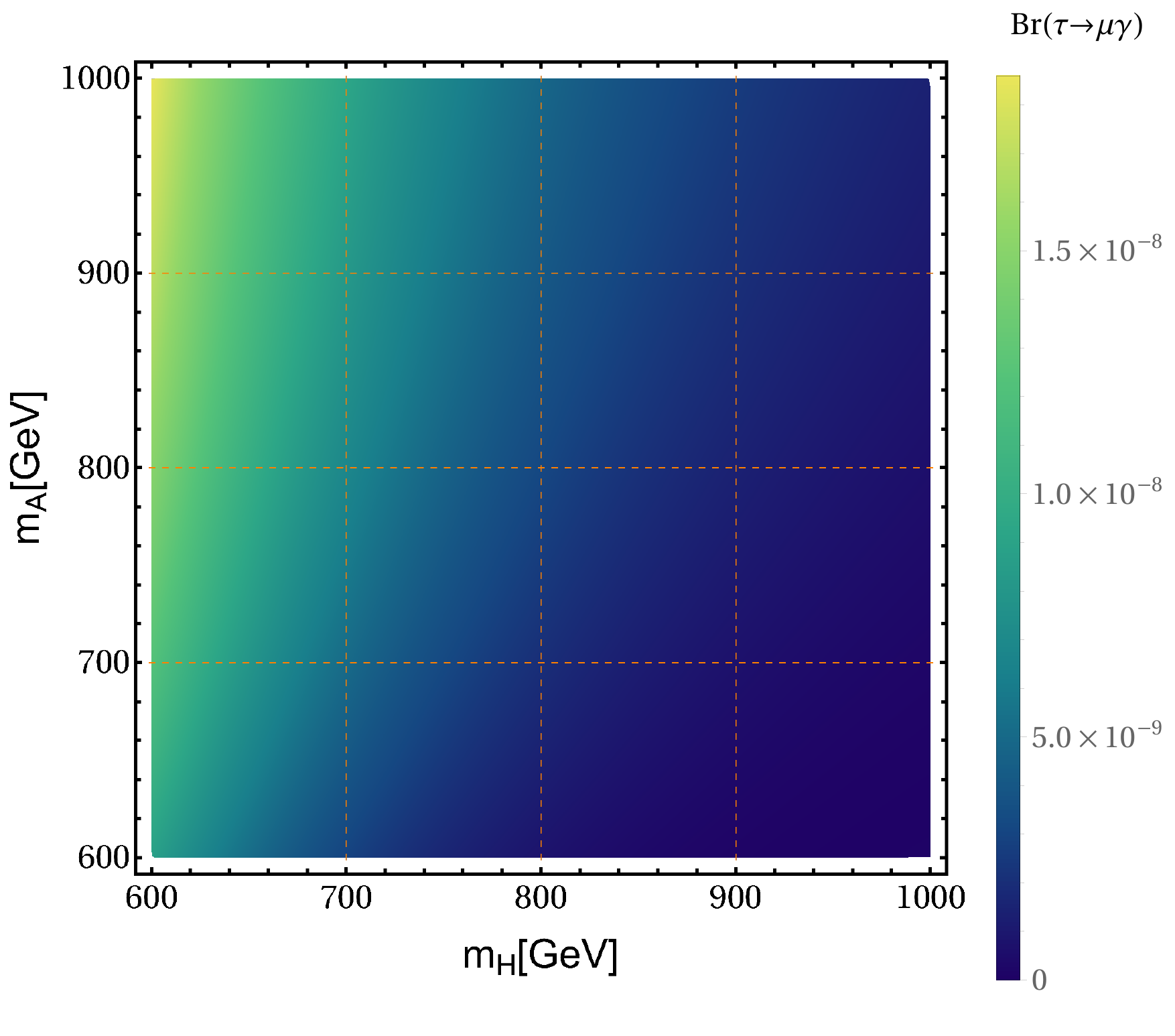}
\vspace{-0.75cm}
\caption{Allowed region in the $m_H-m_A$ plane consistent with the charged lepton flavor-violating constraints. The Yukawa couplings of the FCN leptonic Yukawa interactions have been set to be equal to $10^{-2}$ and $5\times 10^{-3}$ for the left and right plots, respectively.}
\label{clfv}
\end{figure}}
\vspace{-0.5cm}

\section{\label{conclusion}Conclusions}
We have built a renormalizable theory where the SM gauge symmetry is extended by the inclusion of the 
global $U(1)_X$ symmetry and the $\Sigma (18)\times Z_4$ discrete group which leads to a successful fit of SM fermion masses and mixings. The right-handed neutrinos are responsible for the generation of the tiny active neutrino masses through a type I seesaw mechanism mediated by heavy right handed Majorana neutrinos. The resulting physical parameters are in accordance with the recent experimental dat\Vien{a.} 
\Vien{W}e find values for the effective neutrino mass parameters equal to
$\langle m_{ee}\rangle=1.51\times 10^{-3}\, \mathrm{eV}$ for normal ordering and $\langle m_{ee}\rangle =4.88\times 10^{-2} \, \mathrm{eV}$ for inverted ordering, which are well in accordance with the recent experimental limits on neutrinoless double beta decay. The proposed model also successfully accommodates the recent experimental values of the physical observables of the quark sector, including the six quark masses, the quark mixing angles and the CP violating phase in the quark sector. \Antonio{Furthermore, our model can also accommodate the constraints arising from $K-\bar{K}$, $B_d-\bar{B}_d$ and $B_s-\bar{B}_s$ mixings as well as the constraits arising from charged lepton flavor violation.}
\section*{Acknowledgments}
This research is funded by 
ANID-Chile FONDECYT 1210378.
\Revised{H. N. Long  acknowledges the financial support of the International Centre of Physics at the Institute of Physics, Vietnam Academy of Science and Technology with Grant  number CIP.2021.02.}
\appendix
\section{\label{Si18CG} The Clebsch-Gordan coefficients of $\Sigma(18)$ group}
$\Sigma(18)$ is the simplest non-trivial group of $\Sigma(2N^2)$ with $N=3$ which is isomorphic to $(Z_3 \times Z^\prime_3)\rtimes Z_2$. It has 18 elements, $b^k a^m a^{\prime n}$ for $k=0, 1$ and $m, n=0,1,2$,
where $a, a^{'}$ and $b$ satisfy $a^3=a^{\prime 3}=e$, $b^2=e$ , $a a^{\prime}=a^{\prime} a$ and $bab=a^{'}$. All elements of $\Sigma(18)$ are divided into
nine conjugacy classes with $1_{+0}$,
$1_{+1}$, $1_{+2}$, $1_{-0}$,
$1_{-1}$, $1_{-2}$, $2_{10}$,
$2_{20}$ and $2_{21}$ as its nine irreducible representations. The tensor products between doublets of $\Sigma(18)$ are given by \cite{Ishi}:
\bea
&&
2_{10}\left(%
\begin{array}{c}
 x_1  \\
  x_2  \\
\end{array}%
\right) \otimes 2_{10}\left(%
\begin{array}{c}
 y_1  \\
 y_2  \\
\end{array}%
\right) =1_{+1}(x_1y_2+x_2y_1)\oplus 1_{-1}(-x_1y_2+x_2y_1)
\oplus 2_{20}\left(%
\begin{array}{c}
 x_1y_1  \\
  x_2y_2  \\
\end{array}%
\right),\label{CG210210}\crn
&&
2_{20}\left(%
\begin{array}{c}
 x_1  \\
  x_2  \\
\end{array}%
\right) \otimes 2_{20}\left(%
\begin{array}{c}
 y_1  \\
 y_2  \\
\end{array}%
\right) =1_{+2}(x_1y_2+x_2y_1)\oplus 1_{-2}(-x_1y_2+x_2y_1)
\oplus 2_{10}\left(%
\begin{array}{c}
 x_1y_1  \\
  x_2y_2  \\
\end{array}%
\right),\label{CG220220}\crn
&&
2_{21}\left(%
\begin{array}{c}
 x_1  \\
  x_2  \\
\end{array}%
\right) \otimes 2_{21}\left(%
\begin{array}{c}
 y_1  \\
 y_2  \\
\end{array}%
\right) =1_{+0}(x_1y_2+x_2y_1)\oplus 1_{-0}(-x_1y_2+x_2y_1)
\oplus 2_{21}\left(%
\begin{array}{c}
 x_2y_2  \\
  x_1y_1  \\
\end{array}%
\right),\label{CG221221}\crn
&&2_{20}\left(%
\begin{array}{c}
 x_1  \\
  x_2  \\
\end{array}%
\right) \otimes 2_{10}\left(%
\begin{array}{c}
 y_1  \\
 y_2  \\
\end{array}%
\right) =1_{+0}(x_1y_1+x_2y_2)\oplus 1_{-0}(x_1y_1-x_2y_2)
\oplus 2_{21}\left(%
\begin{array}{c}
 x_1y_2  \\
  x_2y_1  \\
\end{array}%
\right), \nonumber \\
&& 2_{21}\left(%
\begin{array}{c}
 x_1  \\
  x_2  \\
\end{array}%
\right) \otimes 2_{10}\left(%
\begin{array}{c}
 y_1  \\
 y_2  \\
\end{array}%
\right) =1_{+2}(x_1y_2+x_2y_1)\oplus 1_{-2}(-x_1y_2+x_2y_1)
\oplus 2_{10}\left(%
\begin{array}{c}
 x_2y_2  \\
  x_1y_1  \\
\end{array}%
\right), \crn
&&
2_{21}\left(%
\begin{array}{c}
 x_1  \\
  x_2  \\
\end{array}%
\right) \otimes 2_{20}\left(%
\begin{array}{c}
 y_1  \\
 y_2  \\
\end{array}%
\right) =1_{+1}(x_1y_1+x_2y_2)\oplus 1_{-1}(x_1y_1-x_2y_2)
\oplus 2_{20}\left(%
\begin{array}{c}
 x_1y_2  \\
  x_2y_1  \\
\end{array}%
\right), \label{CG22} \hspace{0.85 cm}
\eea
where  $x_i, y_i\, (i=1, 2)$ are
the components of two different representations.

The tensor products between singlets and doublets of $\Sigma(18)$ are obtained as \cite{Ishi}:
\bea
&&1_{\pm 0} (x)\otimes 2_{21}\left(%
\begin{array}{c}
 y_1  \\
 y_2  \\
\end{array}%
\right) =2_{21}\left(%
\begin{array}{c}
 x y_1  \\
 x y_2  \\
\end{array}%
\right), \hs 1_{\pm 1} (x)\otimes 2_{21}\left(%
\begin{array}{c}
 y_1  \\
 y_2  \\
\end{array}%
\right) =2_{20}\left(%
\begin{array}{c}
 x y_2  \\
 x y_1  \\
\end{array}%
\right), \crn
&& 1_{\pm 2} (x)\otimes 2_{21}\left(%
\begin{array}{c}
 y_1  \\
 y_2  \\
\end{array}%
\right) =2_{10}\left(%
\begin{array}{c}
 x y_1  \\
 x y_2  \\
\end{array}%
\right), \hs 1_{\pm 0} (x)\otimes 2_{20}\left(%
\begin{array}{c}
 y_1  \\
 y_2  \\
\end{array}%
\right) =2_{20}\left(%
\begin{array}{c}
 x y_1  \\
 x y_2  \\
\end{array}%
\right), \crn
&& 1_{\pm 1} (x)\otimes 2_{20}\left(%
\begin{array}{c}
 y_1  \\
 y_2  \\
\end{array}%
\right) =2_{10}\left(%
\begin{array}{c}
 x y_2  \\
 x y_1  \\
\end{array}%
\right), \hs 1_{\pm 2} (x)\otimes 2_{20}\left(%
\begin{array}{c}
 y_1  \\
 y_2  \\
\end{array}%
\right) =2_{21}\left(%
\begin{array}{c}
 x y_2  \\
 x y_1  \\
\end{array}%
\right),\crn
&&1_{\pm 0} (x)\otimes 2_{10}\left(%
\begin{array}{c}
 y_1  \\
 y_2  \\
\end{array}%
\right) =2_{10}\left(%
\begin{array}{c}
 x y_1  \\
 x y_2  \\
\end{array}%
\right), \hs 1_{\pm 1} (x)\otimes 2_{10}\left(%
\begin{array}{c}
 y_1  \\
 y_2  \\
\end{array}%
\right) =2_{21}\left(%
\begin{array}{c}
 x y_1  \\
 x y_2  \\
\end{array}%
\right), \crn
&& 1_{\pm 2} (x)\otimes 2_{10}\left(%
\begin{array}{c}
 y_1  \\
 y_2  \\
\end{array}%
\right) =2_{20}\left(%
\begin{array}{c}
 x y_2  \\
 x y_1  \\
\end{array}%
\right).\label{CG12}
\eea
The tensor products between singlets of $\Sigma(18)$ are obtained as \cite{Ishi}:
\bea
&& 1_{\pm 0}(x) \otimes 1_{\pm 0} (y) = 1_{+ 0} (xy), \hs 1_{\pm 1}(x) \otimes 1_{\pm 1} (y) = 1_{+ 2} (xy), \crn
&&1_{\pm 2}(x) \otimes 1_{\pm 2} (y) = 1_{+ 1} (xy),\hs 1_{\pm 1}(x) \otimes 1_{\pm 0} (y) = 1_{+ 1} (xy), \crn
&& 1_{\pm 2}(x) \otimes 1_{\pm 0} (y) = 1_{+ 2} (xy), \hs 1_{\pm 2}(x) \otimes 1_{\pm 1} (y) = 1_{+ 0} (xy), \crn
&& 1_{\pm 0}(x) \otimes 1_{\mp 0} (y) = 1_{-0} (xy), \hs 1_{\pm 1}(x) \otimes 1_{\mp 1} (y) = 1_{- 2} (xy), \crn
&& 1_{\pm 2}(x) \otimes 1_{\mp 2} (y) = 1_{-1} (xy), \hs 1_{\pm 1}(x) \otimes 1_{\mp 0} (y) = 1_{- 1} (xy), \crn
&& 1_{\pm 2}(x) \otimes 1_{\mp 0} (y) = 1_{- 2} (xy),\hs 1_{\pm 2}(x) \otimes 1_{\mp 1} (y) = 1_{- 0} (xy). \label{CG11}
\eea
The rules to conjugate of all the representations of $\Sigma(18)$ are given by:
\bea
&&1_{+0}^*(x^*)=1_{+0}(x^*),\,\, 1_{+1}^*(x^*)=1_{+2}(x^*),\,\, 1_{+2}^*(x^*)=1_{+1}(x^*),\crn
&& 1_{-0}^*(x^*)=1_{-0}(x^*),\,\, 1_{-1}^*(x^*)=1_{-2}(x^*),\,\, 1_{-2}^*(x^*)=1_{-1}(x^*),\\
&&2^*_{10}\left(%
\begin{array}{c}
 x^*_1  \\
  x^*_2  \\
\end{array}%
\right)=2_{20}\left(%
\begin{array}{c}
 x^*_1  \\
  x^*_2  \\
\end{array}%
\right), \hs 2^*_{20}\left(%
\begin{array}{c}
 x^*_1  \\
  x^*_2  \\
\end{array}%
\right)=2_{10}\left(%
\begin{array}{c}
 x^*_1  \\
  x^*_2  \\
\end{array}%
\right), \crn
 && 2^*_{21}\left(%
\begin{array}{c}
 x^*_1  \\
  x^*_2  \\
\end{array}%
\right)=2_{21}\left(%
\begin{array}{c}
 x^*_1  \\
  x^*_2  \\
\end{array}%
\right). \eea
\section{\label{potential} Renormalizable Higgs potential invariant under $\mathbf{\mathrm{G}}$ symmetry}
The general renormalizable potential invariant under $\mathrm{G}$ 
symmetry is a sum of the following components\footnote{Here, we have used the notation: $\mathcal{V}(a_1\rightarrow a_2,b_1\rightarrow b_2,\cdots)
\equiv \mathcal{V}(a_1, b_1,\cdots)\!\!\!\mid_{\{a_1=a_2, b_1=b_2,\cdots \}}$.}:
\bea \mathcal{V}_{\mathrm{total}}&=& \mathcal{V}(H)+ \mathcal{V}(\phi)+ \mathcal{V}(\phi^{'})+
\mathcal{V}(\varphi)+\mathcal{V}(\varphi^{'})+ \mathcal{V}(\chi)+ \mathcal{V}(\rho)+ \mathcal{V}(H,\phi) + \mathcal{V}(H,\phi^')\crn
&+& \mathcal{V}(H, \varphi)+ \mathcal{V}(H, \varphi^')+\mathcal{V}(H, \chi)+\mathcal{V}(H, \rho)+\mathcal{V}(\phi, \phi^') +  \mathcal{V}(\phi, \varphi) +\mathcal{V}(\phi, \varphi^')\crn
&+& \mathcal{V}(\phi,\chi)+ \mathcal{V}(\phi, \rho) +\mathcal{V}(\phi^', \varphi) +\mathcal{V}(\phi^', \varphi^')
+\mathcal{V}(\phi^', \chi) +\mathcal{V}(\phi^', \rho)+\mathcal{V}(\varphi, \varphi^')\crn
&+&\mathcal{V}(\varphi, \chi)+\mathcal{V}(\varphi, \rho) +\mathcal{V}(\varphi^', \chi)
+\mathcal{V}(\varphi^', \rho) +\mathcal{V}(\chi, \rho)+\mathcal{V}_{\mathrm{three}}+\mathcal{V}_{\mathrm{four}},\label{Vtotal}\eea
where
\bea
&&\mathcal{V}(H)=\mu_{H}^2 H^\+ H +\lambda^H ({H}^\+H)_{1_{+0}}({H}^\+H)_{1_{+0}},\label{VH}\\
&&\mathcal{V}(\phi)=\mu^2_\phi \phi^\+\phi +\la^{\phi}_1 (\phi^\+\phi)_{1_{+0}}(\phi^\+\phi)_{1_{+0}}
+\la^{\phi}_2 (\phi^\+\phi)_{1_{-0}}(\phi^\+\phi)_{1_{-0}}+\la^{\phi}_3 (\phi^\+\phi)_{2_{21}}(\phi^\+\phi)_{2_{21}}, \label{Vphi}\\
&& \mathcal{V}(\phi^')=\mathcal{V}(\phi\rightarrow \phi^'), \hs
\mathcal{V}(\varphi)=\mathcal{V}(H\rightarrow \varphi), \hs
\mathcal{V}(\varphi^')=\mathcal{V}(\phi\rightarrow \varphi^'), \hs
\mathcal{V}(\chi)=\mathcal{V}(\phi\rightarrow \chi), \label{Vchi}\\
&&\mathcal{V}(\rho)=\mathcal{V}(H\rightarrow \rho), \label{Vrho}\\
&&\mathcal{V}(H, \phi) =\la^{H \phi}_{1}(H^\+ H)_{1_{+0}} (\phi^\+ \phi)_{1_{+0}} +\la^{H \phi}_{2} (H^\+ \phi)_{2_{21}} (\phi^\+ H)_{2_{21}}\crn
&&\hspace{1.4 cm} \, +\la^{H \phi}_{3} (H^\+ \phi)_{2_{21}} (H^\+ \phi)_{2_{21}}+\la^{H \phi}_{4} (\phi^\+ H)_{2_{21}} (\phi^\+ H)_{2_{21}}\Vien{,} \hs 
\mathcal{V}(H, \phi^') =\mathcal{V}(H, \phi\rightarrow \phi^'), \label{VHphip}\\
&&\mathcal{V}(H, \varphi) =\la^{H \varphi}_1 (H^\+ H)_{1_{+0}} (\varphi^\+ \varphi)_{1_{+0}}
+ \la^{H \varphi}_2 (H^\+ \varphi)_{1_{-1}} (\varphi^\+ H)_{1_{-2}}, \label{VHvarphi}\\
&&\mathcal{V}(H, \varphi^') =\la^{H \varphi^'}_{1}(H^\+ H)_{1_{+0}} (\varphi^{'\+} \varphi^')_{1_{+0}} +\la^{H \varphi^'}_{2} (H^\+ \varphi^')_{2_{10}} (\varphi^{'\+} H)_{2_{20}}, \label{VHvarphip}\\
&&\mathcal{V}(H, \chi) =\la^{H \chi}_1(H^\+ H)_{1_{+0}} (\chi^\+ \chi)_{1_{+0}} +\la^{H \chi}_2 (H^\+ \chi)_{2_{21}} (\chi^\+ H)_{2_{21}}, \label{VHchi}\\
&&\mathcal{V}(H, \rho) =\la^{H \rho}_1(H^\+ H)_{1_{+0}} (\rho^\+ \rho)_{1_{+0}} +\la^{H \rho}_2 (H^\+ \rho)_{1_{+2}} (\rho^\+ H)_{1_{+1}}, \label{VHrho}\\
&&\mathcal{V}(\phi, \phi^') =\la^{\phi \phi^'}_1(\phi^\+ \phi)_{1_{+0}} (\phi^{'\+} \phi^')_{1_{+0}} +\la^{\phi \phi^'}_2(\phi^\+ \phi)_{1_{-0}} (\phi^{'\+} \phi^')_{1_{-0}}+\la^{\phi \phi^'}_3 (\phi^\+ \phi)_{2_{21}} (\phi^{'\+} \phi^')_{2_{21}}\crn
&&\hspace{1.4 cm} +\, \la^{\phi \phi^'}_4(\phi^\+ \phi^')_{1_{+0}} (\phi^{'\+} \phi)_{1_{+0}} +\la^{\phi \phi^'}_5(\phi^\+ \phi^')_{1_{-0}} (\phi^{'\+} \phi)_{1_{-0}}+\la^{\phi \phi^'}_6 (\phi^\+ \phi^')_{2_{21}} (\phi^{'\+} \phi)_{2_{21}}, \label{Vphiphip}\\
&&\mathcal{V}(\phi, \varphi)=\la^{\phi \varphi}_1(\phi^\+ \phi)_{1_{+0}} (\varphi^\+ \varphi)_{1_{+0}} + \la^{\phi \varphi}_2 (\phi^\+ \varphi)_{2_{20}} (\varphi^\+ \phi)_{2_{10}}, \label{Vphivarphi}\\
&&\mathcal{V}(\phi, \varphi^') =\la^{\phi \varphi^'}_1(\phi^\+ \phi)_{1_{+0}} (\varphi^{'\+} \varphi^')_{1_{+0}} +\la^{\phi \varphi^'}_2(\phi^\+ \phi)_{1_{-0}} (\varphi^{'\+} \varphi^')_{1_{-0}}+\la^{\phi \varphi^'}_3 (\phi^\+ \phi)_{2_{21}} (\varphi^{'\+} \varphi^')_{2_{21}}\crn
&&\hspace{1.4 cm} +\, \la^{\phi \varphi^'}_4(\phi^\+ \varphi^')_{1_{+2}} (\varphi^{'\+} \phi)_{1_{+1}} +\la^{\phi \varphi^'}_5(\phi^\+ \varphi^')_{1_{-2}} (\varphi^{'\+} \phi)_{1_{-1}}+\la^{\phi \varphi^'}_6 (\phi^\+ \varphi^')_{2_{10}} (\varphi^{'\+} \phi)_{2_{20}}, \label{Vphivarphip}\\
&&\mathcal{V}(\phi, \chi) =\Vien{\mathcal{V}(\phi, \phi^'\rightarrow \chi),}\hs \mathcal{V}(\phi, \rho)\Vien{=\mathcal{V}(H\rightarrow \phi, \varphi^'\rightarrow\rho),}\hs \mathcal{V}(\phi^', \varphi)\Vien{=\mathcal{V}(\phi\rightarrow \phi^', \varphi), }\\ 
&&\mathcal{V}(\phi^', \varphi^')\Vien{=\mathcal{V}(\phi\rightarrow\phi^', \varphi^'), }
\hs \mathcal{V}(\phi^', \chi)\Vien{=\mathcal{V}(\phi\rightarrow \chi, \phi^'),}
\hs \mathcal{V}(\phi^', \rho)\Vien{=\mathcal{V}(H\rightarrow\phi^', \varphi^'\rightarrow\rho), }
\label{Vphiprho} \\
&&\mathcal{V}(\varphi, \varphi^')\Vien{=\mathcal{V}(\phi\rightarrow\varphi, \varphi\rightarrow\varphi^'), }
\hspace{0.03cm} \mathcal{V}(\varphi, \chi)\Vien{=\mathcal{V}(H\rightarrow\varphi, \varphi^'\rightarrow\chi), }
\hspace{0.03cm} \mathcal{V}(\varphi, \rho)\Vien{=\mathcal{V}(H\rightarrow \varphi, \varphi\rightarrow\rho), \hspace{0.2cm}}
\\
&&\mathcal{V}(\varphi^', \chi)\Vien{=\mathcal{V}(\phi\rightarrow\chi, \varphi^'),}
\hs \mathcal{V}(\varphi^', \rho)\Vien{=\mathcal{V}(H\rightarrow\varphi^', \chi\rightarrow\rho),} 
\hs \mathcal{V}(\chi, \rho)\Vien{=\mathcal{V}(H\rightarrow\chi, \varphi^'\rightarrow\rho) ,} 
\\
&&\mathcal{V}_{\mathrm{three}}=\la^{H \phi \phi^'}_1 (H^\+\phi)_{2_{21}}(\phi^{'\+}\phi^')_{2_{21}}+\la^{H \phi \phi^'}_2 (\phi^\+H)_{2_{21}}(\phi^{'\+}\phi^')_{2_{21}}+\la^{H \phi \varphi^'}_1 (H^\+\phi)_{2_{21}}(\varphi^{'\+}\varphi^')_{2_{21}}\crn
&&\hspace{1.5 cm}+\la^{H \phi \varphi^'}_2 (\phi^\+H)_{2_{21}}(\varphi^{'\+}\varphi^')_{2_{21}}+ \la^{H \phi \chi}_1 (H^\+\phi)_{2_{21}}(\chi^{\+}\chi)_{2_{21}}+\la^{H \phi \chi}_2 (\phi^\+H)_{2_{21}}(\chi^{\+}\chi)_{2_{21}}, \label{Vthree}\\
&&V_{\mathrm{four}} =\lambda^{\varphi \varphi^' \chi \rho}_1 (\varphi^\+ \varphi^')_{\underline{2}_{20}} (\chi^\+ \rho)_{\underline{2}_{10}}+\lambda^{\varphi \varphi^' \chi \rho}_2 (\varphi^{'\+}\varphi)_{\underline{2}_{10}} (\rho^\+ \chi)_{\underline{2}_{20}}.  \label{Vfour}
\eea
It is noted that all the other renormalizable three-and four-scalar interactions 
are forbidden by one/some of the model symmetries.
\newpage
\section{\label{minimumeq} The potential minimum condition}
 \vspace{-1.4 cm}
\bea
\mu_H^2 v_H +\lambda^{H\phi\varphi}(v_1 + v_2) v_{\varphi^'}^2 +  2 \lambda^{H\phi} (v_1 + v_2)^2 v_H +
 \lambda^{H\phi\phi^'} (v_1 + v_2) v^{'2}\hspace{0.4 cm}&&\crn
 +\, 2 v_H (\lambda^{H\varphi} v_{\varphi}^2 + 2 \lambda^{H\varphi^'} v^2_{\varphi^{'}} + \lambda^{H\chi} v_{\chi}^2 + \lambda^{H} v_H^2 +
    4 \lambda^{H\phi^'} v^{'2} + \lambda^{H\rho} v_\rho^2)=0, \hspace{0.4 cm}&& \label{eq1}\\
    \mu_\phi^2 v_1 + 2 \lambda^{\phi} v_1 (2 v_1^2 + v_2^2)
+ 2 \lambda^{\phi\varphi} v_1 v_{\varphi}^2 +
 3 \lambda^{\phi\varphi^{'}} v_1 v_{\varphi}^{'2} + 3 \lambda^{\phi\varphi^'} v_2 v^2_{\varphi^{'}}\hspace{0.4 cm}&& \crn
 +\, \lambda^{\phi\chi} v_1 v_\chi^2 +
 \lambda^{H\phi\varphi^'} v^2_{\varphi^{'}} v_H + 2 \lambda^{H\phi} v_1 v_H^2 + 2 \lambda^{H\phi} v_2 v_H^2 +
 5 \lambda^{\phi\phi^'} v_1 v^{'2} \hspace{0.4 cm}&&\crn
 +\, \lambda^{\phi\phi^'} v_2 v^{'2} + \lambda^{H\phi\phi^'} v_H v^{'2} +
 2 \lambda^{\phi\rho} v_1 v_\rho^2=0,\hspace{0.4 cm}&& \label{eq2} \\
\mu_\phi^2 v_2 + 2 \lambda^{\phi} v_2 (2 v_2^2 + v_1^2)
+ 2 \lambda^{\phi\varphi} v_2 v_{\varphi}^2 +
 3 \lambda^{\phi\varphi^{'}} v_1 v^2_{\varphi^{'}} + 3 \lambda^{\phi\varphi^'} v_2 v^2_{\varphi^{'}} \hspace{0.4 cm}&&\crn
 +\, \lambda^{\phi\chi} v_2 v_\chi^2 +
 \lambda^{H\phi\varphi^'} v^2_{\varphi^{'}} v_H + 2 \lambda^{H\phi} v_1 v_H^2 + 2 \lambda^{H\phi} v_2 v_H^2 +
 5 \lambda^{\phi\phi^'} v_2 v^{'2} \hspace{0.40 cm}&& \crn
 +\, \lambda^{\phi\phi^'} v_1 v^{'2} + \lambda^{H\phi\phi^'} v_H v^{'2} +
 2 \lambda^{\phi\rho} v_2 v_\rho^2=0,\hspace{0.4 cm}&& \label{eq3} \\
 2 \mu^2_{\phi^{'}} + \lambda^{\phi\phi^'} (5 v_1^2 + 2 v_1 v_2 + 5 v_2^2) + 4 \lambda^{\phi^'\varphi} v_\varphi^2 +
 12 \lambda^{\phi^'\varphi^'} v^2_{\varphi^{'}} + 5 \lambda^{\phi^'\chi} v_\chi^2 \hspace{0.40 cm}&&\crn
+\, 2 v_H \left[\lambda^{H\phi\phi^'} (v_1 + v_2) + 4 \lambda^{H\phi^'} v_H\right] + 12 \lambda^{\phi^'} v^{'2} +
 4 \lambda^{\phi^'\rho} v_\rho^2=0, \hspace{0.4 cm}&& \label{eq4} \\
 2 \mu_\varphi^2 v_\varphi+4 v_\varphi \left[\lambda^{\phi\varphi} (v_1^2 + v_2^2) + \lambda^{\varphi} v_\varphi^2 + 2 \lambda^{\varphi\varphi^'} v_\varphi^2 +
\lambda^{\varphi\chi} v_\chi^2 + \lambda^{H\varphi} v_H^2 + 2 \lambda^{\phi^'\varphi} v^{'2}\right]\hspace{0.4 cm}&&\crn
+\, 2 \lambda^{\varphi\varphi^'\chi\rho} v_{\varphi^{'}} v_\chi v_\rho + 4 \lambda^{\varphi\rho} v_\varphi v_\rho^2=0, \hspace{0.4 cm}&&\label{eq5} \\
2 \mu^2_{\varphi^{'}} v_{\varphi^'}+v_{\varphi^'} \left\{3 \lambda^{\phi\varphi^'} (v_1 + v_2)^2 + 4 \lambda^{\varphi\varphi^'} v_\varphi^2 + 12 \lambda^{\varphi^'} v^2_{\varphi^{'}} +
    3 \lambda^{\varphi^'\chi} v_\chi^2 + 12 \lambda^{\phi^'\varphi^'} v^{'2} \hspace{0.4 cm}\right. &&\crn
 +\,\left. 2 v_H \left[\lambda^{H\phi\varphi^'} (v_1 + v_2) + 2 \lambda^{H\varphi^'} v_H\right] \right\}
    +  \lambda^{\varphi\varphi^'\chi\rho}v_\varphi v_\chi v_\rho + 4 \lambda^{\varphi^'\rho} v_{\varphi^'} v_\rho^2=0, \hspace{0.4 cm}&&\label{eq6}\\
2 v_\chi \left[ \lambda^{\phi\chi} (v_1^2 + 4 v_2^2) + 2 \lambda^{\varphi\chi} v_\varphi^2 +
    3 \lambda^{\varphi^{'}\chi} v^2_{\varphi^{'}} + 4 \lambda^{\chi} v_\chi^2 + 2 \lambda^{H\chi} v_H^2 +
    5 \lambda^{\phi\phi^{'}\chi} v^{'2}\right]\hspace{0.4 cm} &&\crn
+\, 2 v_\chi\mu_\chi^2 +2 \lambda^{\varphi\varphi^{'}\chi\rho} v_\varphi v_{\varphi^'} v_\rho + 4 \lambda^{\chi\rho} v_\chi v_\rho^2=0, \hspace{0.4 cm}&& \label{eq7}\\
 2 v_\rho\left[\lambda^{\phi\rho} (v_1^2 + v_2^2) + \lambda^{\varphi\rho} v_\varphi^2 + 2 \lambda^{\varphi^'\rho} v^2_{\varphi^{'}} +
\lambda^{\chi\rho} v_\chi^2 + \lambda^{H\rho} v_H^2 + 2 \lambda^{\phi\phi^'\rho} v^{'2} + \lambda^{\rho} v_\rho^2\right]\hspace{0.4 cm}&&\crn
+\,v_\rho \mu_\rho^2+\lambda^{\varphi\varphi^'\chi\rho} v_\varphi v_{\varphi^'} v_\chi =0, \hspace{0.4 cm}&&\label{eq8} \\
\mu_H^2 + 2 \left[\lambda^{H\phi} (v_1 + v_2)^2 + \lambda^{H\varphi} v_\varphi^2 + 2 \lambda^{H\varphi^'} v^2_{\varphi^{'}}+
\lambda^{H\chi} v_\chi^2 + 3 \lambda^{H} v_H^2 + 4 \lambda^{H\phi^'} v^{'2} + \lambda^{H\rho} v_\rho^2\right] > 0, \hspace{0.4 cm}&& \label{ieq1sq}\\
\mu_\phi^2 + 2 \lambda^{\phi} (6 v_1^2 + v_2^2) + 2 \lambda^{\phi\varphi} v_\varphi^2 +
 3 \lambda^{\phi\varphi^'} v^2_{\varphi^{'}}+ \lambda^{\phi\chi} v_\chi^2 + 2 \lambda^{H\phi} v_H^2 +
 5 \lambda^{\phi\phi^'} v^{'2} + 2 \lambda^{\phi\rho} v_\rho^2 > 0, \hspace{0.45 cm}&& \label{ieq2sq} \\
\mu_\phi^2 + 2 \lambda^{\phi} (v_1^2 + 6v_2^2) + 2 \lambda^{\phi\varphi} v_\varphi^2 +
 3 \lambda^{\phi\varphi^'} v^2_{\varphi^{'}} + \lambda^{\phi\chi} v_\chi^2 + 2 \lambda^{H\phi} v_H^2 +
 5 \lambda^{\phi\phi^'} v^{'2} + 2 \lambda^{\phi\rho} v_\rho^2 > 0, \hspace{0.4 cm}&& \label{ieq3sq} \\
2 \mu^2_{\phi^'} + \lambda^{\phi\phi^'} (5 v_1^2 + 2 v_1 v_2 + 5 v_2^2) +
   4\lambda^{\phi^'\varphi} v_\varphi^2 + 12 \lambda^{\phi^'\varphi^'} v^2_{\varphi^{'}} + 5 \lambda^{\phi^'\chi} v_\chi^2 + 36 \lambda^{\phi^'} v^{'2} +
   4 \lambda^{\phi^'\rho} v_\rho^2\hspace{0.4 cm}&&\crn
  +\, 2 v_H \left[\lambda^{H\phi\phi^'} (v_1 + v_2) + 4 \lambda^{H\phi^'} v_H \right]  > 0, \hspace{0.4 cm}&& \label{ieq4sq} \\
\mu_\varphi^2 +
   2 \left[\lambda^{\phi\varphi} (v_1^2 + v_2^2) + 3 \lambda^{\varphi} v_\varphi^2 + 2 \lambda^{\varphi\varphi^'} v^2_{\varphi^{'}} +
\lambda^{\varphi\chi} v_\chi^2 + \lambda^{H\varphi} v_H^2 + 2 \lambda^{\phi^'\varphi} v_H^2 + \lambda^{\varphi\rho} v_\rho^2\right] > 0, \hspace{0.4 cm} && \label{ieq5sq} \\
2 \mu^2_{\varphi^'} + 3 \lambda^{\phi\varphi^'} (v_1 + v_2)^2 + 4 \lambda^{\varphi\varphi^'} v_\varphi^2 +
 36 \lambda^{\varphi^'} v^2_{\varphi^{'}} + 3 \lambda^{\varphi^'\chi} v_\chi^2 + 12 \lambda^{\phi\phi^'\varphi^'} v^{'2} +
 4 \lambda^{\varphi^'\rho} v_\rho^2 \hspace{0.4 cm}&&\crn
+ 2 v_H \left[\lambda^{H\phi\varphi^'} (v_1 + v_2) + 2 \lambda^{H\varphi^'} v_H\right]  > 0, \hspace{0.4 cm}&& \label{ieq6sq}\\
 \mu_\chi^2 + \lambda^{\phi\chi} (v_1^2 + 4 v_2^2) + 2 \lambda^{\varphi\chi} v_\varphi^2 +
   3 \lambda^{\varphi^'\chi} v^2_{\varphi^{'}} + 12 \lambda^{\chi} v_\chi^2 + 2 \lambda^{H\chi} v_H^2 +
   5 \lambda^{\phi\phi^'\chi} v^{'2} + 2 \lambda^{\chi\rho} v_\rho^2 > 0, \hspace{0.4 cm}&& \label{ieq7sq}\\
\mu_\rho^2 +
   2 \left[\lambda^{\phi\rho} (v_1^2 + v_2^2) + \lambda^{\varphi\rho} v_\varphi^2 + 2 \lambda^{\varphi^'\rho} v^2_{\varphi^{'}}+ \lambda^{\chi\rho} v_\chi^2 + \lambda^{H\rho} v_H^2 + 2 \lambda^{\phi\phi^'\rho} v^{'2} + 3 \lambda^{\rho} v_\rho^2\right] > 0. \hspace{0.4 cm}&& \label{ieq8sq}   \eea
 \vspace{-2.0 cm}
 \section{\label{betaexpression}The explicit expressions of $\beta_H, \beta_{\phi}, \beta_{\phi^'}, \beta_{\varphi}, \beta_{\varphi^'}, \beta_{\chi}, \beta_{\rho}$ and $\beta_{H\phi}$}
\vspace{-0.75 cm}
 \bea \beta_H&=& \left(\mu_\phi^2+2 \lambda^{\phi\varphi}v_\varphi^2\right)v_1 v_2 (v_1 - v_2) (v_1 + v_2)^2\crn
 &+& \lambda^{\phi\varphi^'}\left(6  v_1^5+9  v_1^4 v_2 + 3  v_1^3 v_2^2  - 3 v_1^2 v_2^3 -
 9 v_1 v_2^4  - 6 v_2^5 \right) v^2_{\varphi^{'}}  \crn
 &-&\left(\mu_H^2+2 \lambda^{H\varphi}+4\lambda^{H\varphi^'}+2\lambda^{H\chi}\right)\left(2  v_1^3 -v_1^2 v_2  + v_1 v_2^2 - 2  v_2^3 \right)v_H^2\crn
 &+&\lambda^{\phi\phi^'}\left(2  v_1^5  + 7 v_1^4 v_2  + 5 v_1^3 v_2^2 -
 5 v_1^2 v_2^3 - 7 v_1 v_2^4-
 2  v_2^5 \right)v^{'2}\crn
 &+& \lambda^{\phi\chi}\left(7  v_1^3  + 7  v_1^2 v_2  +
 2 v_1 v_2^2  + 2 v_2^3 \right) v_1 v_2 v_\chi^2
 - 8\lambda^{H\phi^'}\left(2  v_1^3-v_1^2 v_2+v_1 v_2^2-2  v_2^3 \right) v_H^2 v^{'2} \crn
 &+& 2 (v_1 - v_2) \left[\lambda^{\phi\rho} v_1 v_2 (v_1 + v_2)^2 -
    \lambda^{H\rho} (2 v_1^2 + v_1 v_2 + 2 v_2^2) v_H^2 \right] v_\rho^2,\label{betaH}\\
\beta_\phi&=&\left(\mu_\phi^2+ 2 \lambda^{\phi\varphi} v_\varphi^2+ 4 \lambda^{\phi\phi^'} v^{'2}+
 2 \lambda^{\phi\rho} v_\rho^2\right) (v_2-v_1)  +
 \lambda^{\phi\chi}\left(4  v_2 - v_1 \right) v_\chi^2,  \label{betaphi}\\
    \beta_{\phi^'}&=&2 \mu^2_{\phi^'} + \lambda^{\phi\phi^'} (5 v_1^2 + 2 v_1 v_2 + 5 v_2^2) + 4 \lambda^{\phi^'\varphi} v_\varphi^2 +
 12 \lambda^{\phi^'\varphi^'} v^2_{\varphi^{'}} \crn
 &+& 4 \lambda^{\phi^'\rho} v_\rho^2+5 \lambda^{\phi^'\chi} v_\chi^2 + 2 v_H \left[\lambda^{H\phi\phi^'} (v_1 + v_2) + 4 \lambda^{H\phi^'} v_H\right],  \label{betaphip}\\
    \beta_{\varphi}&=&v_\varphi\left\{\mu_\varphi^2  +
 2 \left[\lambda^{\phi\varphi} (v_1^2 + v_2^2) + 2 \lambda^{\varphi\varphi^'} v^2_{\varphi^{'}} + \lambda^{\varphi\chi} v_\chi^2 +
\lambda^{H\varphi} v_H^2 + 2 \lambda^{\phi^'\varphi} v^{'2}\right]\right\} \crn
&+& \lambda^{\varphi\varphi^' \chi\rho} v_\varphi v_\chi v_\rho +
 2 \lambda^{\varphi\rho} v_\varphi v_\rho^2,  \label{betavarphi}\\
    \beta_{\varphi^'}&=&4 \mu^2_{\varphi^'} v_{\varphi^'} + 2 \lambda^{\varphi\varphi^' \chi\rho} v_\varphi v_\chi v_\rho +
 8 \lambda^{\varphi^' \rho} v_{\varphi^'} v_\rho^2 + 2 v_{\varphi^'} \left\{3 \lambda^{\phi\varphi^'} (v_1 + v_2)^2  \right.\crn
 &+&\left.   4 \lambda^{\varphi\varphi^'} v_\varphi^2+ 3 \lambda^{\varphi^'\chi} v_\chi^2 + 2 v_H \left[\lambda^{H\phi\varphi^'} (v_1 + v_2) + 2 \lambda^{H\varphi^'} v_H \right] +
    12 \lambda^{\phi^'\varphi^'} v^{'2}\right\} ,  \label{betavarphip}\\
     \beta_{\chi}&=&v_\chi \left[\mu_\chi^2 + \lambda^{\phi\chi} (v_1^2 + 4 v_2^2) + 2 \lambda^{\varphi\chi} v_\varphi^2 +
    3 \lambda^{\varphi^'\chi} v^2_{\varphi^'} + 2 \lambda^{H\chi} v_H^2 + 5 \lambda^{\phi\phi^'\chi} v^{'2}\right] \crn
    &+& \lambda^{\varphi\varphi^'\chi\rho} v_\varphi v_{\varphi^'} v_\rho + 2 \lambda^{\chi\rho} v_\chi v_\rho^2,  \label{betachi}\\
    \beta_{\rho}&=&\left\{\mu_\rho^2 +
    2 \left[\lambda^{\phi\rho} (v_1^2 + v_2^2) + \lambda^{\varphi\rho} v_\varphi^2 + 2 \lambda^{\varphi^'\rho} v^2_{\varphi^'} +
       \lambda^{\chi\rho} v_\chi^2 + \lambda^{H\rho} v_H^2 + 2 \lambda^{\phi^'\rho} v^{'2}\right]\right\} v_\rho \crn
       &+&\lambda^{\varphi\varphi^'\chi\rho} v_\varphi v_{\varphi^'} v_\chi ,  \label{betarho}\\
    \beta_{H\phi}&=&(\mu_\phi^2+ 2 \lambda^{\phi\varphi} v_\varphi^2)v_1 v_2(v_2^2-v_1^2)  - 3\lambda^{\phi\varphi^'}\left(2  v_1^4 + v_1^3v_2 - v_1 v_2^3 - 2 v_2^4 \right) v^2_{\varphi^'} \crn
    &-& \lambda^{\phi\chi}(7  v_1^2 + 2 v_2^2) v_1 v_2 v_\chi^2 -\lambda^{H\phi\varphi^'}( 2  v_1^3 - v_1^2 v_2+ v_1 v_2^2 - 2 v_2^3) v^2_{\varphi^'} v_H  \crn
    &-& \lambda^{\phi\phi^'}(2 v_1^4 + 5 v_1^3 v_2 - 5 v_1 v_2^3 - 2 v_2^4)v^{'2}- \lambda^{H\phi\phi^'}(2  v_1^3- v_1^2 v_2+v_1 v_2^2-2  v_2^3 ) v_H v^{'2} \crn
    &+& 2 \lambda^{\phi\rho} v_1 v_2(v_2^2 -
 v_1^2)v_\rho^2. \label{betaHphi}
\eea


\begin{thebibliography}{99}

\bibitem{U1x1} T. Appelquist, B. A. Dobrescu, A. R. Hopper, Phys. Rev. D \textbf{68}, 035012 (2003) [arXiv: hep-ph/0212073].
\bibitem{U1x2} N. Okada and S. Okada, Phys. Rev. D \textbf{95}, 035025 (2017) [arXiv: 1611.02672 [hep-ph]].
\bibitem{U1x3} A Das, P.S. B. Dev and N. Okada, Phys. Lett. B \textbf{799}, 135052 (2019) [arXiv: 1906.04132 [hep-ph]].

\bibitem{U1X0} A. Davidson,  Phys. Rev. D \textbf{20}, 776 (1979).
\bibitem{U1X1} R. N. Mohapatra and R. E. Marshak, Phys. Rev. Lett. \textbf{44}, 1316 (1980).
\bibitem{U1X2} R. E. Marshak and R. N. Mohapatra, Phys. Lett. \textbf{91B}, 222 (1980).
\bibitem{U1X3} C. Wetterich, Nucl. Phys. B \textbf{187}, 343 (1981).
\bibitem{U1X4} A. Masiero, J. F. Nieves and T. Yanagida, Phys. Lett. \textbf{116 B}, 11 (1982).
\bibitem{U1X5} W. Buchmuller, C. Greub and P. Minkowski, Phys. Lett. B \textbf{267}, 395 (1991).
\bibitem{U1X6} S. Iso, N. Okada and Y. Orikasa, Phys. Lett. B \textbf{676}, 81 (2009) [arXiv: 0902.4050 [hep-ph]].
\bibitem{U1X7} S. Iso, N. Okada and Y. Orikasa, Phys. Rev. D \textbf{80}, 115007 (2009) [arXiv: 0909.0128 [hep-ph]].
\bibitem{U1X8} M . Abbas, S. Khalil, J. High Energy Phys. \textbf{04}, 056 (2008) [arXiv: 0707.0841 [hep-ph]].
\bibitem{U1X9} W. Emam, S. Khalil, Eur.Phys.J.C \textbf{55}, 625 (2007) [arXiv: 0704.1395 [hep-ph]].
\bibitem{U1X10} T. Basak and T. Mondal, Phys. Rev. D \textbf{89}, 063527 (2014) [arXiv: 1308.0023 [hep-ph]].
\bibitem{U1X11} N. Sahu and U. A. Yajnik, Phys. Lett. B \textbf{635}, 1116 (2006) [arXiv: hep-ph/0509285].
\bibitem{U1X12} W. Rodejohann and C. E. Yaguna, JCAP \textbf{1512}, 032 (2015) [arXiv: 1509.04036 [hep-ph]].
\bibitem{U1X13} J. Guo, Z. Kang, P. Ko, and Y. Orikasa, Phys. Rev. D \textbf{91}, 115017 (2015) [arXiv: 1502.0050 [hep-ph]].
\bibitem{U1X14} A. El-Zant, S. Khalil, and A. Sil, Phys. Rev. D \textbf{91}, 035030 (2015) [arXiv: 1308.0836 [hep-ph]].
\bibitem{U1X15} S. Khalil, H. Okada, Phys. Rev. D \textbf{79}, 083510 (2009) [arXiv: 0810.4573 [hep-ph]].
\bibitem{U1X16} S. Khalil, J.Phys.G \textbf{35}, 055001 (2008) [arXiv: hep-ph/0611205].

\bibitem{U1X17} T. Higaki, R. Kitano, R. Sato, J. High Energy Phys. \textbf{07}, 044 (2014) [arXiv: 1405.0013 [hep-ph]].



\bibitem{U1X18} F. F. Deppisch, W. Liu and M. Mitra, J. High Energy Phys. \textbf{1808}, 181 (2018) [arXiv: 1804.04075 [hep-ph]].
\bibitem{U1X19} P. S. B. Dev, R. N. Mohapatra, Y. Zhang, J. High Energy Phys. \textbf{03}, 122 (2018) [arXiv: 1711.07634 [hep-ph]].
\bibitem{U1X20} T. Hasegawa, N. Okada and O. Seto, Phys. Rev. D \textbf{99}, 095039 (2019) [arXiv: 1904.03020 [hep-ph]].

\bibitem{S31} K. Kang, S. K. Kang, J. E. Kim, and P. Ko, Mod. Phys. Lett. A {\bf 12}, 1175 (1997) [arXiv: hep-ph/9611369].
 \bibitem{S32} M. Fukugita, M. Tanimoto, and T. Yanagida, Phys. Rev. D {\bf 57}, 4429 (1998) [arXiv: hep-ph/9709388].
 \bibitem{S33} M. Fujii, K. Hamaguchi and T. Yanagida, Phys. Rev. D {\bf 65}, 115012 (2002) [arXiv: hep-ph/0202210].
 \bibitem{S34} S.-L. Chen, M. Frigerio, and E. Ma, Phys. Rev. D {\bf 70}, 073008 (2004) [arXiv: hep-ph/0404084].
 \bibitem{S35} J. E. Kim and J.-C. Park, JHEP {\bf 0605}, 017 (2006) [arXiv: hep-ph/0512130].
 \bibitem{S36} Y. Koide, Eur. Phys. J. C {\bf 50}, 809 (2007) [arXiv: hep-ph/0612058].
\bibitem{S37} D. Meloni, S. Morisi and E.  Peinado, J. Phys. G {\bf 38}, 015003 (2011) [arXiv:1005.3482 [hep-ph]].
\bibitem{S3DLNV} P. V. Dong, H. N. Long, C. H. Nam, and V. V. Vien, Phys. Rev. D \textbf{85}, 053001 (2012) [arXiv: 1111.6360 [hep-ph]].
\bibitem{S3VL} V. V. Vien and H. N. Long,  J. Exp. Theor. Phys. \textbf{118}, 869 (2014) [arXiv: 1404.6119 [hep-ph]].

\bibitem{Hernandez:2013hea1} A.~E.~C\'arcamo Hern\'andez, R.~Martinez and F.~Ochoa,
Eur. Phys. J. C \textbf{76}, 634 (2016) 
[arXiv:1309.6567 [hep-ph]].
\bibitem{Hernandez:2013hea2} A.~E.~C\'arcamo Hern\'andez, E.~Cata\~no Mur and R.~Martinez,
Phys. Rev. D \textbf{90}, 073001 (2014) 
[arXiv:1407.5217 [hep-ph]].
\bibitem{Hernandez:2014vta} A.~E.~C\'arcamo Hern\'andez, R.~Martinez and J.~Nisperuza,
Eur. Phys. J. C \textbf{75}, 72 (2015) 
[arXiv: 1401.0937 [hep-ph]].
\bibitem{Hernandez:2015zeh} A.~E.~C\'arcamo Hern\'andez, I.~de Medeiros Varzielas and N.~A.~Neill,
Phys. Rev. D \textbf{94}, 033011 (2016) 
[arXiv:1511.07420 [hep-ph]].
\bibitem{Hernandez:2015hrt} A.~E.~C\'arcamo Hern\'andez,
Eur. Phys. J. C \textbf{76}, 503 (2016) 
[arXiv:1512.09092 [hep-ph]].
\bibitem{CarcamoHernandez:2016pdu} A.~E.~C\'arcamo Hern\'andez, S.~Kovalenko and I.~Schmidt,
J. High Energy Phys. \textbf{02}, 125 (2017) 
[arXiv:1611.09797 [hep-ph]].

\bibitem{Hernandez:2015dga} A.~E.~C\'arcamo Hern\'andez, I.~de Medeiros Varzielas and E.~Schumacher,
Phys. Rev. D \textbf{93}, 016003 (2016) 
[arXiv:1509.02083 [hep-ph]].

\bibitem{Arbelaez:2016mhg} C.~Arbel\'aez, A.~E.~C\'arcamo Hern\'andez, S.~Kovalenko and I.~Schmidt,
Eur. Phys. J. C \textbf{77}, 422 (2017) 
[arXiv:1602.03607 [hep-ph]].


\bibitem{CarcamoHernandez:2018vdj} A.~E.~C\'arcamo Hern\'andez, J.~Vignatti and A.~Zerwekh,
J. Phys. G \textbf{46}, 115007 (2019) 
[arXiv: 1807.05321 [hep-ph]].

\bibitem{Garces:2018nar} E.~A.~Garc\'es, J.~C.~G\'omez-Izquierdo and F.~Gonzalez-Canales,
Eur. Phys. J. C \textbf{78}, 812 (2018) 
[arXiv:1807.02727 [hep-ph]].

\bibitem{Pramanick:2019oxb} S.~Pramanick,
Phys. Rev. D \textbf{100}, 035009 (2019) 
[arXiv:1904.07558 [hep-ph]].

\bibitem{CarcamoHernandez:2020pxw} A.~E.~C\'arcamo Hern\'andez, Y.~Hidalgo Vel\'asquez, S.~Kovalenko, H.~N.~Long, N.~A.~P\'erez-Julve and
    V.~V.~Vien, \Revised{ Eur. Phys. J.  \textbf{C 81},  No. 2 (2021) 191,}  arXiv:2002.07347[hep-ph]. 


\bibitem{Garcia-Aguilar:2020vsy} J.~D.~Garc\'\i{}a-Aguilar and J.~C.~G\'omez-Izquierdo,
arXiv: 2010.15370 [hep-ph].


\bibitem{Tp2} M.-C. Chen and K. T. Mahanthappa, Phys. Lett. B \textbf{652}, 34 (2007) [arXiv: 0705.0714 [hep-ph]].
\bibitem{Tp3} G.-J. Ding, Phys. Rev. D \textbf{78}, 036011 (2008) [arXiv: 0803.2278 [hep-ph]].
\bibitem{Tp4} D. A. Eby, P. H. Frampton, X. -G. He and T. W. Kephart, Phys. Rev. D 84, 037302 (2011) [arXiv: 1103.5737 [hep-ph]].
\bibitem{Tp5} P. H. Frampton, C. M. Ho, T. W. Kephart, Phys. Rev. D \textbf{89} (2014) 027701 [arXiv: 1305.4402 [hep-ph]].
\bibitem{CarcamoHernandez:2019vih} A.~E.~C\'arcamo Hern\'andez, Y.~Hidalgo Vel\'asquez and N.~A.~P\'erez-Julve,
Eur. Phys. J. C \textbf{79}, 828 (2019) 
[arXiv:1905.02323 [hep-ph]].

	
\bibitem{Tpvla2019} V. V. Vien, H. N. Long, A. E. Cárcamo Hernández, Mod. Phys. Lett. A \textbf{34}, 1950005 (2019) [arXiv: 1812.07263
    [hep-ph]].


\bibitem{D41} W. Grimus, A.S. Joshipura, S. Kaneko, L. Lavoura, M. Tanimoto, J. High Energy Phys. \textbf{0407}, 078 (2004) [arXiv:
    hep-ph/0407112].
 \bibitem{D42} H. Ishimori \emph{et al.}, Phys. Lett. B \textbf{662}, 178 (2008) [arXiv: 0802.2310 [hep-ph]].
 \bibitem{D43} A. Adulpravitchai, A. Blum, C. Hagedorn, J. High Energy Phys. \textbf{0903}, 046 (2009) [arXiv: 0812.3799 [hep-ph]].
\bibitem{D4VL} V. V. Vien and H. N. Long, Int. J. Mod. Phys. A, \textbf{28}, 1350159 (2013) [arXiv: 1312.5034 [hep-ph]].
\bibitem{D4V} V. V. Vien, Mod. Phys. Lett. A \textbf{29}, 1450122 (2014).
\bibitem{D4VLq} V. V. Vien and H. N. Long, J. Korean Phys. Soc. \textbf{66}, 1809 (2015) [arXiv: 1408.4333 [hep-ph]].
\bibitem{CarcamoHernandez:2020ney} A.~E.~C\'arcamo Hern\'andez, C.~O.~Dib and U.~J.~Salda\~na-Salazar,
Phys. Lett. B \textbf{809}, 135750 (2020) 
[arXiv: 2001.07140 [hep-ph]].


  \bibitem{Q61} K. S. Babu and J. Kubo, 
Phys.Rev. D \textbf{71}, 056006 (2005)  [arXiv: hep-ph/0411226 [hep-ph]].

\bibitem{Q62} Y. Kajiyama, E. Itou and J. Kubo, 
    decay,
Nucl. Phys. B \textbf{743}, 74 (2006) [arXiv: hep-ph/0511268].
\bibitem{Q63} Y. Kajiyama, 
J. High Eenergy Phys. \textbf{04}, 007 (2007) [arXiv: hep-ph/0702056].
\bibitem{Q64} N. Kifune, J. Kubo and A. Lenz, 
    Symmetry,
Phys.Rev. D \textbf{77}, 076010 (2008) [arXiv: 0712.0503 [hep-ph]].
\bibitem{Q65} K. Babu and Y. Meng, 
Phys.Rev. D \textbf{80}, 075003 (2009) [arXiv: 0907.4231 [hep-ph]].
\bibitem{Q66} K. Kawashima, J. Kubo and A. Lenz, 
Phys.Lett. B \textbf{681}, 60 (2009) [arXiv: 0907.2302 [hep-ph]].
\bibitem{Q67} K. Babu, K. Kawashima and J. Kubo, 
Phys.Rev. D \textbf{83}, 095008 (2011) [arXiv:1103.1664 [hep-ph]].
\bibitem{Q68} J. C. Gmez-Izquierdo, F. Gonzalez-Canales and M. Mondragn, Int. J. Mod. Phys. A \textbf{32}, 1750171 (2017) [arXiv: 1705.06324
    [hep-ph]].
\bibitem{Q69} T. Araki and Y. F. Li, Phys. Rev. D \textbf{85}, 065016 (2012) [arXiv: 1112.5819 [hep-ph]].
\bibitem{A401} F. Feruglio, C. Hagedorn, Y. Lin and L. Merlo, Nucl. Phys. B \textbf{775}, 120 (2007) [arXiv: hep-ph/0610165].
\bibitem{A41} K. S. Babu, E. Ma and J. W. F. Valle, Phys. Lett. B \textbf{552}, 207 (2003) [arXiv: hep-ph/0206292].
\bibitem{A42} G. Altarelli and F. Feruglio, Nucl. Phys. B \textbf{720}, 64 (2005) [arXiv: hep-ph/0504165].
\bibitem{A43} E. Ma, Phys. Rev. D {\bf73}, 057304 (2006) [arXiv: hep-ph/0601225].
\bibitem{A44} X. G. He, Y. Y. Keum and R. R. Volkas, J. High Energy Phys. {\bf 0604}, 039 (2006) [arXiv: hep-ph/0601001].
\bibitem{A45} S. Morisi, M. Picariello, and E. Torrente-Lujan, Phys. Rev. D {\bf75}, 075015 (2007) [arXiv: hep-ph/0702034].
\bibitem{A46} F. Bazzocchi, S. Kaneko and S. Morisi, J. High Energy Phys. \textbf{0803}, 063 (2008) [arXiv: 0707.3032 [hep-ph]].
\bibitem{A47} F. Bazzocchi, M. Frigerio, and S. Morisi, Phys. Rev. D {\bf78}, 116018 (2008) [arXiv: 0809.3573 [hep-ph]].
\bibitem{A48} G. Altarelli, F. Feruglio and C. Hagedorn, J. High Energy Phys.\textbf{0803}, 052 (2008) [arXiv: 0802.0090 [hep-ph]].
\bibitem{A49} M. Hirsch, S. Morisi and J. W. F. Valle, Phys. Rev. D {\bf 78}, 093007 (2008) [arXiv: 0804.1521 [hep-ph]].
\bibitem{A410} E. Ma, Phys. Lett. B {\bf671}, 366 (2009) [arXiv: 0808.1729 [hep-ph]].
\bibitem{A411} G. Altarelli and D. Meloni, J. Phys. G {\bf 36}, 085005 (2009) [arXiv: 0905.0620 [hep-ph]].
\bibitem{A412} Y. Lin, Nucl. Phys. B {\bf 813}, 91 (2009) [arXiv: 0804.2867 [hep-ph]].
\bibitem{A413} Y. H. Ahn and C. S. Chen, Phys. Rev. D {\bf 81}, 105013 (2010) [arXiv: 1001.2869 [hep-ph]].
\bibitem{A414} J. Barry and W. Rodejohann, Phys. Rev. D {\bf 81}, 093002 (2010);  Erratum: Phys.Rev. D \textbf{81} 119901 (2010)  [arXiv:
    1003.2385 [hep-ph]].
\bibitem{A4Dong2010} P. V. Dong, L. T. Hue, H. N. Long and D. V. Soa, Phys. Rev. D \textbf{81}, 053004 (2010) [arXiv: 1001.4625 [hep-ph]].
\bibitem{A415} G. J. Ding and D. Meloni, Nucl. Phys. B {\bf 855}, 21 (2012) [arXiv: 1108.2733 [hep-ph]].
\bibitem{Ishimori12b} H. Ishimori \emph{et al.,} Prog. Theor. Phys. Suppl. \textbf{183}, 1 (2010)  [arXiv:1003.3552 [hep-th]].
\bibitem{A4VL2015} V. V. Vien, H. N. Long, Int. J. Mod. Phys. A \textbf{30}, 1550117 (2015) [arXiv: 1405.4665 [hep-ph]].
\bibitem{A4Hue} T. Phong Nguyen, L. T. Hue, D. T. Si and T. T. Thuc, Prog. Theor. Exp. Phys. \textbf{2020}, 033B04 (2020) [arXiv:1711.05588
    [hep-ph]].
\bibitem{Q8Dev2010} S. Dev and S. Verma, Mod. Phys. Lett. A \textbf{25}, 2837 (2010) [arXiv: 1005.4521 [hep-ph]].
\bibitem{S3BLGM2019} J. C. G$\mathrm{\acute{o}}$mez-Izquierdo and M. Mondrag$\mathrm{\acute{o}}$n, Eur. Phys. J. C \textbf{79}, 285
    (2019).
\bibitem{D4BL2020} V. V. Vien, J. Phys. G \textbf{47}, 055007 (2020).
\bibitem{Q6BL2020} V. V. Vien, Nucl. Phys. B \textbf{956}, 115015 (2020).

\bibitem{Ishi} H. Ishimori \emph{et. al.,} Prog. Theor. Phys. Suppl. \textbf{183}, 1 (2010) [arXiv: 1003.3552 [hep-th]].
\bibitem{HeU1X} X.-G. He, Y.-Y. Keum and R. R. Volkas, J. High Energy Phys. \textbf{0604}, 039 (2006) [arXiv: 0601001 [hep-ph]].



\bibitem{AME07} A. Mondrag$\acute{o}$n, M. Mondrag$\acute{o}$n, and E. Peinado, Phys.Rev.D \textbf{76}, 076003 (2007) [arXiv: 0706.0354
    [hep-ph]].
\bibitem{PJ08} P. M. Ferreira, Joao P. Silva, Phys.Rev.D \textbf{78}, 116007 (2008) [arXiv: 0809.2788 [hep-ph]].
\bibitem{Kubo13} J. Kubo, Fortsch.Phys. \textbf{61}, 597 (2013) [arXiv: 1210.7046 [hep-ph]].
\bibitem{PPich17} A. Pe$\mathrm{\tilde{n}}$uelas, A. Pich, J. High Energy Phys. \textbf{12}, 084 (2017) [arXiv: 1710.02040 [hep-ph]].
\bibitem{Bento18} Miguel P. Bento, Howard E. Haber, J. C. Rom$\mathrm{\tilde{a}}$o, Jo$\mathrm{\tilde{a}}$o P. Silva, J. High Energy Phys.
    \textbf{10}, 143 (2018) [arXiv: 1808.07123 [hep-ph]].
\bibitem{Diaz19} M. A. Arroyo-Ure$\tilde{\mathrm{n}}$a, J. Lorenzo D\'iaz-Cruz, Bryan O. Larios-L\'opez and M. A. P\'erez-de Le\'on, Chin.
    Phys. C \textbf{45}, 023118 (2021) 
[arXiv: 1901.01304 [hep-ph]].
\bibitem{WU19} W. Rodejohann, U. Salda$\mathrm{\tilde{n}}$a-Salazar, J. High Energy Phys. \textbf{07}, 036 (2019) [arXiv: 1903.00983
    [hep-ph]].

\bibitem{PDG2020} P.A. Zyla \emph{et al.} (Particle Data Group), Prog. Theor. Exp. Phys. \textbf{2020}, 083C01 (2020).


\bibitem{nubound} S. Roy Choudhury and S. Choubey, JCAP \textbf{1809}, 017 (2018) [arXiv: 1806.10832 [astro-ph.CO]].



\bibitem{KamLAND} A. Gando \emph{et al}. (KamLAND-Zen Collaboration), Phys. Rev. Lett. \textbf{117}, 082503 (2016) [arXiv: 1605.02889
    [hep-ph]].
\bibitem{Agostini18} M. Agostini \emph{et al.} (GERDA Collaboration), Phys. Rev. Lett. \textbf{120},132503 (2018) [arXiv:1803.11100 [nucl-ex]].
\bibitem{MAJO} C. E. Aalseth \emph{et al.} (Majorana Collaboration), Phys. Rev. Lett. \textbf{120}, 132502 (2018) [arXiv: 1710.11608 [nucl-ex]].
\bibitem{EXO1} M. Auger \emph{et al}. (EXO-200 collaboration), JINST \textbf{7}, P05010 (2012).
\bibitem{EXO2} J. B. Albert \emph{et al}. (EXO-200 collaboration), Nature \textbf{510}, 229 (2014).
\bibitem{Albert18} J. B. Albert \emph{et al.} (EXO-200 Collaboration), Phys. Rev. Lett. \textbf{120}, 072701 (2018) [arXiv: 1707.08707 [hep-ex]].
\bibitem{CUORE1} C. Alduino \emph{et al.}(CUORE collaboration), JINST \textbf{11}, P07009 (2016).
\bibitem{CUORE2} C. Alduino \emph{et al.}(CUORE collaboration), Phys. Rev. Lett. \textbf{120},132501 (2018) [arXiv: 1710.07988 [nucl-ex]].
\bibitem{Xing:2019vks} Z.~z.~Xing,
  Phys.\ Rept.\  {\bf 854}, 1 (2020) 
  [arXiv: 1909.09610 [hep-ph]].
\bibitem{Tanabashi:2018oca} M.~Tanabashi {\it et al.} [Particle Data Group],
  Phys.\ Rev.\ D {\bf 98}, 030001 (2018).

\bibitem{Dedes:2002er}
A.~Dedes and A.~Pilaftsis,
Phys. Rev. D \textbf{67}, 015012 (2003)
doi:10.1103/PhysRevD.67.015012
[arXiv:hep-ph/0209306 [hep-ph]].


\bibitem{Aranda:2012bv}
A.~Aranda, C.~Bonilla and J.~L.~Diaz-Cruz,
Phys. Lett. B \textbf{717}, 248-251 (2012)
doi:10.1016/j.physletb.2012.09.011
[arXiv:1204.5558 [hep-ph]].


\bibitem{Lindner:2016bgg}
M.~Lindner, M.~Platscher and F.~S.~Queiroz,
Phys. Rept. \textbf{731}, 1-82 (2018)
doi:10.1016/j.physrep.2017.12.001
[arXiv:1610.06587 [hep-ph]].

\bibitem{Ivan2019}  I. Esteban \emph{et al.,} J. High Energy Phys. \textbf{01}, 106 (2019)  [arXiv: 1811.05487 [hep-ph]].

\end{thebibliography}
\end{document}